\documentclass[fleqn,usenatbib]{mnras}

\usepackage[T1]{fontenc}
\usepackage{ae,aecompl}

\usepackage{graphicx}	% Including figure files
\usepackage{amsmath}	% Advanced maths commands

\usepackage{amssymb}	% Extra maths symbols
\usepackage{mwe}
\usepackage{todonotes}
\usepackage{xcolor}
\usepackage{hyperref}

\usepackage{siunitx}
\usepackage{enumitem}
\usepackage{lscape}
          \usepackage{pdflscape}
          \usepackage{morefloats}
			\usepackage{booktabs} 
              \usepackage{multirow}
              \usepackage{array}
              \usepackage{longtable}
               \usepackage{tabu}
                \usepackage{threeparttable} %table with notes
                 \usepackage{threeparttablex} %longtable with notes
                 \usepackage{rotating}
                  \usepackage{xtab}
                   \usepackage{multicol}
                   \usepackage{adjustbox}
                    
\usepackage{upgreek}
\usepackage{units}
\usepackage{caption}

\newcommand{\kms}{km~s\ensuremath{^{-1}}}
\newcommand{\msun}{$M_{\odot}$}

\newcommand{\angstrom}{\mbox{\normalfont\AA}}

\newenvironment{dedication}
    {\vspace{6ex}\begin{quotation}\begin{center}\begin{em}\begin{large}}
    {\par\end{large}\end{em}\end{center}\end{quotation}}

\DeclareRobustCommand{\ion}[2]{%
\relax\ifmmode
\ifx\testbx\f@series
{\mathbf{#1\,\mathsc{#2}}}\else
{\mathrm{#1\,\mathsc{#2}}}\fi
\else\textup{#1\,{\mdseries\textsc{#2}}}%
\fi}
\title {Centrally concentrated molecular gas driving galactic-scale ionised gas outflows in star-forming galaxies}

\author[L. M. Hogarth et al.]{\Large \parbox{\textwidth}{
L. M. Hogarth$^{1}$\thanks{E-mail: lucy.hogarth.18@ucl.ac.uk},
A. Saintonge$^{1}$\thanks{E-mail: a.saintonge@ucl.ac.uk}, 
L. Cortese$^{2,3}$,
T. A. Davis$^{4}$, 
S. M. Croom$^{3,5}$,
J. Bland-Hawthorn$^{3,5}$,
S. Brough$^{3,8}$,
J. J. Bryant$^{3,5,6}$,
B. Catinella$^{2,3}$,
T. J. Fletcher$^{1}$,
B. Groves$^{2,3,12}$
J. S. Lawrence$^{11}$,
\'A. R. L\'opez-S\'anchez$^{3,6,9}$,
M. S. Owers$^{9,10}$,
S. N. Richards$^{7}$,
G. W. Roberts-Borsani$^{13,1}$,
E. N. Taylor$^{14}$,
J. van de Sande$^{3,5}$,
N. Scott$^{3,5}$}
\\
\\
$^{1}$University College London, Department of Physics and Astronomy, Gower Street, London, WC1E 6BT, UK \\
$^{2}$International Center for Radio Astronomy Research, University of Western Australia, 35 Stirling Highway, Crawley, WA, 6009, Australia \\
$^{3}$ARC Centre of Excellence for All Sky Astrophysics in 3 Dimensions (ASTRO 3D) \\
$^{4}$School of Physics and Astronomy, Cardiff University, Queens Buildings, The Parade, Cardiff CF24 3AA, UK\\
$^{5}$Sydney Institute for Astronomy University of Sydney NSW 2006, Australia\\
$^{6}$Australian Astronomical Optics, AAO-USydney, School of Physics, University of Sydney, NSW 2006, Australia\\
$^{7}$SOFIA Science Center, USRA, NASA Ames Research Center, Building N232, M/S 232-12, P.O. Box 1, Moffett Field, CA 94035-0001, USA\\
$^{8}$School of Physics, University of New South Wales, NSW 2052, Australia\\
$^{9}$Department of Physics and Astronomy, Macquarie University, NSW 2109, Australia\\
$^{10}$Astronomy, Astrophysics and Astrophotonics Research Centre, Macquarie University, Sydney, NSW 2109, Australia\\
$^{11}$Australian Astronomical Optics - Macquarie, Macquarie University, NSW 2109, Australia\\
$^{12}$Research School of Astronomy \& Astrophysics, Australian National University, Weston Creek, ACT, 2611, Australia\\
$^{13}$Department of Physics \& Astronomy, University of California, Los Angeles, 475 Portola Plaza, CA 90095, USA\\
$^{14}$Centre for Astrophysics and Supercomputing, Swinburne University of Technology, Hawthorn, VIC 3122, Australia}

\date{Accepted XXX. Received YYY; in original form ZZZ}

\pubyear{2020}

\begin{document}
\label{firstpage}
\pagerange{\pageref{firstpage}--\pageref{lastpage}}
\maketitle

% Abstract of the paper
\begin{abstract}
We perform a joint-analysis of high spatial resolution molecular gas and star-formation rate (SFR) maps in main-sequence star-forming galaxies experiencing galactic-scale outflows of ionised gas. Our aim is to understand the mechanism that determines which galaxies are able to launch these intense winds. We observed CO(1\textrightarrow0) at 1\arcsec\ resolution with ALMA in 16 edge-on galaxies, which also have 2\arcsec\ spatial resolution optical integral field observations from the SAMI Galaxy Survey. Half the galaxies in the sample were previously identified as harbouring intense and large-scale outflows  of ionised gas (``outflow-types''), the rest serve as control galaxies. The dataset is complemented by integrated CO(1\textrightarrow0) observations from the IRAM 30-m telescope to probe the total molecular gas reservoirs. We find that the galaxies powering outflows do not possess significantly different global gas fractions or star-formation efficiencies when compared with a control sample. However, the ALMA maps reveal that the molecular gas in the outflow-type galaxies is distributed more centrally than in the control galaxies. For our outflow-type objects, molecular gas and star-formation is largely confined within their inner effective radius ($\rm r_{eff}$), whereas in the control sample the distribution is more diffuse, extending far beyond $\rm r_{eff}$. We infer that outflows in normal star-forming galaxies may be caused by dynamical mechanisms that drive molecular gas into their central regions, which can result in locally-enhanced gas surface density and star-formation.
\end{abstract}


%%%%%%%%%%%%%%%%%%%%%%%%%%%%%%%%%%%%%%%%%%%%%%%%%%

%%%%%%%%%%%%%%%%% BODY OF PAPER %%%%%%%%%%%%%%%%%%

\section{Introduction}
\label{sec: intro}

Theoretical models of galaxy evolution and numerical simulations rely on intense, galactic-scale outflows in order to regulate star-formation and produce galaxies that match observations, for example the sizes of galactic disks and bulges and the slope of the mass-metallicity relation \citep[e.g.][]{guedes11, dave11}. In these models, the intensity of stellar feedback is assumed, fine-tuned or left as an unconstrained parameter. Current simulations, therefore, either make very specific predictions as to how mass outflow rates scale with galaxy properties, or require observational input to assist in the fine-tuning of the parameters. Either way, strong constraints derived from observations of outflows in galaxies of varying masses, star-formation activity and redshift are a vital ingredient. 

\par Outflows appear to be common at all redshifts studied so far, and in particular for galaxies with extreme star-formation activity or powerful active galactic nuclei \citep[AGN, e.g. ][]{veilleux05,cicone14}. For galaxies with more ``normal" levels of star-formation activity, there is growing evidence that outflows of ionised and neutral gas are common as long as certain conditions are met. In particular, detailed studies based on Sloan Digital Sky Survey \citep[SDSS, ][]{york00, alam15}, galaxies suggest a positive correlation between the strength of outflows and quantities such as stellar mass and star-formation rate surface density \citep[$\Sigma _{\rm SFR}$, ][]{chen10}. In particular, a critical threshold of $\Sigma _{\rm SFR} \approx 0.01$\ \msun~yr$^{-1}$~kpc$^{-2}$ is often reported as being necessary for an outflow to be launched \citep{heckman03}. This $\Sigma _{\rm SFR}$ threshold has also been found to hold on resolved scales \citep[see ][]{newman12, davies19, roberts-borsani20}. Using the NaD line as a tracer of cool metal-enriched gas, \citet{roberts-borsani19} have shown that outflows are systematic in massive galaxies ($\rm M_{*}>10^{10}$\msun), as long as they have $\Sigma _{\rm SFR}>0.01$\ \msun~yr$^{-1}$~kpc$^{-2}$ and disk inclinations lower than $50\deg$ (the latter for geometric rather than physical reasons). 

These results leave us with two important follow-on questions: 

\begin{enumerate}[leftmargin=0.25in]

\item Why is it that some galaxies can reach this $\Sigma _{\rm SFR}$ threshold and launch winds while other similar galaxies (i.e. matched in key quantities such as inclination, mass, total SFR) do not? 

\item Are these ionised and neutral gas winds energetic enough to also affect the cold star-forming interstellar medium (ISM) and thus satisfy the requirements set out by numerical simulations? 

\end{enumerate}

We have designed an observational programme to address both of these questions, with the key objectives of targeting normal star-forming galaxies and combining observations of both ionised and molecular gas. This was achieved by following-up galaxies from the SAMI optical integral field survey \citep{croom12, bryant15} with IRAM 30-m and ALMA observations, using the CO(1\textrightarrow0) emission line as a molecular gas tracer. The details of the sample selection and observations are given in Section~\ref{sec: selectionanddata}. In this paper, we focus on the first of the two key questions described above, namely we use the molecular gas observations to investigate what leads to some, but not all, star-forming galaxies being able to launch large-scale ionised gas winds. These results are presented in Section~\ref{sec: results} and discussed in Section~\ref{sec: discussion}. In a subsequent paper, we will address the question of whether the kind of feedback detected via ionised gas outflows is efficient enough to regulate star-formation by affecting the cold ISM and/or driving molecular gas outflows. 

Throughout this paper we adopt a standard $\Lambda$CDM cosmology with $H_0=70\ $\kms~Mpc$^{-1}$, $\Omega _{m_0}=0.3$, $\Omega _{\Lambda}=0.7$ and a \citet{chabrier03} Initial Mass Function (IMF.) 

\begin{figure*}
    \centering
    \includegraphics[scale=0.38]{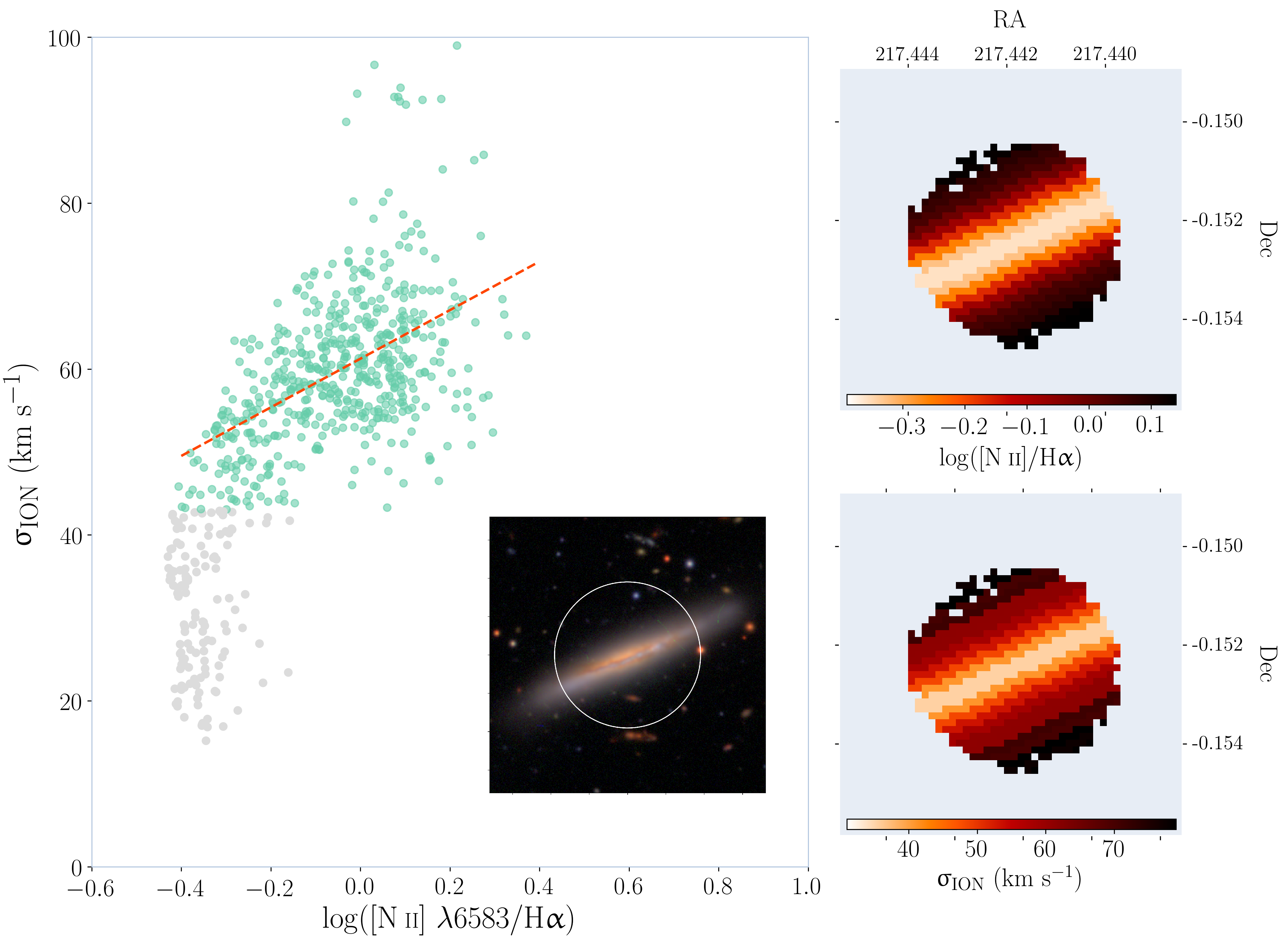}
    \caption{Diagnostic plots used to reveal large-scale outflows in SAMI galaxies. \textbf{Left}: Diagnostic performed on face-on and edge-on galaxies. For galaxies harbouring galactic-scale outflows, we expect a positive correlation between the velocity dispersion of the ionised gas and diagnostic shock ratios for high velocity dispersions. The velocity dispersion map is derived from the simultaneous fitting of the seven strong optical emission lines within the SAMI wavelength range \citep[see][]{scott18}. Spaxels with high velocity dispersions that correlate with increasing [\ion{N}{ii}]$\ \lambda 6583$/H$\upalpha$ ratio confirm the presence of outflowing material driving shocks through the ISM. The velocity dispersion and [\ion{N}{ii}]/H$\upalpha$ ratio is plotted for each spaxel in GAMA593680 (green and grey points represent the high and low velocity dispersion values respectively). The inset plot illustrates the SAMI footprint on the HSC \citep[Hyper Suprime-Cam Subaru Strategic Program; ][]{aihara19} optical image of the object. \textbf{Right}: Diagnostic performed on edge-on galaxies. For edge-on galaxies, a further diagnostic can be performed by looking for increasing line ratios and velocity dispersions above and below the plane of the disk. In outflow-type objects, there should be evidence for extra-planar gas with increasing velocity dispersion ($\upsigma _{\rm ION}$) and higher[\ion{N}{ii}]/H$\upalpha$ ratio moving away from the plane of the disk. The SAMI data for these two values have been value averaged by distance from the plane of the disk to better visualise this effect.} 
    \label{fig:diagnostics}
\end{figure*}

\section{Sample selection and data}
\label{sec: selectionanddata}

The galaxies in this study are selected from the SAMI Galaxy Survey, an optical integral field  spectroscopic survey comprised of $\gtrsim 3000$ spatially resolved galaxies at $z \lesssim 0.1$ \citep{croom12, bryant15, green18, scott18}. The survey is ideal to use for the identification of galaxies with large-scale outflows, chiefly due to its large sample size, field-of-view of the observations (15\arcsec\ diameter - roughly $5 - 10$ kpc in our objects) and wavelength coverage, as well as its high spectral and spatial resolutions ($\approx$ $30$ \kms\ and $2$\arcsec\ - $1 - 2$ kpc - respectively). 

\par For the SAMI Galaxy Survey, \citet{ho16} developed two diagnostics to identify galaxies harbouring large-scale galactic winds (referred to as ``outflow-types'' throughout this paper); these techniques are illustrated in Figure \ref{fig:diagnostics}. The first diagnostic exploits the shock excitation created by fast-moving, outflowing gas. This results in both high velocity dispersion and increased emission line ratios of [\ion{N}{ii}]$\ \lambda 6583$, [\ion{S}{ii}]$\ \lambda \lambda 6717,\ 6731$ and [\ion{O}{i}]$\ \lambda 6300$ to H$\upalpha$. Taken separately, elevated emission line ratios and high velocity dispersion could indicate beam smearing of AGN photoionisation, but only shocks from high-velocity winds can create the positive correlation between high emission line ratios and velocity dispersion \citep{krumholz16}.  

\par For edge-on galaxies (i.e. $i \gtrapprox 70^{\circ}$) there is a second diagnostic that unambiguously identifies galaxies with strong winds: extra-planar emission from gas excited by the outflowing material can be detected by gas velocity dispersion and line ratios that increase with height above the disk plane (see Figure \ref{fig:diagnostics}). The ability to identify galaxies harbouring galactic-scale outflows allows us to draw a sample of main-sequence galaxies with this characteristic, and follow-up with observations with the IRAM-30m telescope and the ALMA array to obtain information about their molecular gas content. The full catalogue of our objects is given in Table~\ref{tab: GALcat}, with the integrated CO(1\textrightarrow0) observations made with ALMA and IRAM given in Tables~\ref{tab: ALMAmol} \& \ref{tab: IRAMmol} respectively.

\subsection{IRAM-30m sample and observations}
\label{sec: IRAM}

From the sample of 15 edge-on SAMI Galaxy Survey galaxies identified by \citet{ho16} as having large-scale winds, we select the 11 objects with $\log (\rm M_{*} /M_{\odot}) > 9.2$ in order to avoid low metallicity objects, where $\rm \upalpha_{CO}$ (i.e. the CO-$\rm H_2$ conversion function) could be large. The xCOLD GASS survey \citep{saintonge17} has shown that below this stellar mass limit the detectability of CO(1\textrightarrow0) emission lines significantly drops in similar observations. We also selected a further 4 face-on outflow-type candidates from \citet{ho16} identified using the first diagnostic alone. 

\par In 2015, we obtained integrated CO(1\textrightarrow0) fluxes from the IRAM 30-metre telescope for all of these galaxies, using the Eight Mixer Receiver \citep[EMIR;][]{carter12} and the Fast Fourier Transform Spectrometer (FTS). This set up gives us access to 8~GHz of bandwidth for each of the two linear polarisations. The observations were conducted in wobbler-switching mode. At the frequency of the CO(1\textrightarrow0) line, the telescope has a beam size of 22\arcsec, encompassing the entire area observed by the SAMI observation. 

\par Due to excellent weather conditions, the telescope time allocation allowed us to target an additional 13 galaxies from the SAMI Galaxy Survey; a range of peculiar galaxies were chosen, such as objects with counter-rotating or misaligned gas-stellar velocity fields. These additional galaxies are not analysed in this paper, but their IRAM CO(1\textrightarrow0) observations are released here alongside our main sample.

\begin{table*}
  \centering
 
  \caption{Galaxy catalogue measurements $^{a}$} 
	\renewcommand{\arraystretch}{0.8}
    \begin{tabular}{l c c c c c c c}
	
		\hline
	GAMA ID  & RA$_{\rm J2000}$ & DEC$_{\rm J2000}$ & $z_{\rm GAMA}$ & $\rm M_*$ [$\log _{10}$ \textit{\msun}] & SFR [$\log _{10}$ $\rm {yr}^{-1}$] & $\upalpha _{\rm CO}$ $^{b}$\\[2pt]
    		\hline
              \hline
    \noalign{\smallskip}
	106389$^{\star}{}^{\dagger}$            &  215.90105  &  1.00760   &  0.04009  &  $10.20 \pm 0.11$  &  $0.11 \pm 0.06$ & $3.11  \pm 1.18$\\[1pt]
	209807$^{\ast}{}^{\dagger}$             &  135.02106  &  0.07966   &  0.05386  &  $10.81 \pm 0.10$  &  $0.52 \pm 0.05$ & $2.55  \pm 0.97$ \\[1pt]
	228432$^{\star}{}^{\dagger}$            &  217.38573  &  1.11739   &  0.02975  &  $9.36  \pm 0.12$  &  $0.01 \pm 0.07$ & $10.61 \pm 4.03$\\[1pt]
	238125$^{\star}{}^{\dagger}$            &  213.32891  &  1.66440   &  0.02588  &  $9.56  \pm 0.13$  & $-0.45 \pm 0.06$ & $6.12  \pm 2.32$ \\[1pt]
	239249$^{\dagger}$                      &  217.01837  &  1.63906   &  0.02901  &  $9.36  \pm 0.11$  & $-0.89 \pm 0.01$ & $5.86  \pm 2.22$ \\[1pt]
	239376$^{\star}$                        &  217.52015  &  1.53685   &  0.02714  &  $9.60  \pm 0.12$  & $-0.61 \pm 0.08$ & $5.40  \pm 2.05$  \\[1pt]
    31452$^{\star}{}^{\dagger}$             &  179.86349  &  -1.15511  &  0.02024  &  $9.44  \pm 0.12$  & $-0.09 \pm 0.04$ & $5.59  \pm 2.13$ \\[1pt]
    348116$^{\star}{}^{\ast}$  	            &  140.29345  &  2.20123   &  0.05041  &  $10.62 \pm 0.10$  & $-0.22 \pm 0.09$ & $4.10  \pm 1.56$\\[1pt]
    376121$^{\ast}{}^{\dagger}$             &  132.11778  &  1.39726   &  0.05149  & $11.03  \pm 0.12$  & $-0.02 \pm 0.06$ & $4.85  \pm 1.84$\\[1pt]
    383259$^{\ast}{}^{\dagger}$             &  140.67041  &  2.11154   &  0.05715  &  $10.74 \pm 0.11$  & $0.80  \pm 0.10$ & $2.11  \pm 0.80$\\[1pt]
    417678$^{\star}{}^{\dagger}{}^{\ast}$	&  132.73822  &  2.34617   &  0.03944  &  $10.13 \pm 0.12$  & $0.38  \pm 0.05$ & $2.78  \pm 1.06$\\[1pt]
	486834$^{\dagger}{}^{\ast}$             &  221.74483  &  -1.78889  &  0.04349  &  $9.74  \pm 0.12$  & $-0.40 \pm 0.11$ & $4.57  \pm 1.74$\\[1pt]
	496966$^{\star}{}^{\ast}$               &  212.59187  &  -1.11499  &  0.05417  &  $10.37 \pm 0.11$  & $-0.10 \pm 0.07$ & $2.61  \pm 0.99$\\[1pt]
	567624$^{\star}{}^{\dagger}$            &  212.55950  &  -0.57853  &  0.02578  &  $9.32  \pm 0.12$  & $-0.87 \pm 0.31$ & $8.48  \pm 3.22$\\[1pt]
	570227$^{\star}{}^{\ast}$	            &  222.80168  &  -0.45688  &  0.04339  &  $10.67 \pm 0.11$  & $-0.41 \pm 0.06$ & $4.52  \pm 1.72$\\[1pt]
	574200$^{\star}{}^{\dagger}$            &  134.52337  &  -0.02115  &  0.02856  &  $9.34  \pm 0.12$  & $-0.20 \pm 0.06$ & $6.38  \pm 2.42$\\[1pt]
	593680$^{\star}{}^{\dagger}$            &  217.44190  &  -0.15239  &  0.03000  &  $10.41 \pm 0.11$  & $0.21  \pm 0.08$ & $2.89  \pm 1.10$\\[1pt]
	618220$^{\dagger}{}^{\ast}$             &  214.73902  &  0.36561   &  0.05331  &  $10.61 \pm 0.11$  & $-0.13 \pm 0.08$ & $3.79  \pm 1.44$\\[1pt]
	618906$^{\star}{}^{\dagger}{}^{\ast}$	&  217.35942  &  0.39756   &  0.05650  &  $10.57 \pm 0.10$  & $-0.19 \pm 0.08$ & $1.20  \pm 0.45$\\[1pt]
	618935$^{\star}$                        &  217.55202  &  0.33357   &  0.03446  &  $9.78  \pm 0.12$  & $-0.39 \pm 0.06$ & $6.62  \pm 2.52$ \\[1pt]
	619098$^{\star}$                        &  218.05118  &  0.22324   &  0.03556  &  $9.31 \pm 0.12$   & $-0.49 \pm 0.07$ & $7.33 \pm 2.78$ \\[1pt]
	623679$^{\star}$                        &  139.98309  &  0.64128   &  0.05641  &  $10.22 \pm 0.11$  & $0.01  \pm 0.07$ & $3.55  \pm 1.35$\\[1pt]
	\hline 
	\noalign{\smallskip}
	
	209698$^{\dagger}{}^{\ast}$             &  134.61914  &  0.02347   &  0.02855  &  $10.32 \pm 0.16$  & $-0.32 \pm 0.01$ & $6.20  \pm 2.36$\\[1pt]
	209743$^{\dagger}$                      &  134.67676  &  0.19143   &  0.04059  &  $10.16 \pm 0.12$  & $0.00  \pm 0.04$ & $2.69  \pm 1.02$\\[1pt]
    279818$^{\dagger}$                      &  139.43876  &  1.05542   &  0.02727  &  $9.55  \pm 0.12$  & $-0.24 \pm 0.10$ & $4.62  \pm 1.75$\\[1pt]
	322910$^{\dagger}$	                    &  129.39530  &  1.57389   &  0.03094  &  $9.71  \pm 0.12$  & $-0.39 \pm 0.09$ & $2.66  \pm 1.01$\\[1pt]
	346839$^{\dagger}{}^{\ast}$             &  135.23070  &  2.22819   &  0.05856  &  $10.36 \pm 0.13$  & $-1.94 \pm 0.37$ & $5.39  \pm 2.05$\\[1pt]
	371976$^{\dagger}{}^{\ast}$             &  133.68009  &  1.09593   &  0.05796  &  $10.52 \pm 0.14$  & $-1.36 \pm 0.33$ & $2.37  \pm 0.90$\\[1pt]
	41144$^{\dagger}$                       &  184.47038  &  -0.65722  &  0.02964  &  $10.36 \pm 0.12$  & $0.36  \pm 0.08$ & $2.73  \pm 1.04$\\[1pt]
	517302$^{\dagger}$                      &  131.72622  &  2.56007   &  0.02871  &  $10.21 \pm 0.11$  & $-0.37 \pm 0.14$ & $2.15  \pm 0.82$\\[1pt]
	534753$^{\dagger}$                      &  175.02584  &  -0.90141  &  0.02870  &  $10.36 \pm 0.11$  & $0.16  \pm 0.06$ & $2.15  \pm 0.82$\\[1pt]
	570206$^{\dagger}{}^{\ast}$	            &  222.76246  &  -0.52709  &  0.04307  &  $10.51 \pm 0.12$  & $-0.60 \pm 0.08$ & $5.04  \pm 1.92$\\[1pt]
	618151$^{\dagger}{}^{\ast}$             &  214.51701  &  0.27382   &  0.05033  &  $10.50 \pm 0.13$  & $-1.42 \pm 0.27$ & $3.51  \pm 1.33$\\[1pt]
	620034$^{\dagger}$	                    &  222.94282  &  0.28982   &  0.04269  &  $10.23 \pm 0.11$  & $-0.01 \pm 0.04$ & $3.07  \pm 1.17$\\[1pt]
	91996$^{\dagger}{}{\ast}$               &  214.47573  &  0.46141   &  0.05455  &  $10.48 \pm 0.10$  & $-1.02 \pm 0.12$ & $5.61  \pm 2.13$\\[1pt]

							\hline
				              \hline
	\end{tabular}
	\label{tab: GALcat}
	\begin{tablenotes}
	    \item[a] a Objects are divided by a horizontal line into outflow-type objects (above line) and miscellaneous peculiar objects (below line).
	    \item[b] b Variable CO(1\textrightarrow0) conversion factor, $\upalpha _{\rm CO}$ ([$M_{\odot}$ $\rm (K\ \kms)^{-1}$]), calculated using the method outlined in \citet{accurso17a}. 
	    \item[$\star$] $\star$ Marked objects are observed with ALMA.
	    \item[$\dagger$] $\dagger$ Marked objects are observed with the IRAM 30-metre telescope.
	    \item[$\ast$] $\ast$ Marked objects are from ALFALFA. For objects without $\ast$, we have SAMI-\ion{H}{I} data.

	\end{tablenotes}

\end{table*}

\par The data reduction was done using the CLASS software within the GILDAS package\footnote{\url{https://www.iram.fr/IRAMFR/GILDAS/}}. Individual scans are baseline-subtracted using a first order polynomial fit and then combined into a single spectrum re-binned to a spectral resolution of 20~\kms. The integrated CO(1\textrightarrow0) line flux is obtained by adding the signal within a spectral window set by hand to match the line width. In the case of non-detections, we adopt a standard spectral width of 300~\kms\ to measure a 3$\upsigma$ upper limit on the flux. In Table~\ref{tab: IRAMmol}, we give for each galaxy the integrated flux in units of Jy~\kms\ (\rm S$_{\rm CO}$), as well as the central redshift and width of the CO(1\textrightarrow0) line ($\rm z_{\rm CO}$ and $\upsigma_{\rm CO}$, respectively). All these measurements were made using the methods developed for the xCOLD GASS survey \citep[as described in][]{saintonge17}. 

\par The integrated CO(1\textrightarrow0) fluxes ($\rm S_{\rm CO}$) in Jy~\kms\ are converted into luminosities ($\rm L_{\rm CO}$) in K~\kms~pc$^2$ following \citet{solomon97}:
\begin{equation}
    \rm L_{\rm CO} = 3.25 \times 10^7 \rm\ S_{\rm CO}\ {\upnu_{obs}}^{-2}\ {D_L}^2\ (1 + z)^{-3} \ ,
    \label{eq1}
\end{equation}
\noindent where $\upnu_{\rm obs}$ is the observed frequency (GHz), $\rm D_L$ is the luminosity distance (Mpc) and $\rm z$ is the GAMA spectroscopic redshift. The total molecular gas mass is then $\rm M_{\rm H_2}=\upalpha _{\rm CO} \rm\ L_{\rm CO}$, where we use the variable conversion factor ($\upalpha _{\rm CO}$) derived by \citet{accurso17a}. This conversion factor is dependent on metallicity ($12+\log{\rm O/H}$) and the objects' distance off the main-sequence ($\rm \Delta MS$), which requires measurements of our objects' redshifts, stellar masses ($\rm M_*$) and SFRs. We use MAGPHYS $\rm M_*$ and SFR estimates along with emission line ratios from 3\arcsec apertures centred on our objects from SAMI to estimate the metallicity by the \citet{pettini04} calibration. These values and the adopted values for $\upalpha _{\rm CO}$ are given in Table~\ref{tab: GALcat}. The CO(1\textrightarrow0) luminosities and molecular gas mass fractions ($\log \rm M_{\rm H_2}/\rm M_*$) for all the galaxies in the IRAM sample are in Table~\ref{tab: IRAMmol}, with the spectra presented in Appendix~\ref{sec: IRAMspectra}. 

\subsection{ALMA sample and observations}
\label{sec: ALMA}

From the original \citet{ho16} sample, we selected 9 SAMI edge-on galaxies from our IRAM sample to be observed with the ALMA array. From the SAMI Galaxy Survey catalogue, we selected 7 control objects, each matched in stellar mass, inclination and redshift to the outflow-type sample (i.e. $\rm \log \left(M_*/\it M_\odot \rm \right) <10.2$, ellipticity<0.5 and z<0.05). In Cycle 5, we obtained 13.3 hours of observation time to map the CO(1\textrightarrow0) emission in these 16 galaxies. 

Observations were conducted in Band 3 with a synthesised beam of 1\arcsec\ ($\approx 0.5$~kpc - $1$~kpc) and a spectral resolution of $\approx$ 10~\kms. This observational setup was chosen to produce data with a resolution comparable to that of the SAMI observations. The on-source time was between 45 and 53 minutes for each galaxy, to ensure sensitivity to a molecular gas mass surface density of 0.8 \msun~pc$^{-2}$, well into the atomic-gas dominated regime of the Kennicutt-Schmidt relation \citep{bigiel08}. 

\begin{figure*}
    \centering
    \includegraphics[scale=0.31]{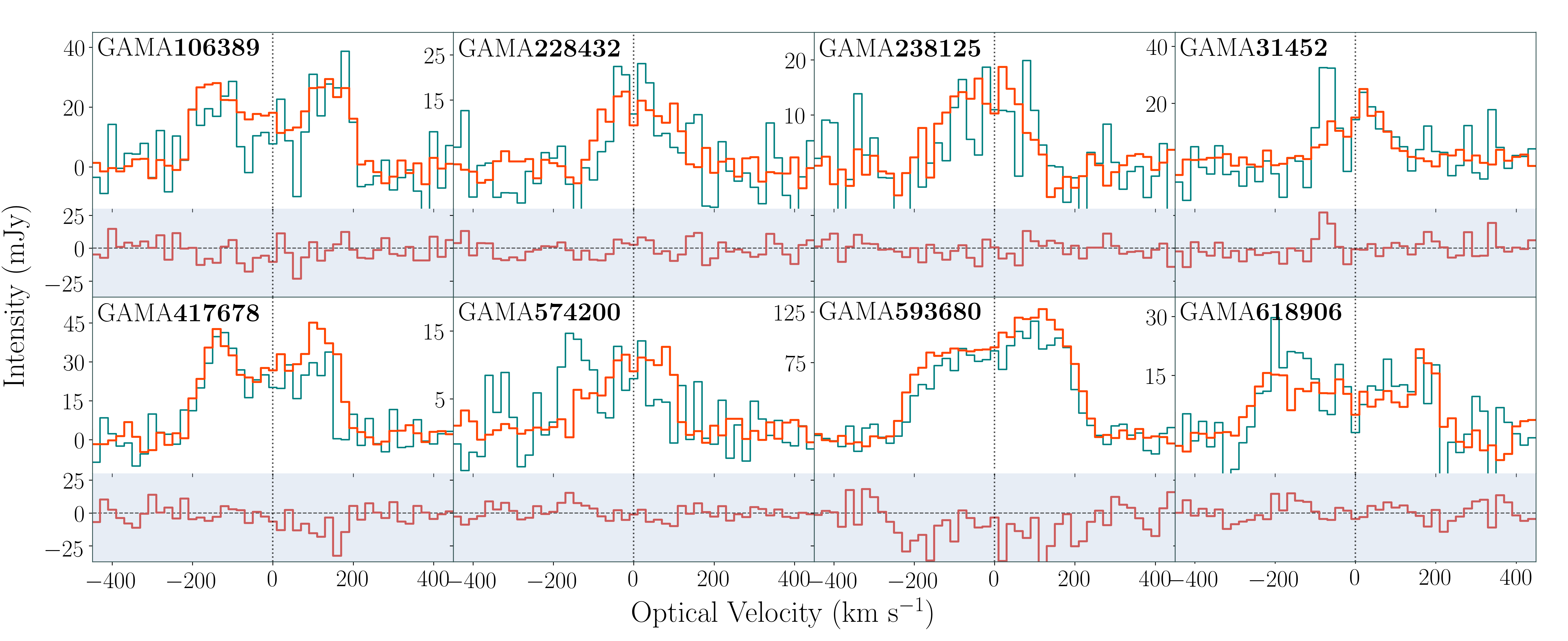}
    \vspace{0.7cm}
    \includegraphics[scale=0.3]{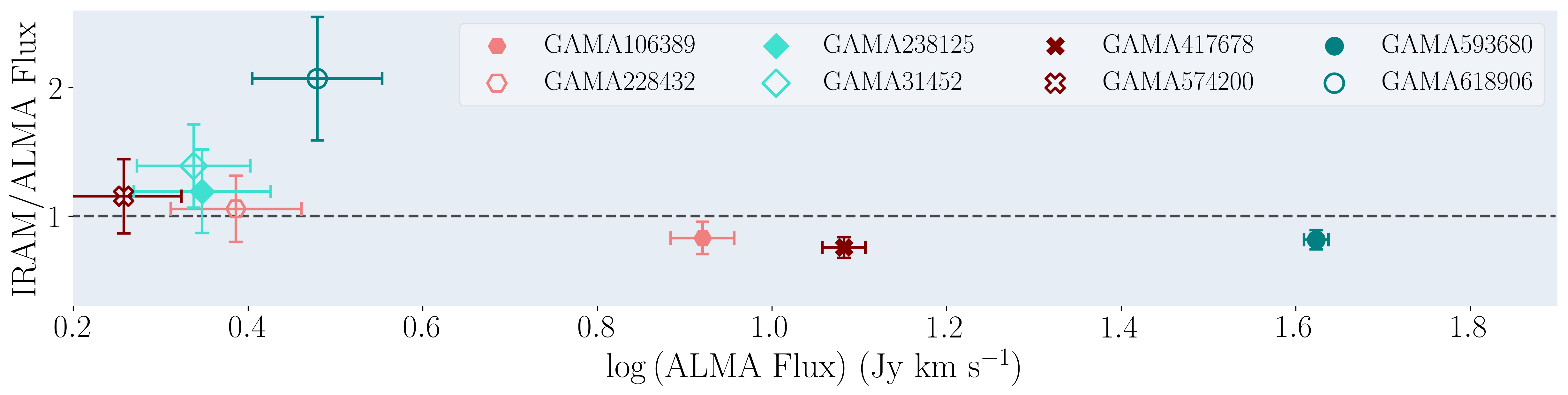}

    \caption{\textbf{Top}: Comparison of IRAM (teal) and ALMA (orange) CO(1\textrightarrow0) spectra for outflow-type galaxies in both observation samples. The IRAM-ALMA flux difference is shown below each plot. All galaxies in the figure are confirmed by the SAMI Galaxy Survey as harbouring galactic-scale outflows of ionised gas. The ALMA and IRAM spectra have been re-binned to a common spectral resolution of $20\rm\ \kms$. \textbf{Bottom}: Comparison of the integrated CO(1\textrightarrow0) fluxes in the IRAM and ALMA objects with a linear 1:1 trend.}
    \label{fig: comp}
\end{figure*}

\begin{table*}
  
  \centering
	\caption{Total CO(1\textrightarrow0) luminosities and $\rm H_2$ molecular gas fractions for ALMA outflow-type objects $\rm ^{a}$}
	\renewcommand{\arraystretch}{0.8}
	\begin{tabular}{l c c c c c c c c} 
		\hline
	GAMA ID  & $\rm {{S/N}_{CO}}^{b}$ & Flag $\rm ^{b}$ & $\upsigma _{\rm CO}$ [$\rm \kms$] $\rm ^{b}$ & $\rm S_{\rm CO}$ [Jy \kms] $\rm ^{b}$ & $\rm z_{\rm CO}$ $\rm ^{b}$ & $\rm L'_{\rm CO10}$ [$10^8 \ \rm K \ \kms\ {pc}^2$] $\rm ^{b}$ & $\log$ ($\rm M_{H_2} / M_*$) $\rm ^{b}$ \\[2pt]
    		\hline
              \hline
    \noalign{\smallskip}

	106389           &  33  &  1  &  169 &  $8.33\pm 0.73$  &  0.04012  &  $6.13 \pm 0.57$  &  $-0.92 \pm 0.20$ \\[1pt]
	228432	         &  15  &  1  &  83  &  $2.43\pm 0.46$  &  0.02984  &  $0.98 \pm 0.19$  &  $-0.34 \pm 0.22$ \\[1pt]
	238125           &  14  &  1  &  77  &  $2.23\pm 0.44$  &  0.02581  &  $0.68 \pm 0.14$  &  $-0.94 \pm 0.23$\\[1pt]
	31452            &  19  &  1  &  59  &  $2.18\pm 0.35$  &  0.02029  &  $0.40 \pm 0.08$  &  $-1.08 \pm 0.22$\\[1pt]
	417678$^{\ast}$  &  53  &  1  &  140 &  $12.09\pm 0.70$ &  0.03946  &  $8.63 \pm 0.59$  &  $-0.75 \pm 0.20$\\[1pt]
	567624	         & <3   &  2  &  $\ldots$   &  $\ldots$              &  $\ldots$        &  $\ldots$               &  $\ldots$\\[1pt]
	574200           & 18   &  1  &  69  &  $1.81\pm 0.30$  &  0.02867  &  $0.67 \pm 0.12$  &  $-0.71 \pm 0.22$\\[1pt]
	593680	         & 100  &  1  &  165 &  $42.04\pm 1.40$ &  0.03003  &  $17.24 \pm 1.16$ &  $-0.71 \pm 0.20$\\[1pt]
	618906$^{\ast}$  & 10   &  1  &  184 &  $3.02\pm 0.56$  &  0.05648  &  $4.45  \pm 0.83$ &  $-1.85 \pm 0.21$\\[1pt]
	\hline 
	\noalign{\smallskip}
	
	239376	         & <3   &  2  &  $\ldots$  &  $\ldots$              &  $\ldots$       & $\ldots$                & $\ldots$   \\[1pt]
	348116$^{\ast}{}^{\star}$	 &  23  & 1  &  196 &  $8.85\pm 1.06$  &  0.05036  &  $10.34 \pm 1.26$ & $-0.99 \pm 0.20$ \\[1pt]
	496966$^{\ast}$  & 12   & 1  &  159 &  $3.82\pm 0.70$  &  0.05413  &  $5.17  \pm 0.95$ & $-1.24 \pm 0.21$\\[1pt]
	570227$^{\ast}{}^{\star}$  &  27   &  1  &  208 &  $8.45\pm 0.91$  &  0.04332  &  $7.30  \pm 0.81$ & $-1.15 \pm 0.20$\\[1pt]
	618935	         &  4   & 1  &  70  &  $0.57\pm 0.21$  &  0.03436  &  $0.31  \pm 0.12$ & $-1.46 \pm 0.26$\\[1pt]
	619098	         & <3   &  2  & $\ldots$   &  $\ldots$        &  $\ldots$     & $\ldots$               & $\ldots$\\[1pt]
	623679$^{\ast}$	 & 10   &  1  &  106 &  $2.75\pm 0.62$  &  0.05635  &  $4.04  \pm 0.92$ & $-1.07 \pm 0.22$\\[1pt]
							\hline
				              \hline
	\end{tabular}
	\label{tab: ALMAmol}
		\begin{tablenotes}
	    \item[a] a Objects are divided by a horizontal line into outflow-type objects (above line) and control (i.e. non-outflow-type) objects (below line).
	    \item[b] b If Flag=2, $\rm {S/N}_{CO}$<3 (with adopted velocity range of 300 $\rm \kms$) for the observation and we do not detect CO(1\textrightarrow0).
	    \item[$\ast$] $\ast$ Marked objects are from ALFALFA. For the unmarked galaxies, we have SAMI-\ion{H}{I} data.
	    \item[$\star$] $\star$ Marked objects may be AGN-contaminated.

	\end{tablenotes}
\end{table*}

\begin{table*}
 \centering
 
  \caption{Total CO(1\textrightarrow0) luminosities and $\rm H_2$ molecular gas fractions for IRAM positive-detection objects $\rm ^a$} 
	\renewcommand{\arraystretch}{0.8}
   \begin{tabular}{l c c c c c c c c}
	
		\hline
	GAMA ID   & $\rm {{S/N}_{CO}}^{b}$ & Flag $\rm ^{b}$ & $\upsigma _{\rm CO}$ [$\rm \kms$] $\rm ^{b}$ & $\rm S_{\rm CO}$ [Jy \kms] $\rm ^{b}$ & $\rm z_{\rm CO}$ $\rm ^{b}$ & $\rm L'_{\rm CO10}$ [$10^8 \ \rm K \ \kms\ {pc}^2$] $\rm ^{b}$ & $\log$ ($\rm M_{H_2} / M_*$) $\rm ^{b}$ \\[2pt]
    		\hline
              \hline
    \noalign{\smallskip}
	106389           & 10  &  1  &  169  &  $6.90 \pm 0.86$ &  0.04015  &  $5.18  \pm 0.61$ &  $-0.99 \pm 0.17$\\[1pt]
	209807$^{\ast}$  & 18  &  1  & 140  &  $14.22 \pm 1.39$ &  0.05380  &  $19.40 \pm 1.81$ &  $-1.12 \pm 0.17$\\[1pt]
	228432           & 8   &  1  & 37   &  $2.57 \pm 0.40$  &  0.02979  &  $1.06  \pm 0.15$ &  $-0.31 \pm 0.18$\\[1pt]
	238125           & 6   &  1  & 90   &  $2.65 \pm 0.50$  &  0.02593  &  $0.82  \pm 0.14$ &  $-0.86 \pm 0.18$\\[1pt]
	239249           & <3  &  2  & $\ldots$ &  $\ldots$     &  $\ldots$ &  $\ldots$         &  $\ldots$\\[1pt]
    31452	         & 7   &  1  & 72   &  $3.02 \pm 0.51$  &  0.02026  &  $0.57  \pm 0.10$ &  $-0.93 \pm 0.18$\\[1pt]
    376121$^{\ast}$  & 8   &  1  & 188  &  $4.28 \pm 0.61$  &  0.05148  &  $5.33  \pm 0.77$ &  $-1.62 \pm 0.17$\\[1pt]
    383259$^{\ast}$  & 81  &  1  &  62  &  $24.06 \pm 1.95$ &  0.05709  &  $37.02 \pm 3.00$ &  $-0.85 \pm 0.17$\\[1pt]
    417678$^{\ast}$  & 23  &  1  &  137 &  $9.12 \pm 0.83$  &  0.03950  &  $6.63  \pm 0.60$ &  $-0.86 \pm 0.17$\\[1pt]
	486834$^{\ast}$  & <3  &  2  &  $\ldots$ &  $\ldots$    &  $\ldots$ &  $\ldots$ &  $\ldots$\\[1pt]
	567624	         & <3  &  2  &$\ldots$ &  $\ldots$      &  $\ldots$ &  $\ldots$ &  $\ldots$\\[1pt]
	574200           & 6   &  1  & 87   &  $2.09 \pm 0.39$  &  0.02834  &  $0.79  \pm 0.14$ &  $-0.63 \pm 0.18$\\[1pt]
	593680	         & 35  &  1  & 170  &  $34.26 \pm 2.91$ &  0.03007  &  $14.33 \pm 1.22$ &  $-0.79 \pm 0.17$  \\[1pt]
	618220$^{\ast}$  & 7   &  1  & 122  &  $2.60 \pm 0.44$  &  0.05335  &  $3.47  \pm 0.58$ &  $-1.49 \pm 0.18$ \\[1pt]
	618906$^{\ast}$	 & 9   &  1  & 181  &  $6.24 \pm 0.86$  &  0.05653  &  $9.38  \pm 1.29$ &  $-1.52 \pm 0.17$\\[1pt]
	\hline 
	\noalign{\smallskip}
	
	209698$^{\ast}$  & 10  &  1  & 84   &  $4.88 \pm 0.62$  &  0.02828  &  $1.85  \pm 0.22$ &  $-1.26  \pm 0.17$ \\[1pt] 
	209743           & 10  &  1  & 109  &  $5.05 \pm 0.66$  &  0.04056  &  $3.88  \pm 0.51$ &  $-1.15 \pm 0.17$\\[1pt]
    279818           & 6   &  1  & 19   &  $1.14 \pm 0.22$  &  0.02720  &  $0.39  \pm 0.08$ &  $-1.29 \pm 0.18$\\[1pt]
	322910	         & 13  &  1  & 20   &  $2.51 \pm 0.28$  &  0.03093  &  $1.12  \pm 0.13$ &  $-1.24 \pm 0.17$\\[1pt]
	346839$^{\ast}$  & <3  &  2  & $\ldots$ &  $\ldots$     &  $\ldots$ &  $\ldots$ &  $\ldots$\\[1pt]
	371976$^{\ast}$  & <3  &  2  & $\ldots$ &  $\ldots$     &  $\ldots$ &  $\ldots$ &  $\ldots$\\[1pt]
	41144            & 22  &  1  & 127  &  $20.28 \pm 1.87$ &  0.02967  &  $8.28  \pm 0.76$ &  $-1.00 \pm 0.17$\\[1pt]log
	517302           & 6   &  1  & 53   &  $1.97 \pm 0.38$  &  0.02873  &  $0.75  \pm 0.15$ &  $-2.00 \pm 0.18$ \\[1pt]
	534753           & 20  &  1  & 96   &  $17.76 \pm 1.68$ &  0.02863  &  $6.79  \pm 0.64$ &  $-1.20 \pm 0.17$\\[1pt]
	570206$^{\ast}$	 & <3  &  2  & $\ldots$ &  $\ldots$     &  $\ldots$ &  $\ldots$ &  $\ldots$\\[1pt]
	618151$^{\ast}$  & <3  &  2  & $\ldots$ &  $\ldots$     &  $\ldots$ &  $\ldots$ &  $\ldots$\\[1pt]
	620034	         & 9   &  1  & 134  &  $4.90 \pm 0.66$  &  0.04277  &  $4.17  \pm 0.56$ &  $-1.12 \pm 0.17$\\[1pt]
	91996$^{\ast}$   & <3  &  2  & $\ldots$ &  $\ldots$     &  $\ldots$ &  $\ldots$ &  $\ldots$\\[1pt]

							\hline
				              \hline
	\end{tabular}
	\label{tab: IRAMmol}
	\begin{tablenotes}
	    \item[a] a Objects are divided by a horizontal line into outflow-type objects (above line) and miscellaneous peculiar objects (below line).
	    \item[b] b If Flag=2, $\rm {S/N}_{CO}$<3 (with adopted velocity range of 300 $\rm \kms$) for the observation and we do not detect CO(1\textrightarrow0).
	    \item[$\ast$] $\ast$ Marked objects are from ALFALFA. For the unmarked galaxies, we have SAMI-\ion{H}{I} data.

	\end{tablenotes}
\end{table*}

The ALMA data were reduced using standard CASA (Common Astronomy Software Applications) pipeline subroutines \citep{mcmullin07}. The calibrated dirty cubes were cleaned using the \texttt{tclean} task over a range of $\approx 50$ channels centred on the peak of CO(1\textrightarrow0) emission using the interactive keyword. Each channel was inspected by eye and cleaning regions selected by hand. To extract the spectra (we note that we do not detect continuum emission in any of our objects), we define apertures by smoothing our cubes over the trimmed channel range with a 2D Gaussian kernel with $\upsigma_{smooth} = 1.5$ spaxels, using the \texttt{Gaussian2DKernel} and \texttt{convolve} sub-routines within the \texttt{astropy.convolution} package \citep{astropy13, astropy18}. We then collapse the smoothed cubes over their spectral axes and set all spaxels with a values over a standard deviation ($1\upsigma$) to 1 and those below to 0 to create a mask which we apply to each trimmed channel in our original cubes as an aperture (we also define a trimmed spatial region based on the collapsed smoothed cube of $\approx 200 \times 200$ spaxels outside which all spaxels are set to 0). We collapse the masked cubes spatially to obtain the spectra given in Figure~\ref{fig: comp}. The optical velocity is centred at 0 \kms\ using the GAMA spectroscopic redshift to determine the systemic velocity of each galaxy. 

To verify that the ALMA maps are not missing any extended flux, we compare in Figure \ref{fig: comp} the integrated IRAM-30m and ALMA CO(1\textrightarrow0) spectra for the 8 galaxies that the samples have in common for which we have positive CO(1\textrightarrow0)-detection. The spectra are rebinned to the same spectral resolution using the \texttt{spectres} package \citep{carnall17}. There is a good agreement between the emission profiles measured with ALMA and IRAM, showing that ALMA has not resolved out significant amounts of flux. For 3 of the 16 galaxies (1 outflow-type and 2 controls) observed with ALMA, we do not detect CO(1\textrightarrow0) emission (i.e. S/N<3).

\begin{figure}
    \centering
    \includegraphics[scale=0.42]{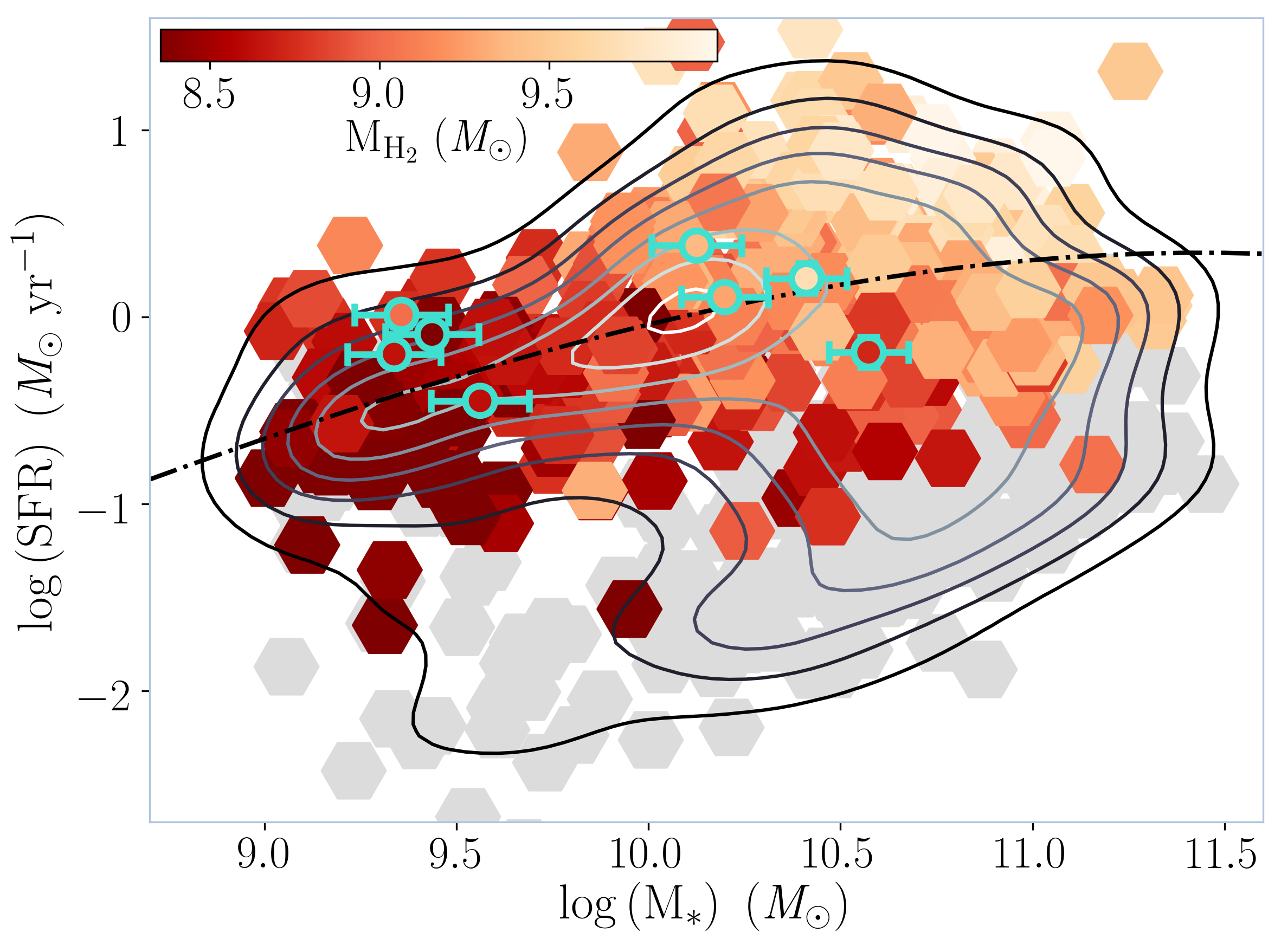}
    \caption{ALMA outflow-type objects plotted in the SFR - stellar mass ($\rm M_{\rm *}$) plane. Pale turquoise circles indicate an object identified as an outflow-type by SAMI (for which we have positive CO(1\textrightarrow0) detection). The xCOLD GASS catalogue is also given for comparison (red hexagons). The observed and catalogue objects are shaded by their molecular gas mass ($\rm M_{\rm H_2}$). Grey hexagons represent the objects in the xCOLD GASS catalogue with no CO(1\textrightarrow0)-detection. The grey contours depict density levels in the xCOLD GASS catalogue and the black dashed line is the main-sequence trend as calculated by \citet{saintonge16}.}
    \label{fig: ALMAsm-sfr}
\end{figure}

The total integrated flux is measured by numerically integrating each global ALMA CO(1\textrightarrow0) emission profile, where we define channels with signal-detection as those with a flux above 3 standard deviations ($3 \upsigma$) of the noise in a line-free region. The uncertainty on this integrated flux is calculated as $\upsigma_{\rm err} = \rm 3 \sqrt{N}\ \upsigma\ dv$, where $\rm N$ is the number of channels with signal>$3 \upsigma$ and $\rm dv$ is the channel width in \kms. The global $\rm L_{\rm CO}$ is calculated as explained in Section~\ref{sec: IRAM}, with values presented in Table \ref{tab: ALMAmol}. As detailed in Section~\ref{sec: IRAM}, we use the variable conversion factor derived by \citet{accurso17a}. This is done to account for the range of stellar mass ($\rm M_*$) and metallicity values in our ALMA outflow-type objects (see Figure~\ref{fig: ALMAsm-sfr}). Again, We use MAGPHYS $\rm M_*$ and SFR estimates and emission line ratios from 3\arcsec apertures centred on our objects from SAMI to estimate the metallicity by the \citet{pettini04} calibration. These measurements and the values of variable $\upalpha _{\rm CO}$ for each object are given in Table~\ref{tab: GALcat}. 

\begin{figure*}
\centering
\begin{minipage}{\textwidth}
  \centering
  \includegraphics[width=\linewidth]{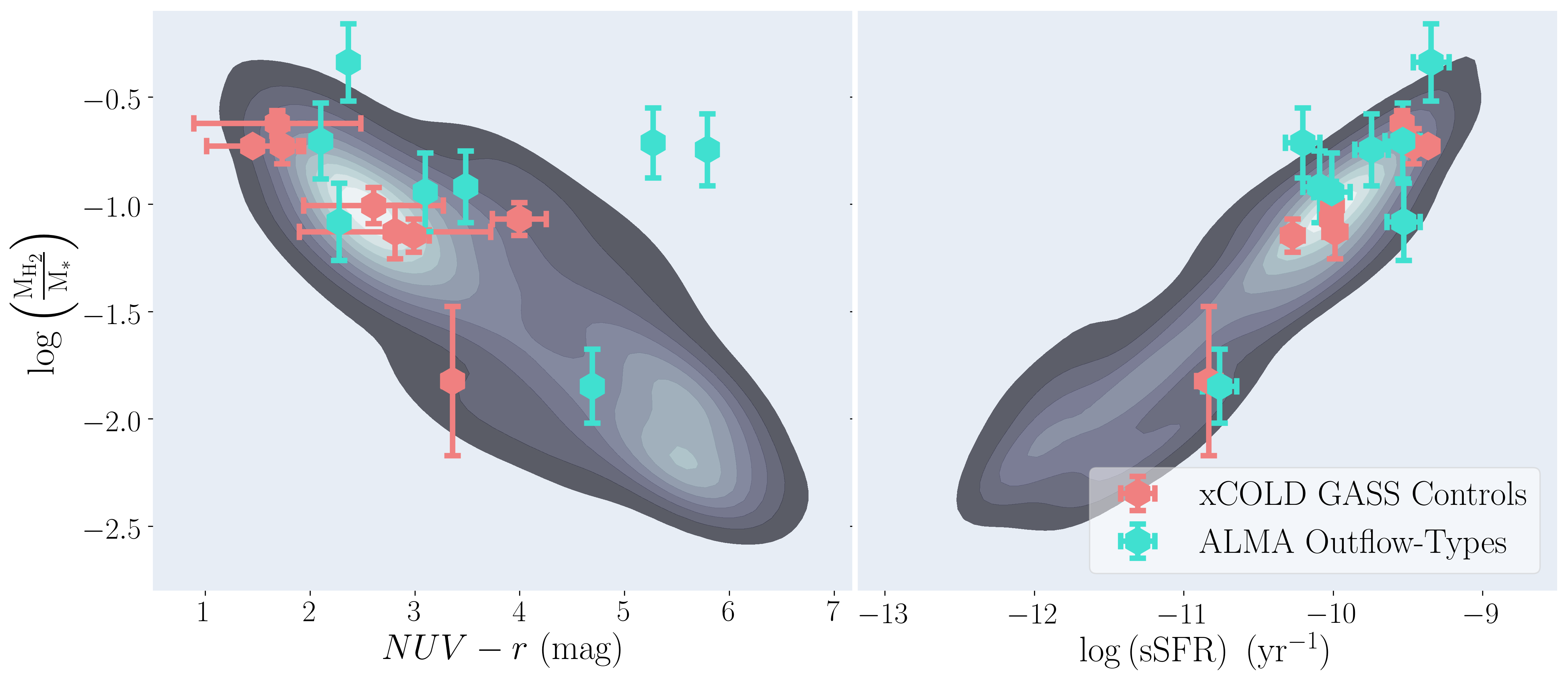}
\end{minipage}%
    \caption{Molecular hydrogen mass ($\rm M_{\rm H_2}$) against both the $NUV-r$ colour index (left panel) and specific SFR (sSFR, right panel) for the ALMA outflow-type sample with positive CO(1\textrightarrow0)-detection (turquoise) and averaged xCOLD GASS controls (red, see text) plotted alongside the xCOLD GASS catalogue (grey density contours).}
    \label{fig: ALMAdiag}
\end{figure*}

\begin{figure*}
\centering
\begin{minipage}{\textwidth}
  \centering
  \includegraphics[width=.9\linewidth]{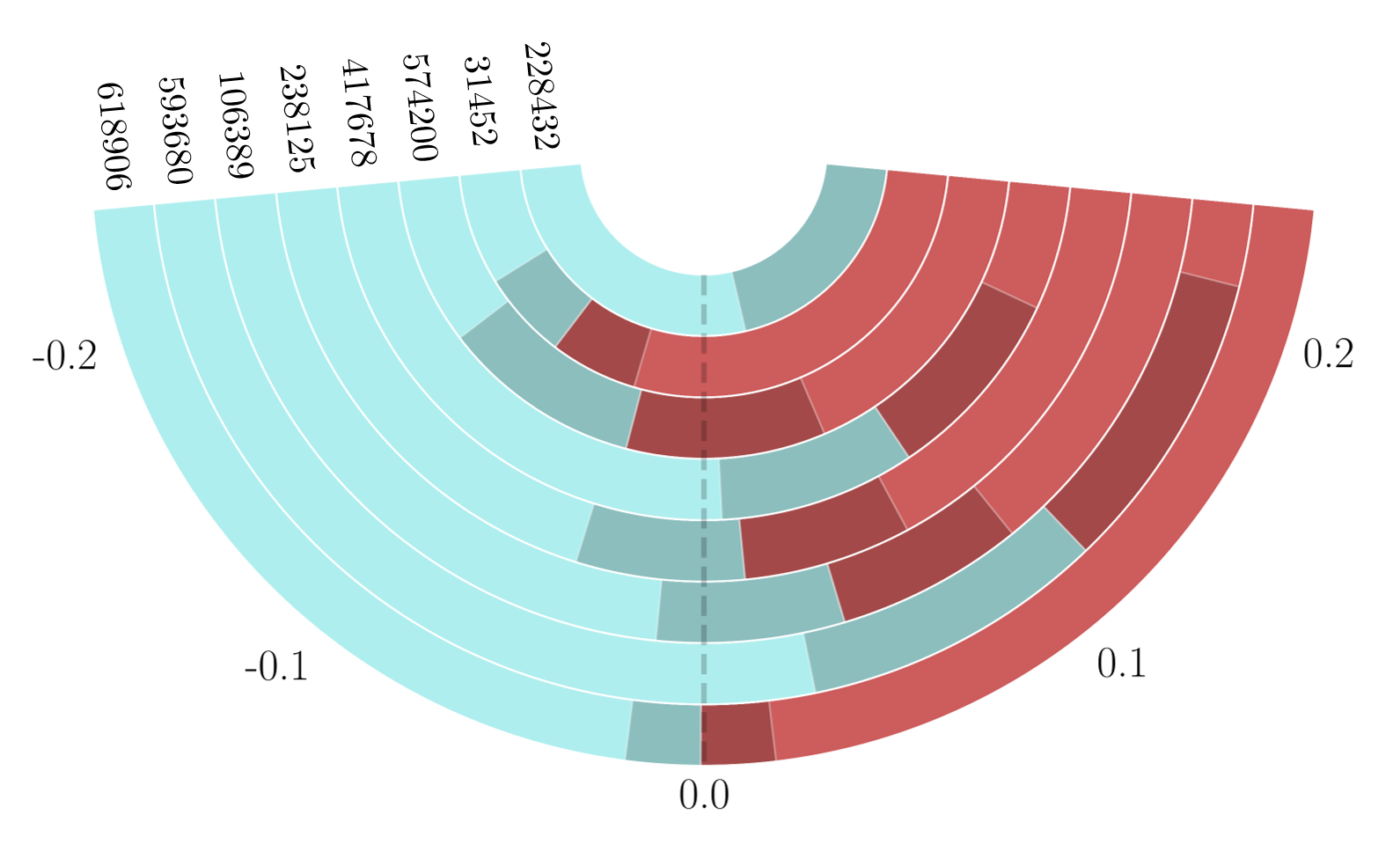}
\end{minipage}

 \caption{Comparison of molecular gas fractions ($f_{\rm M_{H_2}}\ = \ \rm M_{\rm H_2} / \rm M_{*}$) in ALMA outflow-type and xCOLD GASS derived control galaxies (see text). The figure depicts the ``swing'' of the difference in molecular gas fractions $\Delta f_{\rm M_{H_2}}\ = \ f_{\rm M_{H_2},  outflow} \ - \ f_{\rm M_{H_2},  control}$, with each inset disk representing one of the 8 ALMA outflow-type galaxies (with positive CO(1\textrightarrow0)-detection) and its corresponding xCOLD GASS control object. $\Delta f_{\rm M_{H_2}}$ is given by the position of the interface between blue and red regions ($0.67 \times 1\upsigma$ uncertainties are also illustrated by the grey shaded areas). The ``swing'' is given by the relative sizes of the blue (outflow-type) and red (control) regions. The uncertainty on the ``swing'' is given by the shaded areas. The objects are ordered by their sSFR; the object with the highest sSFR value is the innermost ring and the object with the lowest is the outermost.}
    \label{fig: delgasfrac}
\end{figure*}

%===========================================================================================
%===========================================================================================

\section{Results}
\label{sec: results}

The results of this paper are divided into two parts; Section~\ref{subsec: glob} for our integrated results and Section~\ref{subsec: spatresob} for our spatially resolved data. Due to contamination of our ALMA control sample by AGN and CO(1\textrightarrow0) non-detections, there are only 3 viable control objects. We compensate for this by including additional suitable control objects where possible in both our integrated and resolved analyses to create a more robust comparative sample for our outflow-type objects. In Section~\ref{subsec: glob}, we use a control sample derived solely from xCOLD GASS (referred to as the ``xCOLD GASS controls'') for the interpretation of the global/integrated results (i.e. we do not include the original, viable ALMA controls).

\begin{figure}
    \centering
    \includegraphics[scale=0.42]{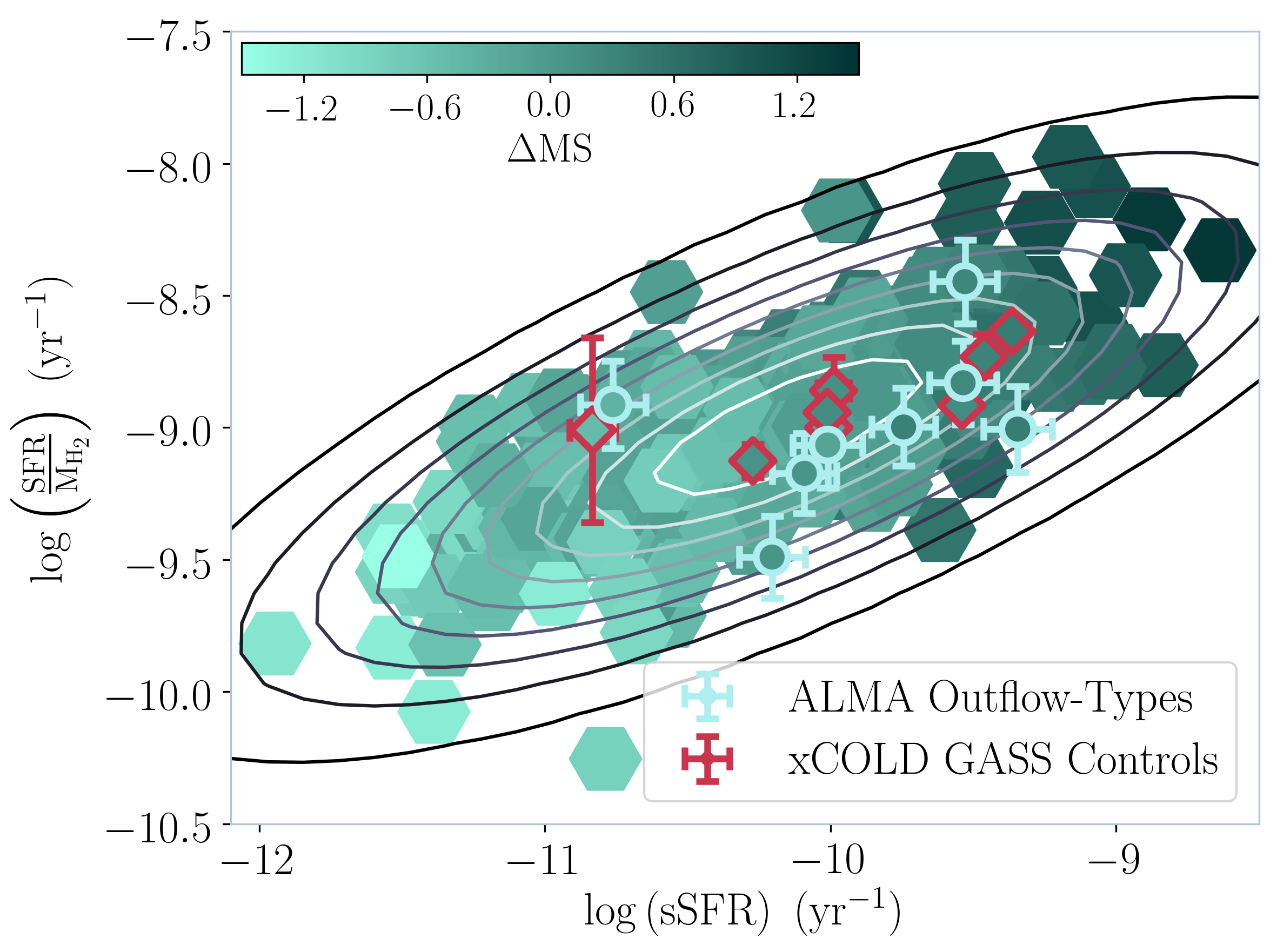}
    \vskip 1pt
    \includegraphics[scale=0.275]{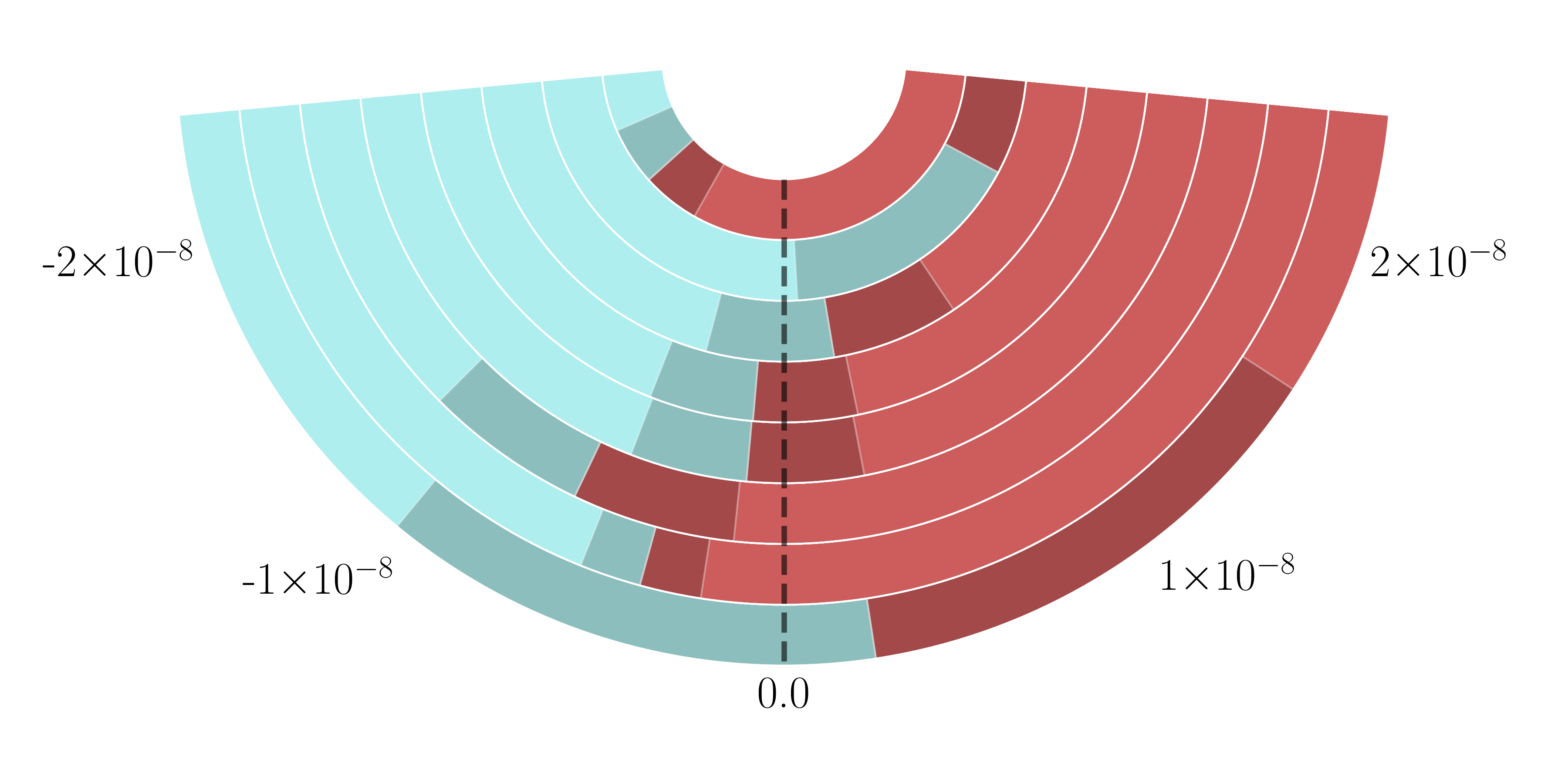}
    \caption{\textbf{Top}: ALMA outflow-type galaxies (light turquoise circles) and xCOLD GASS derived controls (red diamonds) plotted with star-formation efficiency ($\rm SFR/M_{\rm H_2}$) against specific star-formation rate (sSFR). The xCOLD GASS catalogue is also depicted for comparison (green hexagons). The outflow-type, xCOLD GASS control and catalogue objects are shaded by the value $\Delta \rm MS$, which indicates the objects' vertical displacement from the main-sequence trend in the SFR - $\rm M_*$ plane as determined by \citet{saintonge16}. The grey contours represent density levels in the xCOLD GASS catalogue. \textbf{Bottom}: ``Swing'' plot depicting the quantity $\Delta \rm SFE = SFE_{\rm outflow}-SFE_{\rm control}$. The ring structure is equivalent to that used in Figure \ref{fig: delgasfrac} (i.e. ordering by the objects' sSFR values); the turquoise represents the outflow-types and the red the xCOLD GASS controls. The shaded areas represent the $0.67 \times 1\upsigma$ uncertainties.}
    \label{fig: sfre}
\end{figure}

\par For the resolved analysis in Section~\ref{subsec: spatresob}, we use additional spatially resolved data to supplement the 3 viable control objects we have from ALMA. Further CO(1\textrightarrow0) emission maps are not available for additional control galaxies, but we do extract extra controls from the SAMI sample (which we refer to as the ``SAMI controls'') to aid analysis and validate the ALMA control sample. For any examination of our resolved CO(1\textrightarrow0) maps, we use only the three original controls from ALMA (referred to as the ``ALMA controls''). The exact procedures for extracting supplemental controls from xCOLD GASS and SAMI will be detailed further in Sections~\ref{subsec: glob} and \ref{subsec: spatresob}.

\subsection{Global Molecular Gas Contents}
\label{subsec: glob}

\par The results of our integrated analysis are reported in Figures \ref{fig: ALMAsm-sfr} - \ref{fig: sfre}. We extract additional global properties for our objects (e.g. $NUV-r$) from the Galaxy And Mass Assembly (GAMA) DR2/DR3 catalogue \citep{liske15, baldry18} with their respective uncertainties to supplement the interpretation of our results. 

\begin{figure*}
    \centering
    \includegraphics[scale=0.55]{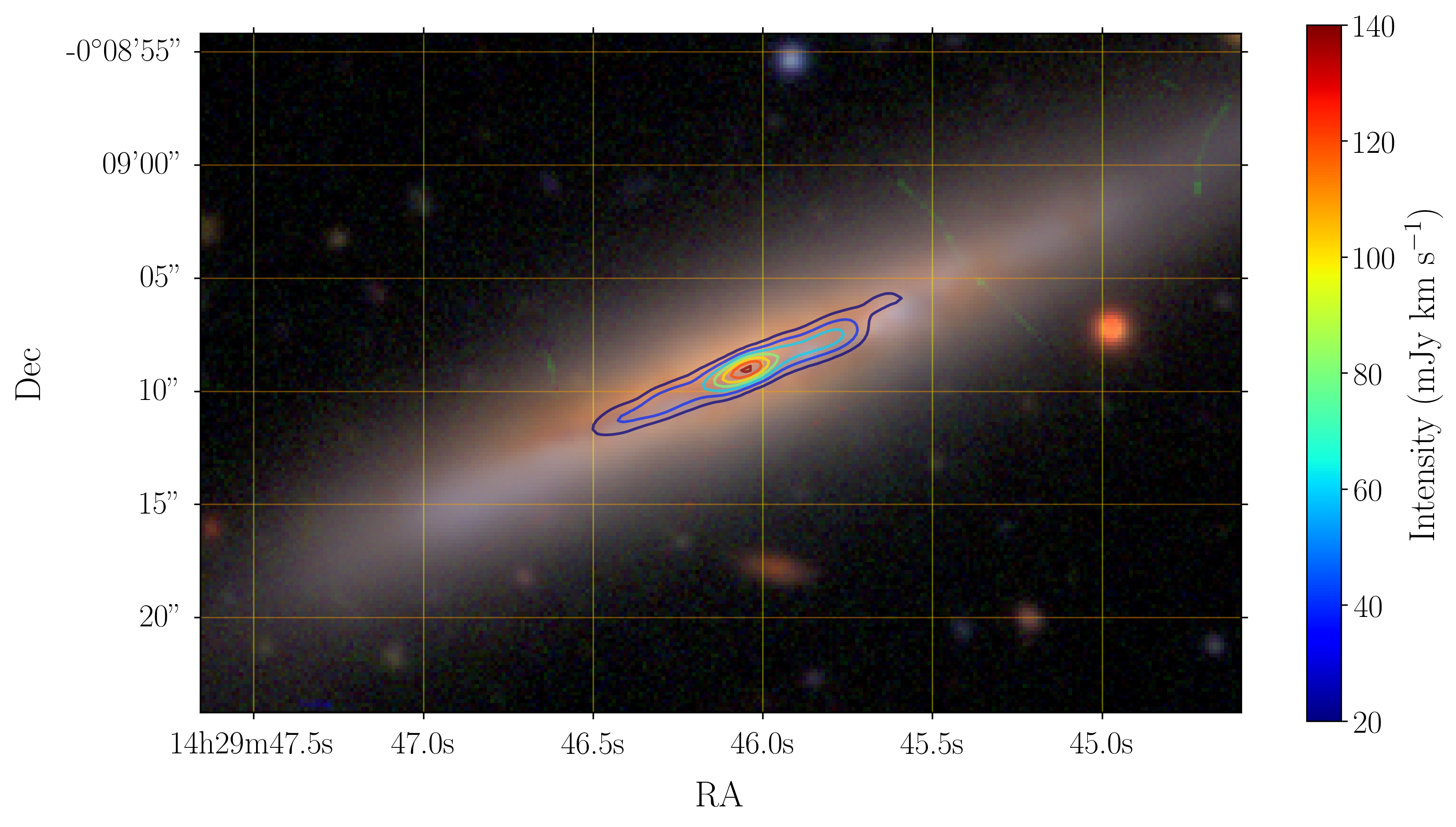}
    \caption{Zeroth moment map for GAMA593680 illustrated by colour-coded contours, drawn over an optical HSC \citep[Hyper Suprime-Cam Subaru Strategic Program; ][]{aihara19} image (using bands \textit{g}, \textit{r}, and \textit{i}) to allow a visual comparison of the extent of the CO(1\textrightarrow0) gas. The horizontal and vertical axes represent the RA and Dec directions respectively.}
    \label{fig: HSCmom0}
\end{figure*}

\par In our integrated analysis we have replaced our observed ALMA control sample with objects derived from the xCOLD GASS catalogue. To compile a viable control sample, we identify the objects from xCOLD GASS with SFR and $\rm M_*$ values within a $0.3 \ \rm dex$ box centred on each galaxy within the ALMA outflow-type sample (see Figure~\ref{fig: ALMAsm-sfr}). We use MAGPHYS SFR and $\rm M_*$ estimates for our outflow-types, which are a sound comparison with xCOLD GASS estimators \citep{saintonge18}. Furthermore, we discount potential objects classed as AGN-hosting based on the BPT diagnostic and limit the objects' inclination to $i>50^{\circ}$. For xCOLD GASS control objects with undetected CO(1\textrightarrow0), we use the 3$\upsigma$ upper limit for their molecular gas mass. From all the xCOLD GASS galaxies identified as prospective controls for each ALMA outflow-type galaxy, we extract $80 \%$ at random (i.e. $0.8 \times n$ objects, where $n$ is the number of xCOLD GASS control objects identified for each galaxy). The median values of $\rm M_*$, SFR, $\rm M_{H_2}$ and $NUV-r$ are used for the xCOLD GASS control object properties and their uncertainties (where the error on the median is estimated as $0.67 \times 1\upsigma$). Given that the outflow-type galaxies identified by SAMI \citep[and defined by][]{ho16} are exceedingly rare in the redshift range spanned by xCOLD GASS, the probability of these xCOLD GASS controls also being ``outflow-types'' is statistically insignificant. The controls assembled from xCOLD GASS are given as the control sample alongside the ALMA outflow-types in Figures \ref{fig: ALMAdiag}, \ref{fig: delgasfrac} \& \ref{fig: sfre}.

\par The integrated properties of the outflow-type galaxies in our ALMA sample are given in Figures \ref{fig: ALMAsm-sfr}, \ref{fig: ALMAdiag} \& \ref{fig: sfre} alongside the xCOLD GASS catalogue. The plots in Figures \ref{fig: ALMAdiag} suggest little difference between the global properties of the ALMA outflow-types and xCOLD GASS controls in terms of their sSFR and gas fractions. The similarity of both samples in these diagnostic figures imply that they do not contain fundamentally different galaxy types (see Section~\ref{sec: discussion}). We will analyse this further in Figures \ref{fig: delgasfrac} \& \ref{fig: sfre}.   

\par In Figure \ref{fig: delgasfrac}, we use a novel ``swing'' plot to assess the difference in molecular gas fractions between the ALMA outflow-types and xCOLD GASS controls. The plot visualises the quantity $\Delta f_{\rm M_{H_2}}\ = \ f_{\rm M_{H_2},  outflow} \ - \ f_{\rm M_{H_2},  control}$ (i.e. the difference in molecular gas fraction between outflow-type $f_{\rm M_{H_2}, outflow}$ and xCOLD GASS control $f_{\rm M_{H_2}, control}$ galaxies), where each ring represents one of the 8 ALMA outflow-xCOLD GASS control galaxy pairs. A positive ``swing'' (i.e. the interface between blue and red regions has a positive angular value) indicates that the outflow-type in the pair has the larger gas fraction compared to its xCOLD GASS control. Otherwise, a negative ``swing'' indicates that the xCOLD GASS control possesses the higher gas fraction. The median ``swing'' position is $0.05 \pm 0.04$, meaning the outflow-types are only marginally more gas-rich compared to their control counterparts. This implies that outflow-type and our xCOLD GASS control galaxies have roughly equivalent reservoirs of material for further star-formation with respect to the stars they have already created.

\par Theory predicts that galaxies launching galactic-scale outflows possess lower gas fractions than galaxies without such violent outflows, due to their intense wind driving out reservoirs of cold molecular gas. However, our sample appears to contradict this conjecture. In order to scrutinise this behaviour, we derive the star-formation efficiency (SFE) for each outflow-type galaxy and xCOLD GASS control (see Figure \ref{fig: sfre}). We define this quantity as $\rm SFR/\rm M_{H_2}$, which has a positive correlation with specific star-formation rate (sSFR). The lower plot in Figure \ref{fig: sfre} is a ``swing'' plot for the quantity $\Delta \rm SFE = \rm SFE_{\rm outflow} - \rm SFE_{\rm control}$. Much as in Figure \ref{fig: delgasfrac}, a positive value indicates that the outflow-type in the outflow-control pair has the greater SFE and vice versa. We find no statistically significant average ``swing'' across the galaxy pairs in this analysis (a median position of $(0.006 \pm 0.03)\ \times \ 10^{-8}\  \rm yr^{-1}$). Our outflow-type galaxies, therefore, do not appear to have a star-formation process that is globally more efficient than our xCOLD GASS control sample. We will consider the physical implications of these findings in Section~\ref{sec: discussion}.

\par We note in our outflow-type sample some anomalies in the physical appearance of GAMA31452 in the HSC optical images in Appendix~\ref{sec: IRAMspectra} (visually, the arms are warped and show clear signs of major disruption). It is included in our sample due to the small sample size, but it may contribute to some level of contamination.

\subsection{Spatially Resolved Observations}
\label{subsec: spatresob}

\begin{figure*}
    \centering
    \includegraphics[scale=0.29]{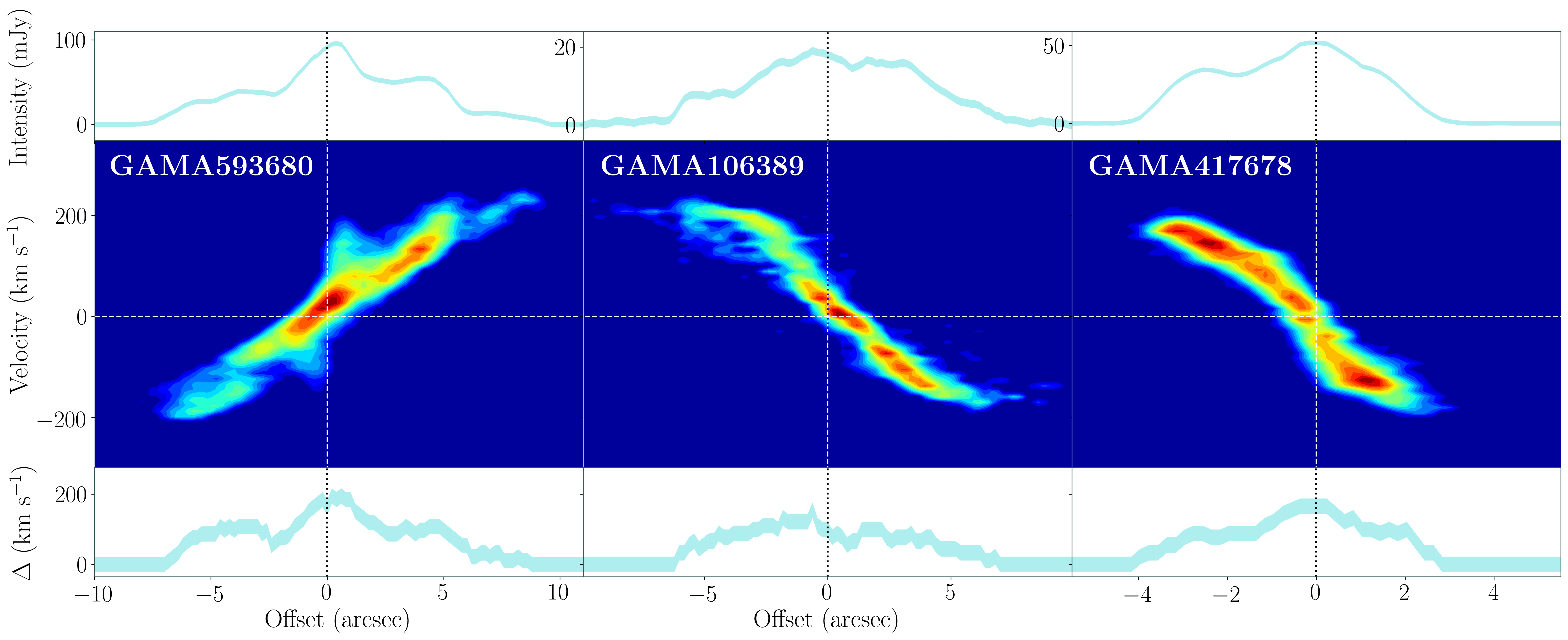}
    \caption{\textbf{From left to right}: PVDs of GAMA593680, GAMA106389 and GAMA417678 with offset on the horizontal axis in arcseconds and optical velocity (\kms) on the vertical axis with respect to the objects' redshift velocity. In each instance, the profile plotted above the PVD is the intensity as a function of offset (i.e. the profile obtained by collapsing over the velocity axis). The uncertainty is assumed as 3 standard deviations of the collapsed data (depicted as the width of the line in the profile). The bottom plots give the width in the velocity direction of the emission ($\Delta$). The width is defined as where the emission is above 2 standard deviations of the noise and the uncertainty is given by the the velocity change over a pixel width. Using a simple model, GAMA593680 (far left) is interpreted as the emission resulting from two, concentric Gaussian rings. GAMA106389 (centre) is best described by a Sersic/exponential profile. GAMA417678 is not in equilibrium and, therefore, has no simple model.}
    \label{fig: pvds}
\end{figure*}

\begin{figure}
    \centering
     \includegraphics[scale=0.4]{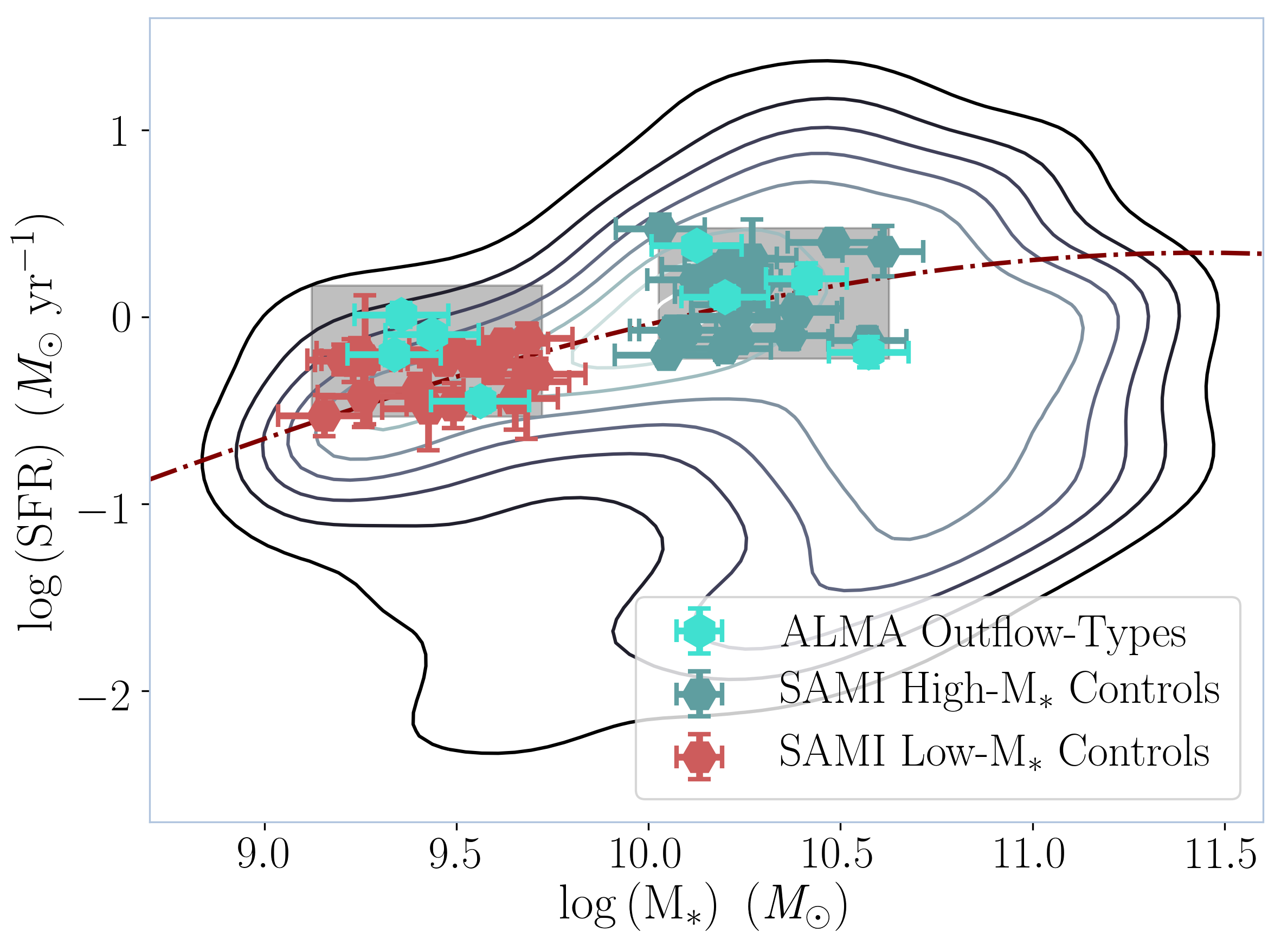}
    \caption{Selection of additional SAMI control sample in the SFR - $\rm M_*$ plane. Turquoise markers indicate ALMA outflow-type objects, which broadly fall into lower and higher stellar mass regions of the SFR - $\rm M_*$ plane (i.e. $\log(\rm M_*) < 10$ and $\log(\rm M_*) > 10$). Two control samples are drawn from the SAMI Galaxy Survey from the regions shaded in grey (which cover 0.7$\times$0.6 dex in the SFR - $\rm M_*$ plane), centred on the mean position of the low-$\rm M_*$ and high-$\rm M_*$ outflow-type objects respectively. The objects selected from SAMI Galaxy Survey are given by the red and dark blue markers in the low- and high-$\rm M_*$ regions respectively. The grey contours depict density levels in the xCOLD GASS catalogue and the black dashed line is the main-sequence trend as calculated by \citet{saintonge16}.} 
    \label{fig: SAMI_sfr}
\end{figure}

\begin{figure*}
    \centering
    \includegraphics[scale=0.28]{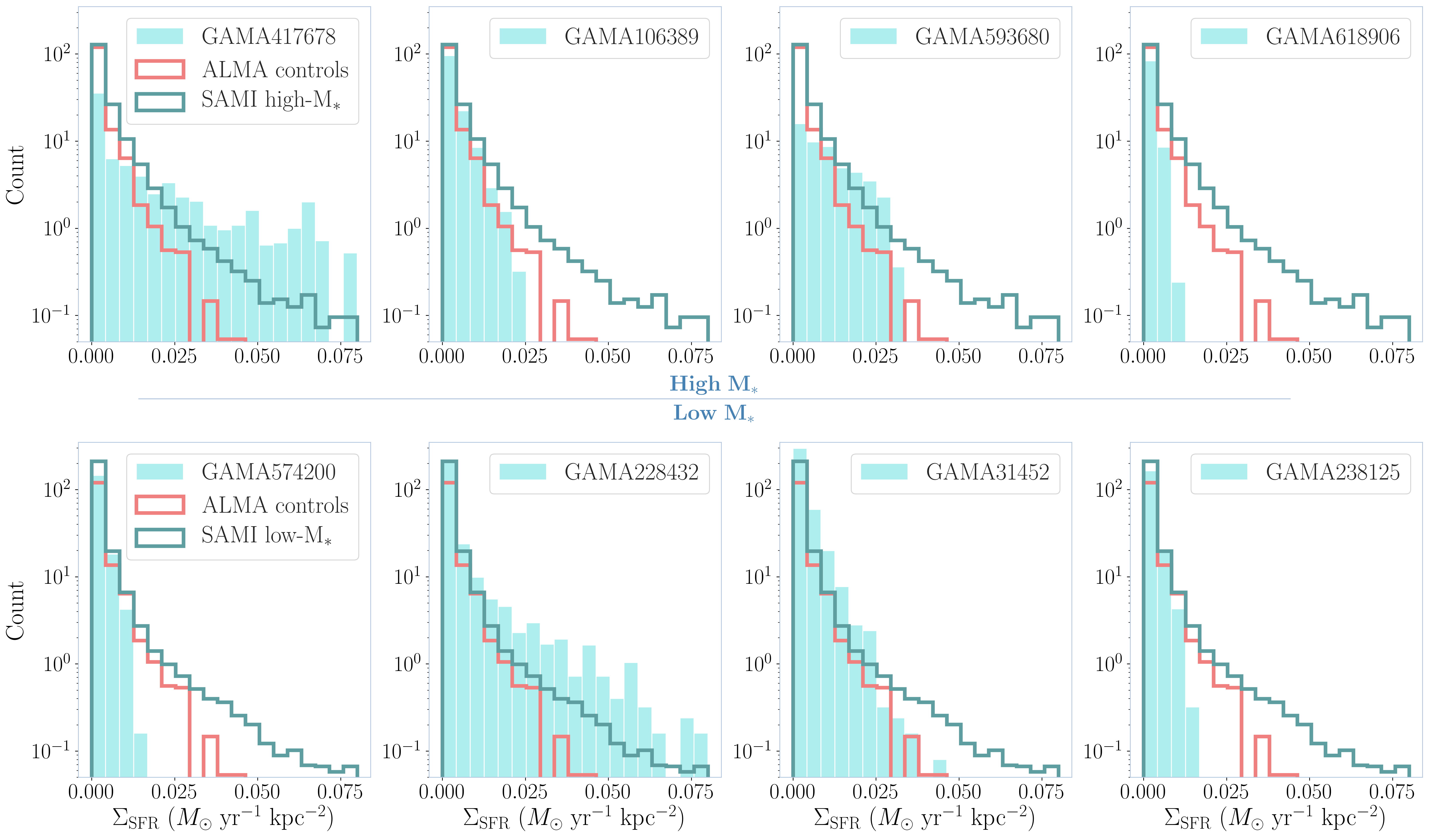}
    \caption{Histograms of SFR density per spaxel ($\Sigma _{\rm SFR}$) for each ALMA outflow-type object (turquoise bars). The objects are separated into two rows according to their stellar masses (high-$\rm M_*$ and low-$\rm M_*$) as described in the text, and ordered on each row by specific SFR (sSFR) from highest to lowest from left to right. The averaged $\Sigma _{\rm SFR}$ distributions of the ALMA control sample (red step bars) and additional SAMI control samples (low- and high-$\rm M_*$, teal step bars) are also given in each panel. We re-bin our spaxel areas to the dispersion of the SAMI PSF ($\upsigma _{\rm PSF}$), which corresponds to a physical scale of $\approx 1 \rm\ kpc$.}
    \label{fig: sfrdhists}
\end{figure*}

\begin{figure*}
    \centering
    \includegraphics[scale=0.28]{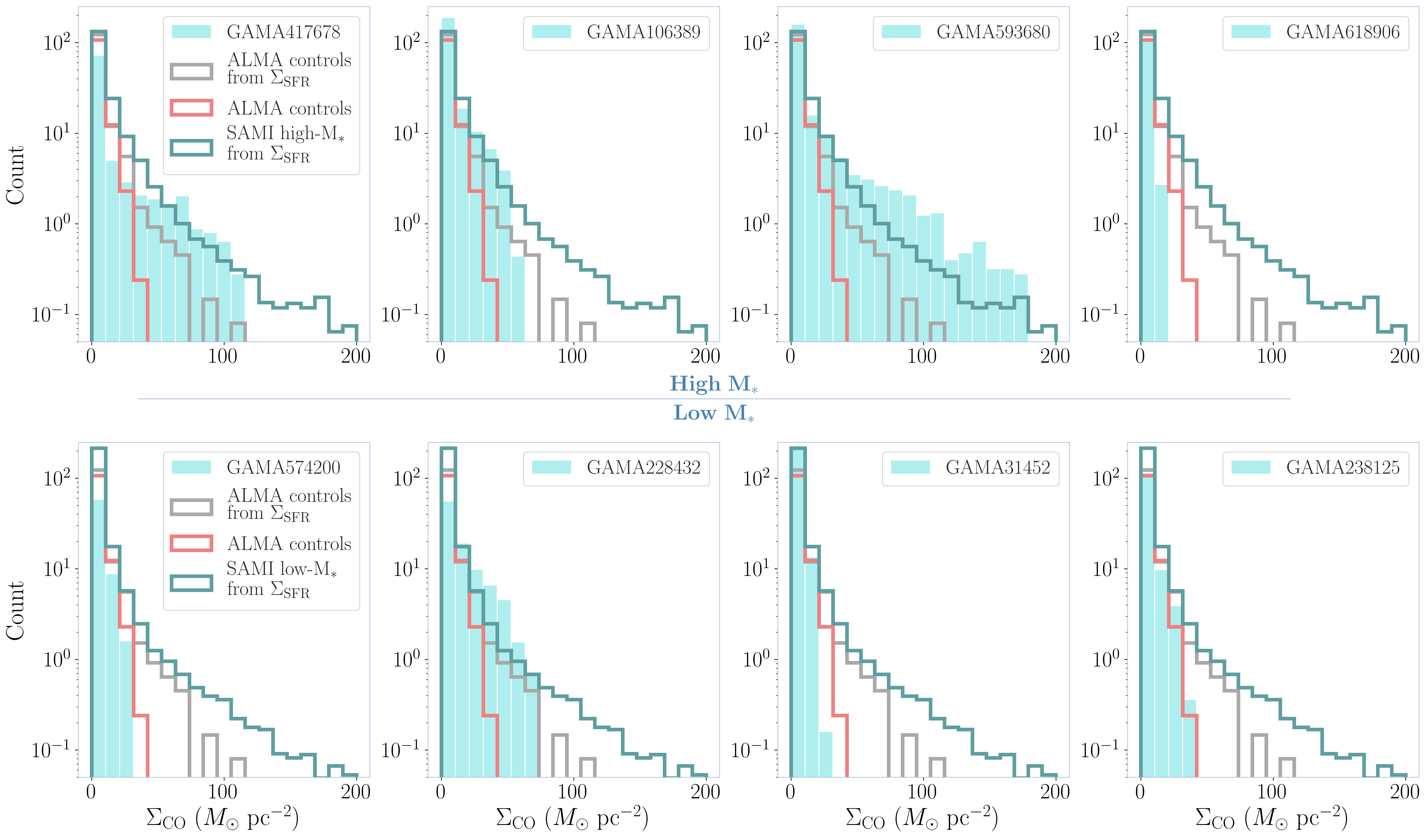}
    \caption{Histograms of molecular gas density per spaxel ($\Sigma _{\rm CO}$) for each ALMA outflow-type object (turquoise bars). The objects are separated into two rows according to their stellar masses (high-$\rm M_*$ and low-$\rm M_*$) as described in the text, and ordered on each row by specific SFR (sSFR) from highest to lowest from left to right. For comparison, the averaged $\Sigma _{\rm CO}$ distribution of the ALMA control sample (red step bars) is given in each panel. Using the $\Sigma_{\rm SFR}$ - $\Sigma_{\rm CO, H_2}$ relation determined by \citet{leroy13}, we also give the transformed ALMA control group and SAMI high-$\rm M_*$/low-$\rm M_*$ from the $\Sigma_{\rm SFR}$ spaxel distribution in Figure~\ref{fig: sfrdhists} (teal step bars). We re-bin our spaxel areas to the dispersion of the SAMI PSF ($\upsigma _{\rm PSF}$), which corresponds to a physical scale of $\approx 1 \rm\ kpc$.}  
    \label{fig: molhists}
\end{figure*}

\begin{figure*}
    \centering
    \includegraphics[scale=0.316]{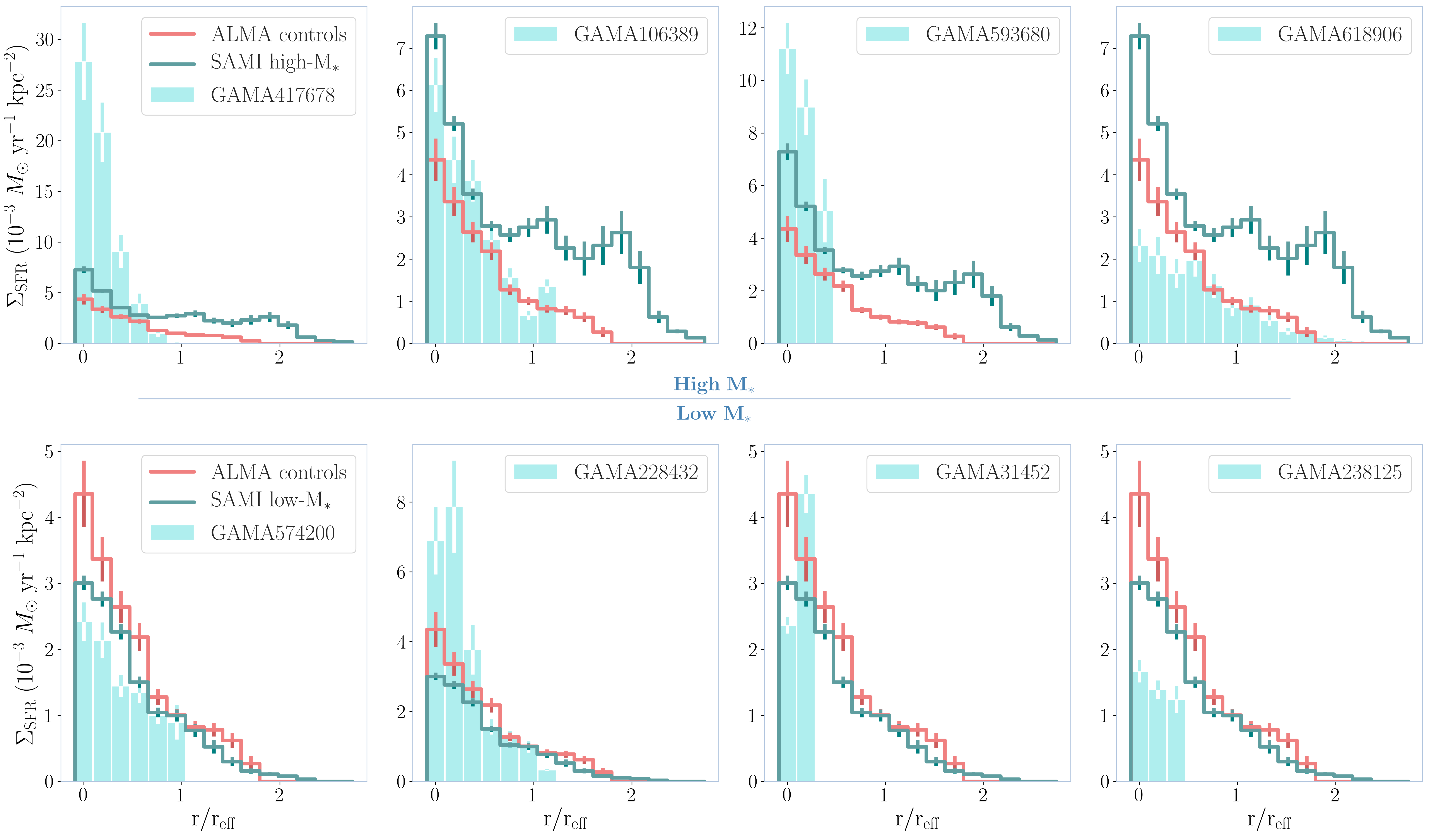}
    \caption{Bar plot depicting the average values of $\Sigma _{\rm SFR}$ spaxels (scaled to the SAMI resolution) binned by their radial distance for each outflow-type object (turquoise bars) from the centre along the plane of the edge-on disk. In each panel, we also give the averaged radial profiles derived from our ALMA control group (red step bars) and additional control groups drawn from SAMI (teal step bars, see text). The radial distances of $\Sigma _{\rm SFR}$ spaxels are normalised by the individual objects' effective radii ($\rm r_{eff}$) and the bars are binned at intervals of 1 kpc over the median $\rm r_{eff}$ of all objects in the outflow-type and ALMA control samples to approximate the width of SAMI PSF). The uncertainties represent 3 standard errors (3$\upsigma$)in the mean spaxel value of each distance bin. The outflow-type objects are separated into high-$\rm M_*$ and low-$\rm M_*$ subgroups in the figure, with the radial distributions of the high-$\rm M_*$ objects on the top row and the low-$\rm M_*$ on the bottom row (presented with the SAMI high-$\rm M_*$ and SAMI low-$\rm M_*$ additional control groups respectively). Each row is ordered by the objects' specific SFR (sSFR) from highest to lowest from left to right.}
    \label{fig: radhistsSFR}
\end{figure*}

\begin{figure*}
   \centering
    \includegraphics[scale=0.316]{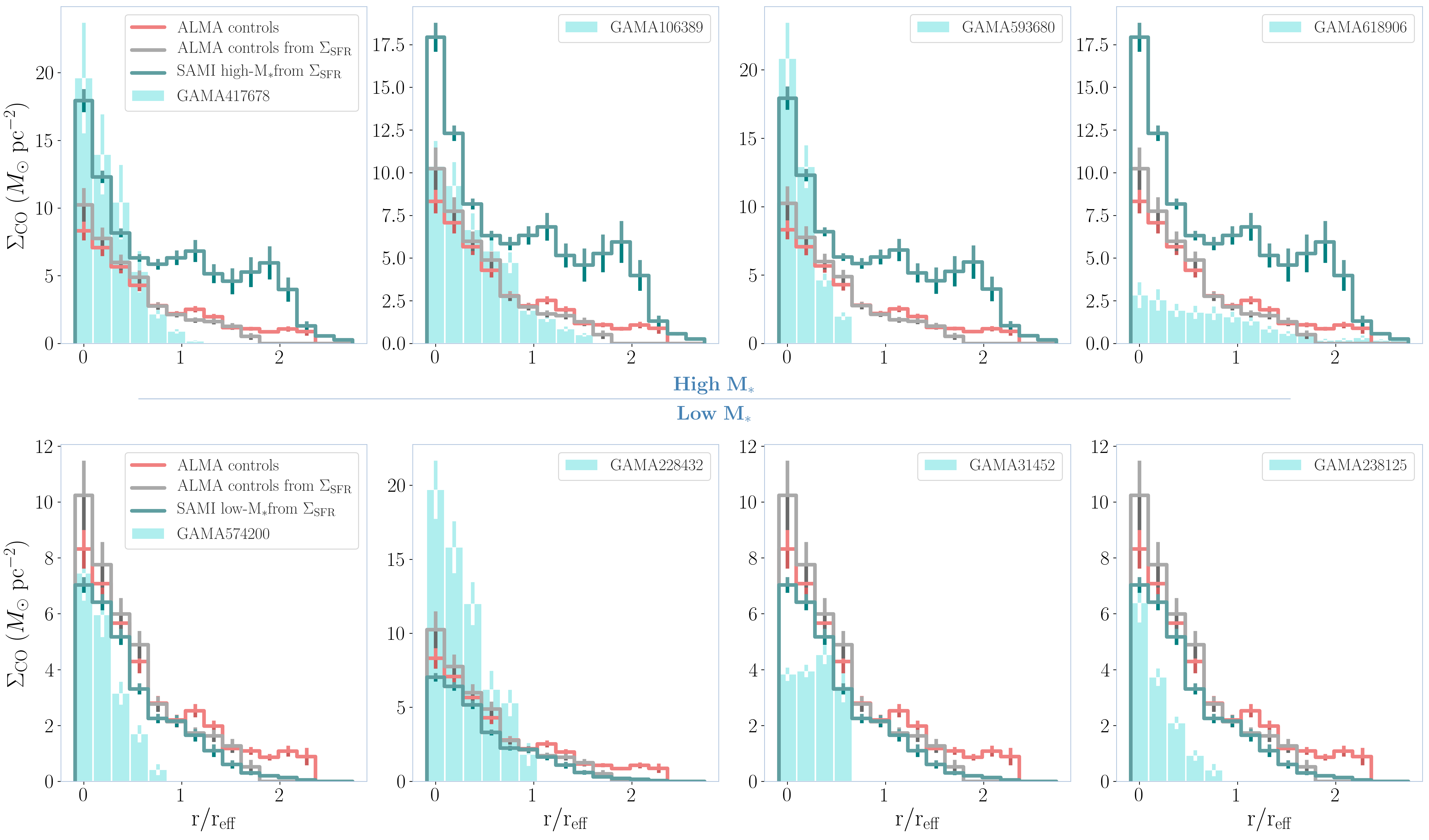} 

   \caption{Bar plot depicting the average values of $\Sigma _{\rm CO}$ spaxels (scaled to the SAMI resolution) binned by their radial distance for each outflow-type object (turquoise bars) from the centre along the plane of the edge-on disk. In each panel, we also give the averaged radial profiles derived from our ALMA control group (red step bars). The radial distances of $\Sigma _{\rm CO}$ spaxels are normalised by the individual objects' effective radii ($\rm r_{eff}$) and the bars are binned at intervals of 1 kpc over the median $\rm r_{eff}$ of all objects in the outflow-type and ALMA control samples to approximate the width of SAMI PSF). The uncertainties represent 3 standard errors (3$\upsigma$)in the mean spaxel value of each distance bin. The outflow-type objects are separated into high-$\rm M_*$ and low-$\rm M_*$ subgroups in the figure as in Figure~\ref{fig: radhistsSFR}. Using the $\Sigma_{\rm SFR}$ - $\Sigma_{\rm CO, H_2}$ relation determined by \citet{leroy13}, we also give the transformed ALMA control group and SAMI high-$\rm M_*$/low-$\rm M_*$ from the $\Sigma_{\rm SFR}$ radial profiles in Figure~\ref{fig: radhistsSFR}.} 
    \label{fig: radhistsCO}
\end{figure*}

\begin{figure*}
   \centering
    \includegraphics[scale=0.316]{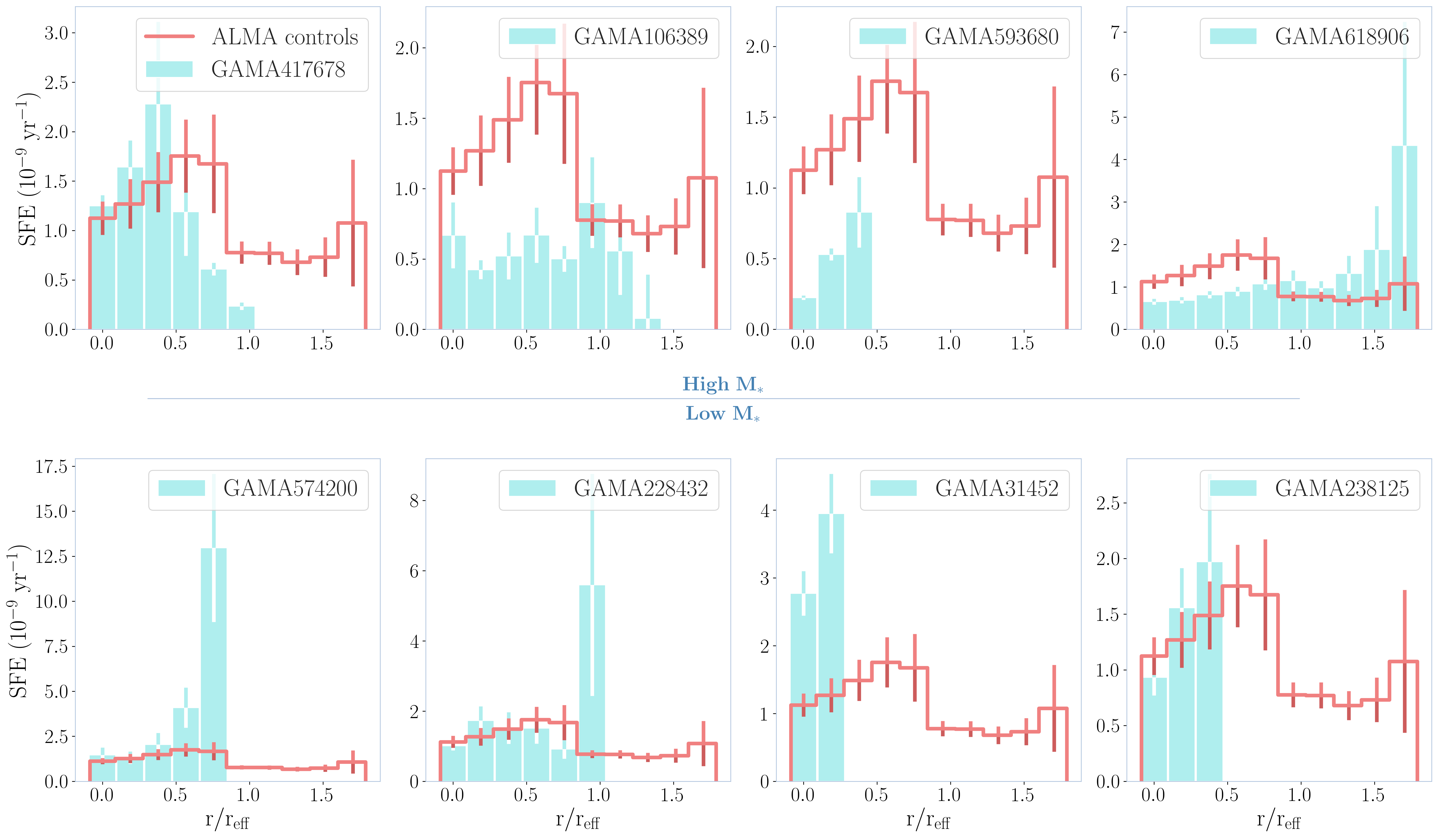} 

   \caption{Bar plot depicting the average values of SFE spaxels (scaled to the SAMI resolution) binned by their radial distance for each outflow-type object (turquoise bars) from the centre along the plane of the edge-on disk. SFE spaxels are calculated from our ALMA CO(1\textrightarrow0) maps degraded to the SAMI resolution and SFR maps derived from SAMI Galaxy Survey data products (see text). In each panel, we also give the averaged radial profiles derived from our ALMA control group (red step bars). The radial distances of SFE spaxels are normalised by the individual objects' effective radii ($\rm r_{eff}$) and the bars are binned at intervals of 1 kpc over the median $\rm r_{eff}$ of all objects in the outflow-type and ALMA control samples to approximate the width of SAMI PSF). The uncertainties represent 3 standard errors (3$\upsigma$)in the mean spaxel value of each distance bin. The outflow-type objects are separated into high-$\rm M_*$ and low-$\rm M_*$ subgroups in the figure as in Figure~\ref{fig: radhistsSFR}.} 
    \label{fig: radhistsSFE}
\end{figure*}

\begin{figure*}
   \centering
    \includegraphics[scale=0.316]{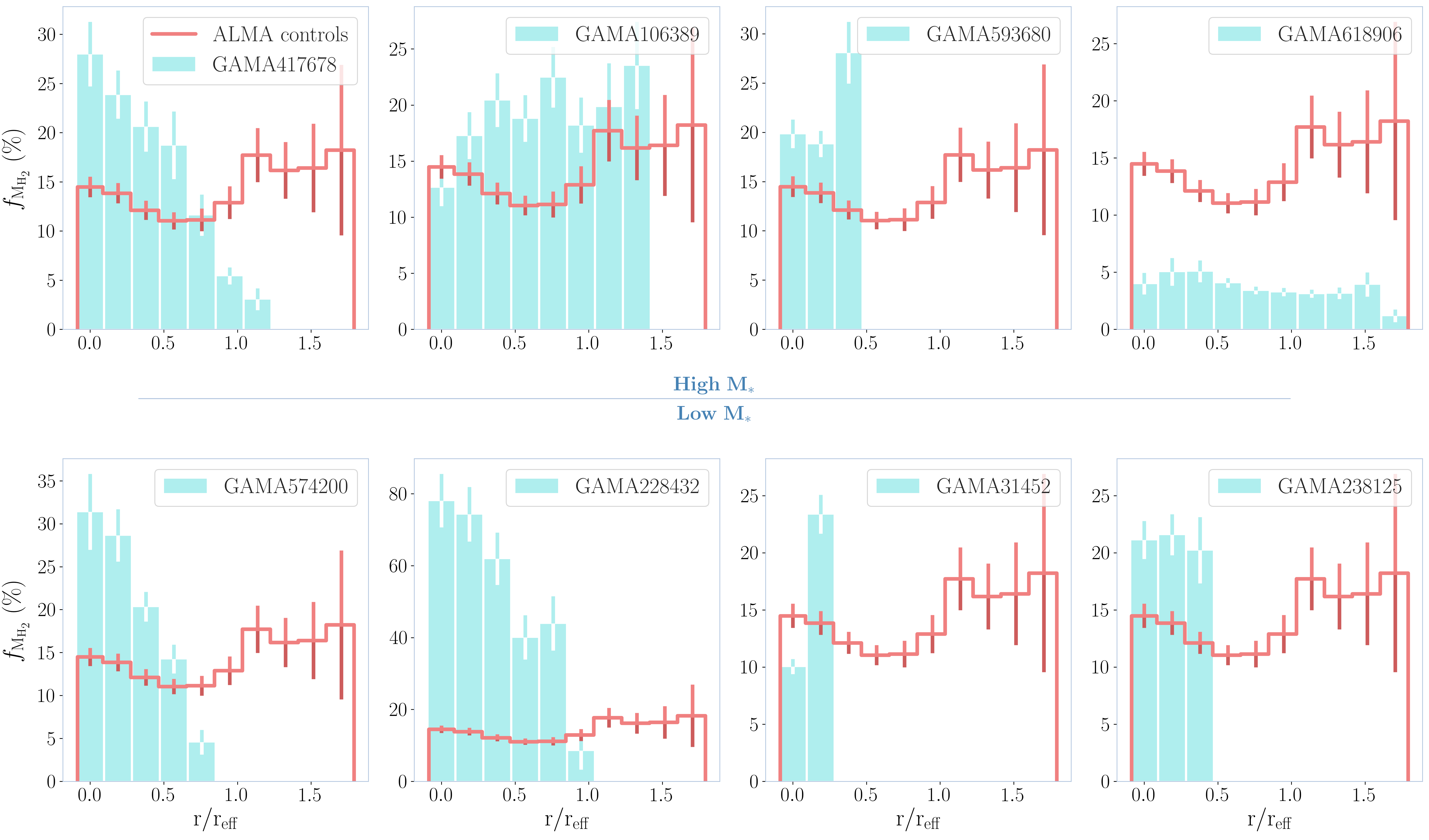} 

   \caption{Bar plot depicting the average values of $f_{\rm M_{H_2}}$ spaxels (scaled to the SAMI resolution) binned by their radial distance for each outflow-type object (turquoise bars) from the centre along the plane of the edge-on disk. $f_{\rm M_{H_2}}$ spaxels are calculated from our ALMA CO(1\textrightarrow0) maps degraded to the SAMI resolution and stellar mass maps by \citet{taylor11}. In each panel, we also give the averaged radial profiles derived from our ALMA control group (red step bars). The radial distances of $f_{\rm M_{H_2}}$ spaxels are normalised by the individual objects' effective radii ($\rm r_{eff}$) and the bars are binned at intervals of 1 kpc over the median $\rm r_{eff}$ of all objects in the outflow-type and ALMA control samples to approximate the width of SAMI PSF). The uncertainties represent 3 standard errors (3$\upsigma$)in the mean spaxel value of each distance bin. The outflow-type objects are separated into high-$\rm M_*$ and low-$\rm M_*$ subgroups in the figure as in Figure~\ref{fig: radhistsSFR}.} 
    \label{fig: radhistsGF}
\end{figure*}

We extract spatially resolved information from our ALMA data cubes in the first instance by constructing moment maps (i.e. by collapsing the cubes over the respective moment axes). Moment maps are extracted using our own subroutines over a tight spatial box ($\sim 22 \arcsec \times\ 22 \arcsec$) and trimmed channel range ($\sim 45 - 50$ channels) around the signal to minimise noise. We derive the zeroth, first and second moment maps, representing 2D spatial maps of the intensity, velocity and velocity dispersion of the gas respectively. In order to adequately cut any extra-planar emission from the cubes, we initially smooth the datacubes both spectrally and spatially with a 2D Gaussian kernel using the \texttt{Gaussian2DKernel} and \texttt{convolve} sub-routines within the \texttt{astropy.convolution} package \citep{astropy13, astropy18}. Spectrally, we smooth by a factor of 4 and spatially, we smooth by of factor of 1.5. From this smoothed cube, we create a mask by setting all values below $3\upsigma$ of the noise-level in the smoothed cube to zero in the original, un-smoothed cube. The final maps are presented in their entirety in Appendix~\ref{sec: ALMAmaps}.

\par The zeroth (intensity) moment represents the integrated values of the spectrum per spaxel and is determined by collapsing the cleaned data cubes over their spectral axes, such that the $i^{th}$ spaxel value in the map is:

\begin{equation}
    M_{0, i} = \Delta v \sum I_i \ ,
    \label{eq2}
\end{equation}

\noindent where $\Delta v$ is the width of a spaxel along the spectral axis and $I_i$ is the intensity of the $i^{th}$ spaxel. We illustrate this moment map for one of our outflow-type objects in Figure \ref{fig: HSCmom0} alongside an HSC optical image. The unmasked intensity spaxels are transformed into molecular gas mass using Equation~\ref{eq1}, again using the variable CO(1\textrightarrow0)-to-$\rm H_2$ conversion factor given by \citet{accurso17a}. The first moment is defined as the intensity-weighted coordinate and describes the intensity-weighted mean velocity of the gas in the unmasked regions of our maps. It is determined by:

\begin{equation}
    M_{1, i} = \frac{\sum I_i \cdot v_i}{M_{0, i}} \ ,
    \label{eq3}
\end{equation}

\noindent where $v_i$ is the coordinate of the $i^{th}$ spaxel along the spectral axis and other symbols are as previously defined. The final moment we calculate is the second moment, interpreted as the intensity weighted dispersion of the coordinate, typically used to describe the width of the spectral lines (which should be equivalent to the velocity dispersion of the gas in the absence of beam smearing). We define this quantity in the unmasked areas of the map as:

\begin{equation}
    M_{2, i} = \sqrt{\frac{\sum I_i \cdot {(v_i - M_{1, i})}^2} {M_{0, i}}} 
    \label{eq4}
\end{equation}

\noindent where all symbols are as previously detailed.

\par Position-velocity diagrams (PVDs) were also extracted from the masked cubes to aid analysis of our objects. We produce PVDs by rotating the CO(1\textrightarrow0) by the position angle of the object (obtained from SAMI Galaxy Survey catalogues) so the galactic plane is parallel with the y-axis of the spaxel maps. A slit is then defined in the spatial plane that extends over the kinematic major axis of the CO(1\textrightarrow0) emission. The slit width is determined adaptively by the size of slit required to contain 95\% of the CO(1\textrightarrow0) emission for each object. We then collapse the cubes in the spatial direction perpendicular to the galactic plane to produce an image in offset - velocity space. We give examples of the output of this this procedure in Figure \ref{fig: pvds}, where we show PVDs of GAMA593680, GAMA106389 and GAMA417678 (using slit widths of $1.6 \arcsec$, $2.4 \arcsec$ and $2.0 \arcsec$ respectively). PVDs for our full ALMA sample are given in Appendix~\ref{sec: PVDs}.

\par In every instance, we find the emission of CO(1\textrightarrow0) to be largely centralised in a thin disk. Furthermore, in the second moment maps in Appendix~\ref{sec: ALMAmaps} we observe that the beam-smeared velocity dispersion does not exceed $\approx 80$~\kms\ in the cases of the outflow-type galaxies, far below that typically measured along lines of sight towards molecular outflows.

\par Figure~\ref{fig: pvds} also demonstrates the gas disk properties and kinematics that we observe in our sample. From a simple visual analysis, we identify three different gas structures from the PVDs and intensity profiles: GAMA593680 (left panel) possesses no traits of the classic Sersic/exponential surface brightness profile and, by eye, would appear to be suggestive of two concentric rings, reminiscent of those caused by bar resonances. The apparent asymmetry may be due to a clumpy spiral structure within the galaxy and displays classic signatures of a central bar; GAMA106389 (centre panel) is more prosaic, having a structure and surface brightness profile suggestive of a Sersic/exponential disk at the centre of the galaxy (there is suggestion of a ``hole'' towards the centre of the distribution); GAMA417678 (right panel) appears far more extended on one side of its kinematic centre than the other, which may be indicative of a disturbed system. If the gas disk is not in equilibrium, it may be difficult to model accurately.

\par As previously discussed, in addition to the 3 viable ALMA control galaxies (GAMA496966, GAMA618935 and GAMA623679), additional control galaxies were selected from the SAMI Galaxy Survey database to strengthen the results of our resolved analysis. In Figure~\ref{fig: SAMI_sfr}, we demonstrate how our ALMA outflow-type objects fall into roughly two groups based on their stellar masses; low-$\rm M_*$ ($\log{\rm M_*} < 10$) and high-$\rm M_*$ ($\log{\rm M_*} > 10$). In order to take account of this apparent dichotomy in our outflow-type objects, we draw two independent control samples from SAMI based on the location of the ALMA low-$\rm M_*$ and high-$\rm M_*$ outflow-types in the SFR-$\rm M_*$ plane (again using MAGPHYS estimates). We select the SAMI objects for our additional control samples based on two regions surrounding the aforementioned low-$\rm M_*$ and high-$\rm M_*$ ALMA outflow-type objects, as depicted in Figure~\ref{fig: SAMI_sfr}. The area of both regions in the SFR-$\rm M_*$ plane are $0.6 \times 0.7$ dex centred on the mean position of the low-$\rm M_*$ and high-$\rm M_*$ outflow-types respectively (both regions have the same area, where the region size approximates the maximum scatter in $\rm M_*$ and SFR). The possible redshifts and ellipticities of the SAMI controls are limited to below the maximum values in the outflow-type sample (the final selection is given in Figure~\ref{fig: SAMI_sfr}). We also discount any AGN-contaminated objects by excluding those with [\ion{N}{ii}]/H$\upalpha$ and [\ion{O}{iii}]/H$\upbeta$ ratios indicative of AGN-excitation in the BPT diagram.

For our resolved analysis, we directly compare resolved CO gas content with resolved SFR maps for each object. SFR maps are determined from SAMI H$\upalpha$ emission maps. We use the emission line maps instead of the SFR map data products provided by SAMI in order to make a less conservative estimate on masked spaxels \citep[see][]{bryant15, medling18}. In our method, we include all spaxels from the SAMI H$\upalpha$ emission maps that have corresponding [\ion{N}{ii}]/H$\upalpha$ and [\ion{O}{iii}]/H$\upbeta$ ratios such that they fall within the star-forming region of the BPT diagnostic diagram (SAMI also require the [\ion{S}{ii}]/H$\upalpha$ and [\ion{O}{i}]/H$\upalpha$ ratios to fall in the star-forming region). Spaxels outside this region are masked. We then correct our masked H$\upalpha$ maps ($\rm {H\upalpha}_{mask}$) for extinction using the same method described in \citet{bryant15, green18, scott18}:

\begin{equation}
    \rm {H\upalpha} _{cor, mask} = \frac{1}{1.53} {\left[\frac{H\upalpha}{H\upbeta}\right]}_{ex} {H\upalpha}_{mask} 
\end{equation}

\medskip

\noindent where $\rm {\left[H\upalpha/H\upbeta\right]}_{ex}$ is the extinction map provided in the SAMI data products. SFR maps are then calculated by $\rm SFR = 7.9\times {10}^{-42}\ L[{H\upalpha}_{cor, mask}]$, where $\rm L[{H\upalpha}_{cor, mask}]$ is the luminosity from $\rm {H\upalpha} _{cor, mask}$.

\par Both our ALMA CO(1\textrightarrow0) maps and SAMI SFR maps are re-binned to match the observational SAMI resolution by scaling the spaxel sizes to the dispersion of the PSF ($\upsigma _{\rm PSF}$), which corresponds to a physical scale of $\approx 1 \rm\ kpc$. We conduct 25 re-bins of each map with 5 different offsets in the re-binning grid in the RA and Dec directions (corresponding to offsets of $0 - 0.8\ \times$ bin size). This is in order to reduce artefacts arising due to the position of the re-binning grid \citep[a similar method is used in][]{zabel20}. The re-binned spaxel areas are converted from angular size ($\uptheta _{\rm spaxel}$) to physical scale using GAMA catalogue redshifts (z) with the \texttt{astropy.cosmology} package \citep{astropy13, astropy18}. Spaxel areas $\rm A_{\rm spaxel}$ are corrected for inclination ($i$) using the GAMA survey ellipticity values from SersicCatAllv07 \citep{kelvin12}, such that the spaxel areas equate to:

\begin{equation}
    \rm A_{\rm spaxel} = \frac{1}{\cos i} {\left[\uptheta _{\rm spaxel} \frac{D_L}{{(1+z)}^{\ 2}}\right]}^2
    \label{eq5}
\end{equation}

\medskip 

\par Figure~\ref{fig: sfrdhists} bins $\Sigma _{\rm SFR}$ maps for our outflow-type, ALMA control and SAMI control samples by spaxel values. Given the diversity of global properties we find in Figure~\ref{fig: SAMI_sfr}, in Figure~\ref{fig: sfrdhists} we give the spaxel distribution for each ALMA outflow-type object in separate panels instead an averaged sample over the group to avoid domination of individual objects in a statistically small sample or the loss of information by assuming these objects will have similar SFR distributions. The low-$\rm M_*$ and high-$\rm M_*$ outflow-type objects are on the bottom and top rows of Figure~\ref{fig: sfrdhists} respectively and ordered on each row by specific SFR (sSFR) from highest to lowest from left to right. In each panel, we also plot the averaged histograms for both the ALMA and SAMI controls (i.e. histograms including spaxels from every object in those samples), which are scaled by the number of objects in the respective groups in order to allow visual comparison. 
 
\par We do not find consistent evidence of higher $\Sigma_{\rm SFR}$ in our ALMA outflow-type objects Figure~\ref{fig: sfrdhists} in relation to the ALMA control sample or additional SAMI controls. In the instances of GAMA417678 and GAMA228132 (a high-$\rm M_*$ and a low-$\rm M_*$ object respectively), there does appear to be a significant elevation in the dense spaxel tail of their $\Sigma_{\rm SFR}$ distribution. However, the majority of the outflow-types do not appear to possess higher $\Sigma_{\rm SFR}$ spaxels compared to the distributions drawn from the ALMA and SAMI control samples.  

\par In Figure~\ref{fig: molhists}, we bin the molecular gas mass density ($\Sigma _{\rm CO}$) for our ALMA outflow-type objects. To construct these $\Sigma _{\rm CO}$ histograms, we obtain CO(1\textrightarrow0) intensity maps by calculating the zeroth moment (Equation~\ref{eq2}) from the ALMA cubes and re-binning to the SAMI resolution by the same method described for the SAMI $\Sigma _{\rm SFR}$ maps. In Figure~\ref{fig: molhists}, we also include $\Sigma _{\rm CO}$ distributions derived from the $\Sigma_{\rm SFR}$ histograms in Figure~\ref{fig: sfrdhists} of our ALMA control and additional SAMI control samples. These distributions are converted into $\rm H_2$ gas density ($\Sigma_{\rm CO,\ H_2}$) using the relation determined by \citet{leroy13}:

\begin{equation}
    \Sigma_{\rm CO,\ H_2} = 10\ M_\odot \rm pc^{-2} \left(\frac{\Sigma_{\rm SFR, \ H\upalpha+24\mu m }}{-2.35\pm0.08}\right)^{\nicefrac{1}{0.95\pm 0.13}}
    \label{eq6}
\end{equation}

\medskip

\noindent using the parameters from \citet{leroy13} from their fit of $\rm H\upalpha+24\mu m$ with $\rm H_2$ gas density. The panel structure in the figure is equivalent to that in Figure~\ref{fig: sfrdhists}. 

\par Again, we do not find a consistent elevation of $\Sigma _{\rm CO}$ in our ALMA outflow-type objects with regard to the ALMA control sample or SAMI control samples. However, as is also the case in Figure~\ref{fig: sfrdhists}, we do find evidence of higher $\Sigma _{\rm CO}$ spaxels compared to the control samples in a portion of our outflow-type objects. Low-$\rm M_*$ object GAMA228432 and high-$\rm M_*$ objects GAMA106389, GAMA417678 and GAMA593680 possess an enhancement of $\Sigma _{\rm CO}$ spaxels with respect to the ALMA control sample (where both GAMA228432 and GAMA417678 were also noted in Figure~\ref{fig: sfrdhists} to have elevated $\Sigma _{\rm SFR}$ spaxels). Only GAMA593680, however, appears to possess significantly enhanced $\Sigma _{\rm CO}$ relative to the SAMI control sample (transformed via Equation~\ref{eq6}). It should be noted that this does not correspond with either of the objects observed to have $\Sigma _{\rm SFR}$ enhancement in Figure~\ref{fig: sfrdhists} (i.e. GAMA417678 and GAMA228432). 

\par In Figure~\ref{fig: radhistsSFR}, we assess the distribution of $\Sigma _{\rm SFR}$ with respect to radius along the plane of the edge-on disk (where the radial positions of the spaxels are normalised by the objects' effective radii, $\rm r_{eff}$). In these bar plots, spaxels are binned by their distance from the centre of the galaxy \citep[the centre defined by the RA and Dec coordinates drawn from the TilingCatv29 catalogue in the GAMA Survey,][]{driver09, baldry10} in the direction along the plane \citep[i.e. determined by the position angle given by the SersicCatAllv07 catalogue in the GAMA Survey,][]{kelvin12}. The sizes of the bars in Figure~\ref{fig: radhistsSFR} are determined by the SAMI resolution in kpc (approximated as the dispersion of a SAMI PSF of $\upsigma_{\rm PSF} \approx 2\arcsec$ at the most distant galaxy) divided by the median $\rm r_{eff}$ across our ALMA outflow-types and ALMA controls. The mean spaxel value in each distance bin is then calculated to give the height of the bars (where the 3$\upsigma$ errors is taken as the uncertainty). Again, we present the distribution for each outflow-type individually and average together the objects in the comparative control groups. The layout of the panels in the figure is equivalent to those in Figure~\ref{fig: sfrdhists} \& \ref{fig: molhists}.

\par In the $\Sigma _{\rm SFR}$ radial distribution, we observe a greater degree of central concentration in our ALMA outflow-type objects with respect to the ALMA control and additional SAMI control samples. With the exception of GAMA618906, the $\Sigma _{\rm SFR}$ radial distributions of our ALMA outflow-type objects are confined to their inner $\lesssim 1.5 \rm\ r_{eff}$. However, we do not see a consistent elevation in mean $\Sigma _{\rm SFR}$ spaxel values with radius in the ALMA outflow-types compared to the control distributions. With regard to the ALMA control and SAMI low-$\rm M_*$ and high-$\rm M_*$ control samples, we observe a greater similarity between the ALMA control radial distribution with that of the SAMI low-$\rm M_*$ controls. The ALMA control sample is comprised of three objects, two of which are in the SAMI low-$\rm M_*$ control sample, with the other lying outside, but between, the SAMI low-$\rm M_*$ and high-$\rm M_*$ regions in the SFR-$\rm M_*$ plane. This may suggest that our ALMA control sample is more suitable as a control group for our low-$\rm M_*$ outflow-type objects. The difference in the SAMI low-$\rm M_*$ and high-$\rm M_*$ radial profiles also supports our treating of these two groups in our outflow-types in isolation. In comparison to the SAMI high-$\rm M_*$ $\Sigma _{\rm SFR}$ radial profile, all our high-$\rm M_*$ outflow-types have a more centralised distribution.         

In Figure~\ref{fig: radhistsCO}, we plot the radial distribution of the $\rm \Sigma _{CO}$ spaxels in our objects. The bar plot panels have the same format as Figure~\ref{fig: radhistsSFR} (i.e. bar width, panel order etc.). As in Figure~\ref{fig: molhists}, we transform the radial $\Sigma _{\rm SFR}$ distribution of the ALMA controls along with the SAMI low-$\rm M_*$ and high-$\rm M_*$ controls given in Figure~\ref{fig: radhistsSFR} into $\Sigma_{\rm CO}$ using Equation~\ref{eq6}. We do this in order to validate our ALMA control sample, which as in the $\Sigma _{\rm SFR}$ radial distribution, is in good agreement with the SAMI low-$\rm M_*$ $\Sigma _{\rm CO}$ radial profile (the similarity between the $\Sigma _{\rm CO}$ distribution derived from our ALMA control maps and that calculated using Equation~\ref{eq6} from the $\Sigma _{\rm SFR}$ distribution also validates our use of the conversion for our additional SAMI control samples). 

\par Again we find a greater central concentration of $\Sigma _{\rm CO}$ in our ALMA outflow-type objects in comparison to our ALMA and SAMI control samples. If we take the high-$\rm M_*$ SAMI control group as the comparative controls for the ALMA high-$\rm M_*$ outflow-type objects (i.e. the top row in Figure~\ref{fig: radhistsCO}), this centralisation is particularly clear. For the majority of our outflow-type objects, the gas distribution lies in the inner $\approx\rm\ r_{eff}$ of their respective disks, except for GAMA106389 and GAMA618906, which have distributions out to $\approx\rm\ 2 r_{eff}$. However, the two aforementioned objects still have gas distributions that are more centrally distributed with regard to the SAMI high-$\rm M_*$ controls. The $\Sigma _{\rm CO}$ distribution for our ALMA outflow-type objects, therefore, strongly resembles that of the $\Sigma _{\rm SFR}$ profiles given in Figure~\ref{fig: radhistsSFR}. Similarly, we do not find higher density gas spaxels with respect to radius consistently in our objects, which we also infer from Figure~\ref{fig: molhists}.    

\par Our global values for SFE and gas fraction ($\rm f_{M_{H_2}}$) suggest little difference between our outflow-type objects and the xCOLD GASS control sample (matched to the outflow-types in SFR and stellar mass) in terms of their integrated gas content. In Figure~\ref{fig: radhistsSFE} \& \ref{fig: radhistsGF}, we analyse the radial distribution of SFE and gas fraction spaxels. As in Figures~\ref{fig: radhistsSFR} \& \ref{fig: radhistsCO}, spaxels are binned with respect to distance from the objects' centre along the plane of the disk. Constructing both the SFE and $\rm f_{M_{H_2}}$ maps requires the direct combination of both SAMI and ALMA data (i.e. $\rm SFE = SFR/M_{\rm H_2}$, $\rm f_{M_{H_2}} \rm = M_{H_2} / M_*$). In order to directly compare both datasets, the ALMA zeroth moment maps (created as previously detailed) are degraded to match the SAMI spatial resolution. The resolution is degraded by convolving the ALMA map with a 2D Gaussian kernel using the \texttt{Gaussian2DKernel} and \texttt{convolve} sub-routines within the \texttt{astropy.convolution} package \citep{astropy13, astropy18}. We define the ALMA spatial resolution as the beam width across its major and minor axes ($\updelta _{\rm beam_{maj}}$ and $\updelta _{\rm beam_{min}}$) rotated with respect to the beam angle, so that we degrade the ALMA maps with a 2D Gaussian kernel of width:

\begin{equation}
    \upsigma _{\rm degrade_{x, y}} = \sqrt{{\upsigma _{\rm PSF}}^2 - {\updelta _{\rm beam_{maj, min}}}^2} \ ,
    \label{eq7}
\end{equation}

\noindent where $\upsigma _{\rm PSF}$ is the SAMI resolution.

\bigskip

\par To construct the SFE maps, we divide SFR and molecular gas mass map spaxels and produce Figure~\ref{fig: radhistsSFE} by the same method used to create the $\Sigma _{\rm SFR}$ and $\Sigma _{\rm CO}$ radial plots in Figure~\ref{fig: radhistsSFR} \& \ref{fig: radhistsCO}. The radial limit of the plot only extend to $2 \rm r_{eff}$ due to the lack of significant SFE spaxels in the outer regions of the objects. For our radial SFE profiles in Figure~\ref{fig: radhistsSFE}, we use the ALMA controls as the control sample. We do not observe any consistent enhancement of SFE in our outflow-type objects with regard to the ALMA controls. From this, we could infer that we are not observing a more efficient process of star-formation in the outflow-type objects compared to that taking place in the ALMA controls.     

Our $\rm f_{M_{H_2}}$ maps require maps of stellar mass, which have been derived from pixel-matched ugriZYJ imaging from VST-KiDS and VISTA-VIKING. The imaging has been reprocessed by the GAMA survey \citep{driver16} and re-projected to match the pixel scale of the SAMI cubes ($0.5 \arcsec {\rm pix}^{-1}$). Stellar masses are calculated from the synthetic stellar population modelling of the per-pixel u-J SEDs, in a similar way to the GAMA stellar mass estimates \citep{taylor11}. We divide the molecular gas mass (again degraded to the SAMI resolution) and stellar mass spaxels and bin radially using the same method described for Figure~\ref{fig: radhistsSFE} to produce a radial distribution of $\rm f_{M_{H_2}}$ for the ALMA outflow-types and ALMA control sample. Similar to the radial profiles plotted for $\Sigma _{\rm SFR}$ and $\Sigma _{\rm CO}$, we observe a more centralised radial distribution of $\rm f_{M_{H_2}}$ in the outflow-types compared to the flatter distribution in the ALMA controls. We also see higher values of $\rm f_{M_{H_2}}$, in particular but not exclusively, within the inner $\approx 0.5 \rm r_{eff}$ of the outflow-types (again, with the exception of GAMA618906). This may suggest that our outflow-types possess higher gas content relative to their stellar mass inside their innermost regions compared to the ALMA control group.            

\section{Discussion}
\label{sec: discussion}

\subsection{Equivalent Outflow-Control Global Gas Content}
\label{subsec: gasrich}

Following Figure~\ref{fig: delgasfrac}, we observe no notable difference in global molecular gas fraction between our outflow-type sample and their xCOLD GASS control objects (see Section~\ref{subsec: glob}). This would appear to contradict expectations that galaxies with intense ionised outflows would be more gas-poor due to the entrainment of their molecular gas in the outflowing stellar wind. This result is also at odds with previous studies that have shown differences in gas content in galaxies harbouring intense ionised outflows and those that do not. However, we note the absence in this analysis of \ion{H}{i} gas content in our outflow-types and controls (as we do not have \ion{H}{i} for all our ALMA outflow-type/control sample). It remains possible, therefore, that the samples contain different total amount of cold gas (defined as \ion{H}{i}+$\rm H_2$). \citet{roberts-borsani20} find indications of lower \ion{H}{i} gas fractions in their outflow-type sample compared to their corresponding control group. This may suggest that the total cold gas fraction in our outflow-type sample may be depleted compared to the controls, while their molecular gas fractions are equivalent. Nevertheless, it is clear from this work that the outflow-type objects do not have a significantly different molecular fraction (i.e. $\rm M_{\rm H_2}/M_{\odot}$) compared to non-outflow-type galaxies with normal levels of star-formation. However, we must also consider that the gas potentially being expelled from our outflow-types may not escape the objects and may instead fall back down to the disk \citep[``galactic fountain model'',][]{shapiro76}. Given the relatively low velocity dispersion measurements we find in our ALMA moment maps (see Appendix~\ref{sec: ALMAmaps}), it is unlikely that the gas possesses outflow velocities greater than circular velocity of the objects (i.e. $\rm \upsigma_{CO} < v_{circ}$), meaning the gas would have insufficient energy to escape the gravitational well of the host galaxy \citep[see][]{roberts-borsani19}. Any outflowing gas, therefore, may be recycled down as an inflow, meaning our integrated gas fraction values would not appear depleted by the presence of outflows in our objects. Differences in integrated gas fraction between our outflow-types and control objects may require higher energy outflows than we observe in our ALMA outflow-type sample. However, it should be noted that our ALMA objects are highly inclined meaning the $\rm \upsigma _{CO}$ values we measure are largely dominated by motions into/out of the plane of the disks. Outflows along the minor axes, therefore, would be difficult to detect in $\rm \upsigma _{CO}$. As the gas we observe is contained within a thin disk, without significant CO(1\textrightarrow0) emission outside the plane, it would seem that these objects are not entraining large quantities of molecular gas.    

In our resolved analysis in Section~\ref{subsec: spatresob}, Figure~\ref{fig: radhistsCO} shows evidence that, despite the similarity in global gas content between our outflow-type and xCOLD GASS control galaxies, the spatial distribution of molecular gas is not equivalent when compared to the resolved ALMA and SAMI control samples. We find that our outflow-type objects possess more central gas distributions and a steeper decline of gas density with radius, with most harbouring their gas within their inner $\approx\rm r_{eff}$. This is compared to the ALMA and SAMI control groups that have less central $\Sigma_{\rm CO}$ distributions with a shallower radial decline. This is consistent with the idea that $\Sigma_{\rm SFR}$ is the primary driver of outflows and that outflows are strongest in the central regions of galaxies \citep[see][]{roberts-borsani20}. However, as shown by Figure~\ref{fig: molhists}, we do not find denser spaxels of $\Sigma_{\rm CO}$ in our outflow-types, nor a higher incidence of dense spaxels, with respect to the ALMA and SAMI control groups. Our radial distribution of molecular gas in Figure~\ref{fig: radhistsCO}, therefore, suggests that the densest regions of gas in our outflow-types are located in the central regions of the objects, whereas in the ALMA and SAMI control samples, the dense regions are less centrally concentrated.

\subsection{Centralised $\Sigma _{\rm SFR}$ and $\Sigma _{\rm CO}$ in Outflow-Type Galaxies}

Figure~\ref{fig: radhistsSFR} illustrates the difference in the distribution of $\Sigma _{\rm SFR}$ with radius between our ALMA outflow-types and our control samples. As previously noted, the outflow-type galaxies appear to possess more centrally distributed star-formation compared to the ALMA and SAMI controls, which display a comparatively shallow decline of $\Sigma _{\rm SFR}$ with radius. One can infer from our findings that the star-formation in outflow-type galaxies occurs in a well-defined, central region, which could induce powerful stellar winds emanating from that central area. This result is in agreement with \citet{roberts-borsani20}, where $\Sigma_{\rm SFR}$ is also found to extend out to $\approx \rm r_{eff}$ in their outflow-type objects. However, we note that we don't find evidence of a consistent elevation of $\Sigma_{\rm SFR}$ spaxels in the outflow-type objects relative to the control samples. Currently, star-formation is the main mechanism theorised to drive supersonic turbulence within the ISM in low redshift galaxies \citep{maclow04}. If this hypothesis holds, cold gas should be entrained by highly turbulent stellar winds within the central regions of our outflow-type galaxies, potentially expelling it along with large-scale outflows of ionised gas. Within this work, we do not find evidence of this by either detecting spatially extended emission in the zeroth moment maps (see Figure~\ref{fig: HSCmom0}) or by kinematic signatures in the PVDs (see Figure~\ref{fig: pvds}). However, using stacking in extraplanar regions in future analyses may reveal the presence of CO gas entrained within the ionised gas ejections from our outflow-type objects. 

\par Figure~\ref{fig: radhistsCO} also implies a centralisation of $\Sigma _{\rm CO}$ spaxels in our outflow-type objects compared to the ALMA and SAMI controls. This is in agreement with analysis conducted by \citet{schruba11} on the CO(2\textrightarrow1) emission of local spiral galaxies within the IRAM HERACLES survey. They report a characteristic exponential decline of CO(2\textrightarrow1) emission in the central region of their galaxy sample with no sharp cut-off. Furthermore, \citet{roberts-borsani20} find signatures of outflow activity, using the \ion{Na}{d} $\lambda \lambda5889,5895\ \angstrom$ neutral gas tracer, to be largely confined to the central $\approx\rm r_{eff}$ of their outflow-type objects from the MaNGA DR15 survey \citep{bundy15}. Outside this central region, they find a steep decline in mass outflow rates and loading factors, implying outflow activity occurs almost exclusively in the centre of galaxies. The centralisation of molecular gas in our outflow-types, and those in other studies, suggests the existence of a mechanism driving gas inwards toward the inner $\approx\rm r_{eff}$ in these objects.

\par We also find in our resolved radial distribution a centralisation of molecular gas fraction spaxels in our outflow-type objects compared to the ALMA control sample. The distribution of outflow-type and ALMA control gas fraction spaxels in Figure~\ref{fig: radhistsGF} suggest that the outflow-types have a central enhancement (within the inner $\approx\rm 0.5 r_{eff}$) of molecular gas relative to the stellar mass, unlike the ALMA controls. In Figure~\ref{fig: radhistsCO} we do not observe a consistent elevation of central molecular gas content in the ALMA outflow-types with respect to the ALMA control sample, but do see a steeper decline with radius. The central enhancement of gas fraction must, therefore, be driven by a greater proportion of gas with respect to the stellar mass in the inner $\approx\rm 0.5 r_{eff}$. Figure~\ref{fig: radhistsGF} further indicates that the gas distribution is more regular in the ALMA control group with regard to the stellar population and suggests there is a greater supply of gas in their outer regions relative to the stellar mass. Furthermore, the similarity of the SFE radial profiles between our outflow-type and ALMA controls implies that there is not a more efficient mechanism operating within galaxies harbouring large-scale galactic outflows and that SFE is fairly uniform throughout their disks. 

\par This enhanced central gas fraction, in combination with the evidence of intense ionised outflows, could be further evidence of a ``galactic fountain'' scenario, where ejected molecular gas is falling back down to the disk (i.e. it is not escaping the gravitational well of the host galaxy). The molecular gas fraction in our outflow-type objects would, therefore, not be depleted by the existence of outflow activity. After falling back to the disk, a separate mechanism may then act to drive the gas inwards to the central regions of the outflow-type object (see Sections~\ref{sec: bars} \& \ref{sec: interact}). In contrast to our findings concerning the molecular gas fraction, \citet{roberts-borsani20} find evidence that the neutral gas fraction is depleted in the central regions of their outflow-type sample (where \ion{H}{i} is measured through stacking of \ion{H}{i} 21~cm observations). The combination of these results suggest that while molecular $\rm H_2$ gas is driven towards the centre of outflow-type galaxies, \ion{H}{i} is simultaneously being propelled towards their outer regions, perhaps tracing the ionised wind. Equally, \ion{H}{i} that does fall inwards may be converted into $\rm H_2$ in the centre of the objects due to high pressures in the central regions. 

\par We can infer from our findings that a mechanism operates in the ALMA outflow-types that is driving gas towards the centre of the objects and stripping their outer regions of the surrounding gas.

\subsubsection{Bar-Driven Gas Transport}
\label{sec: bars}

\par The difference in distributions of $\Sigma _{\rm SFR}$ and $\Sigma _{\rm CO}$ between our outflow-types and their respective control samples is likely indicative of a different mechanism governing the dynamics of gas within the two galaxy classifications. \citet{jogee05} explore the ability of bars to re-distribute angular-momentum in a galaxy, efficiently driving molecular gas into its central regions. Large asymmetric structures, such as bars, are capable of exerting immense torques on the surrounding gas, causing it to lose angular momentum and, consequently, fall inwards.

\par Figure~\ref{fig: pvds} illustrates, with three distinct examples, the wide variety in the morphology and kinematics of the gas disks within our ALMA objects. \citet{jogee05} also note this diversity of CO morphologies within their sample of barred galaxies. In particular, they suggest that an extended molecular gas distribution is indicative of the early stages of bar-driven inflows (meaning a large fraction of the circum-nuclear gas is still along the large-scale stellar bar). By contrast, bar structures in a later stage of their evolution are typically found to hold their circum-nuclear gas within the OILR (outer inner Lindblad resonance) of the bar. This gas is likely to exhibit principally circular motion as opposed to gas in a younger bar structures which tends towards non-circular kinematics. 

\par Unlike \citet{jogee05}, \citet{sheth05} make the distinction that only early-type Hubble barred spirals (i.e. \textit{CO-bright} galaxies) demonstrate the very disparate environment within the central 1-2 kpc compared to the outer regions. This difference between early-type and late-type barred spirals is typically attributed to the higher critical surface densities required for star-formation in Early-types, resulting from steeper rotation curves \citep{ho97}. \citet{ho97} predict, consequently, that early-type barred spirals will have higher central gas concentrations. Furthermore, they theorise that in order for these Early-type spirals to maintain their high density cores, they must have substantially higher mass inflow rates compared to late-type spirals. Our outflow-type objects may, therefore, be more consistent with Late-type barred spirals, possessing a central symmetry sufficient to cause molecular gas to fall inwards but without the high gas densities and SFR observed in early-type spirals.

\par Future kinematical modelling should help elucidate the structure of these galaxies and whether they are consistent with the definition of Hubble late-type.

\subsubsection{Interaction-Driven Gas Transport}
\label{sec: interact}

Galactic interactions are widely accepted to initialise wide-spread disk destabilisation. Large-scale disruption of this kind is also commonly associated with the formation of asymmetric galactic features. These asymmetries established within the merging disks exert strong torques on the adjacent gas, causing it to fall inwards. This is supported by the work of \citet{dimatteo07}, where the gas dynamics of merging galaxy pairs is modelled in detail. They simulate over two hundred merging pairs and observe several key phases of gas contraction and expansion. The final stage of the merger involves a violent inflow of between $\approx 50 - 80 \%$ of the total gas mass and is triggered by the rapid fluctuations in the galactic potential. This gas flows into the central kiloparsec of the merger and can be followed by an intense starburst (which persists for $< 600$ \ Myr). However, \citet{dimatteo07} note that interacting pairs do not always lead to star-formation enhancement. Moreover, they find that high star-formation enhancement is an infrequent result in their simulations (compared to no, low or moderate enhancement) and is associated with the shortest starburst duration times. This scenario is consonant with our results in this study. Furthermore, \citet{dimatteo07} also find relatively small differences between the global SFR values for interacting and non-interacting galaxies \citep[also reported by][]{bloom17}, which supports the integrated analysis presented in Figures \ref{fig: ALMAdiag} \& \ref{fig: sfre}.

\par Again, in order to find evidence of interaction, detailed kinematic analysis is required. From initial observations, Figure~\ref{fig: pvds} appears to suggest that some of the kinematic profiles are indicative of disturbed gas structures. However, our moment maps and PVDs suggest settled, regular gas structures in many of our outflow-type objects. It may be, therefore, that our outflow-type sample is comprised of objects undergoing a variety of processes that drive molecular gas inwards and expel intense ionised winds.  

\subsection{Effect of Timescales}
\label{sec: time}

\par Throughout our resolved analysis in Section~\ref{subsec: spatresob}, we treat our ALMA outflow-types individually due to the wide variety of global (e.g. metallicity and stellar mass) and resolved properties (i.e. some objects possess enhanced star-formation and gas density, while others do not) that we observe in the sample. As discussed previously, we also find a wide variety of kinematical structures in our outflow-type objects in Figure~\ref{fig: pvds}. The diversity of these properties suggests that these objects are either a collection of unrelated galaxies, similar only in their shared proclivity to eject large amounts of ionised gas, or similar objects at different stages in their evolution. For example, in relation to the discussion in Section~\ref{sec: interact}, it is possible that our outflow-types are interacting galaxies but at different stages of their evolution. As detailed by \citet{dimatteo07}, mergers enter phases of gas compression with short-lived, central starbursts and gas expansion with little star-forming activity and diffuse molecular gas distributions. If interaction is the driver of the molecular gas compression and large-scale outflows of ionised gas in our outflow-type objects, it is possible that our outflow-types are at different stages in the compression-expansion cycle, but all still experiencing large-scale ionised outflows. This would be supported by the indications of instability in the PVDs shown in Figure~\ref{fig: pvds}. However, as previously noted, kinematical analysis is required to validate this possibility and whether the diversity of kinematical structures we observe (i.e. exponential, barred and unsettled) can be linked as part of an evolutionary process (different stages of a merger or other process). 

\par In the context of timescales, it is also important to note the similarity of integrated gas content between the outflow-type objects and xCOLD GASS controls matched in stellar mass and SFR (see Section~\ref{subsec: glob}). If our outflow-types are entraining molecular gas in their large-scale ionised outflows, the objects will eventually become significantly gas-depleted with respect to galaxies not hosting galactic-scale outflows. As we do not observe this, we could conclude that molecular gas is not driven out of the objects by the stellar wind or that these objects are in a very early stage of their outflow activity.

\subsection{Dense vs. Diffuse Molecular Gas}
\label{sec: GMC}

The traditional picture of the gas distribution in galaxies is that large amounts exist in Giant Molecular Clouds (GMCs). \citet{solomon79} and \citet{solomon_sanders80} find a typical range of cloud sizes to be 15 - 90~pc \citep[with an average size $\approx$ 40~pc,][]{pety13} and star-formation occurs almost exclusively in GMCs. Our ALMA CO(1\textrightarrow0) maps have a spatial resolution of $\approx$ 1~kpc (once re-binned to the SAMI resolution), approximately an order of magnitude larger than the average size of GMCs. \citet{solomon79} also report a mean $\rm 170~M_{\odot}~{pc}^{-2}$ molecular gas surface density in GMCs. In Figure~\ref{fig: molhists}, $\Sigma _{\rm CO}$ spaxels with a value $> \rm 100~M_{\odot}~pc^{-2}$ are rare and only occur in two objects. Further, the radial averages of $\Sigma _{\rm CO}$ in Figure~\ref{fig: radhistsCO} are consistently $\approx$ an order of magnitude lower than the average value reported by \citet{solomon79}. Both \citet{pety13} and \citet{caldu-primo15} find that molecular gas exists in two distinct phases; a clumpy, compact phase in GMCs and a more diffuse phase. These studies find that gas is divided equally between these two phases, as traced by CO(1\textrightarrow0) emission. Moreover, \citet{caldu-primo15} suggest that this diffuse component is comprised of molecular gas distributed over spatial scales $\approx$ 1~kpc with column densities $\lesssim$ 30 times smaller than those in GMCs. This is more inline with the spatial resolution of our observations and the gas densities that we observe in Figures \ref{fig: molhists} \& \ref{fig: radhistsCO}.

\par We can infer from the aforementioned studies that our CO(1\textrightarrow0) emission maps may trace similar spatial scales to the diffuse component of the molecular gas content in our outflow-type and control objects. The contribution of gas in GMCs will, therefore, be smeared by the ALMA beam, reducing the $\Sigma_{\rm CO}$ measurements in our CO(1\textrightarrow0) maps on small spatial scales. Consequently, there remains a possibility that our ALMA outflow-type objects contain exceptionally dense molecular gas within GMCs that power their galactic-scale outflows of ionised gas. Similarly, the $\approx$ 1~kpc physical resolution of the SAMI maps may also limit our ability to detect intense star-formation pockets.

\section{Summary and Conclusions}

In this paper, we directly compare optical IFU data with CO(1\textrightarrow0) emission tracing cold molecular gas for 9 galaxies identified as harbouring galactic-scale outflows of ionised gas (at $z \lesssim 0.1$). The galaxies were identified using the SAMI Galaxy Survey with criteria outlined in Section~\ref{sec: selectionanddata} and \citet{ho16} for edge-on objects (i.e. inclination, $i \gtrapprox 70^{\circ}$). By combining observations of ionised and molecular gas content and dynamics, we aim to elucidate their coupling in the case of galaxies with powerful, ionised outflows. With CO(1\textrightarrow0) data from both IRAM and ALMA alongside the IFU data from SAMI, we conduct our analyses on both global and resolved scales. Using SAMI, we also identify a further 7 galaxies matched to the outflow-type objects in stellar mass, SFR (with further restrictions on their redshifts and inclinations) that do not possess these intense outflows to act as a control sample for our analysis.

\par In Section~\ref{subsec: glob}, we examine the differences in global gas content and star-formation efficiency (SFE) for our outflow-type objects and a control sample derived from xCOLD GASS (matched to the outflow-type objects in stellar mass and SFR). We then analyse the resolved ALMA and SAMI data in Section~\ref{subsec: spatresob}, by directly comparing the SFR density ($\Sigma _{\rm SFR}$) and $\rm M_{H_2}$ density ($\Sigma _{\rm CO}$) spaxels on scales $\lessapprox 1$~kpc between the outflow-type objects and the controls. We also examine the radial distribution of these $\Sigma _{\rm SFR}$ and $\Sigma _{\rm CO}$ spaxels to assess any structural differences between the galaxies harbouring galactic-scale outflows and those that do not.

\par The most significant findings of this study are summarised as follows:

\begin{itemize}[leftmargin=0.25in]

    \item We find no significant difference in the global properties of our outflow-type galaxies relative to their matched controls from xCOLD GASS, including their molecular gas fractions. This implies that outflow-type galaxies possess the same total amount of molecular gas as the controls with regard to the stars they have already created. 
    
    \item By constructing PVDs as an initial method of kinematical analysis in Section~\ref{subsec: spatresob}, we observe a diversity of structures in our outflow-types (e.g. indications of bars, settled exponential disks and unsettled structures, see Figure~\ref{fig: pvds}). For this reason, the remainder of our analysis treats each outflow-type galaxy individually. We remark, therefore, that galactic-scale outflows in these objects may be instigated by several different processes.
    
    \item Using our resolved maps from ALMA and data from the SAMI Galaxy Survey, we do not detect consistently higher density $\Sigma _{\rm CO}$ and $\Sigma _{\rm SFR}$ spaxels in the outflow-type objects compared to those in their controls. However, we discuss the limitations of our resolution ($\approx$ 1~kpc) in tracing the compact molecular gas phase which drives star-formation in Section~\ref{sec: GMC}. 
    
    \item Binning spaxels of $\Sigma _{\rm CO}$ and $\Sigma _{\rm SFR}$ by radial distance along the edge-on disk (Figure~\ref{fig: radhistsCO} \& \ref{fig: radhistsSFR}), we find that the outflow-type objects have indications of a more centralised distribution of both cold molecular gas and star-formation compared to their controls. The gas and SFR distribution in the outflow-types generally fall within their inner $\approx \rm r_{eff}$ and decline with radius more steeply than the controls. The control samples are comprised of objects with a less centrally structured gas structure and with, consequently, more evenly distributed star-formation.
    
    \item We observe no obvious difference between the distribution of SFE between the outflow-type objects and their controls. 
    
    \item There is an enhancement of molecular gas fraction in the outflow-types within their inner $0.5 \rm~r_{eff}$ compared to their controls (see Figure~\ref{fig: radhistsGF}). The decline of gas fraction with radius in the outflow-types is again steeper than that of the control sample. This in contrast to the ALMA controls that are gas rich in their outer regions with respect to their stellar population.
    
    \item We discuss mechanisms capable of driving large volumes of gas towards the centre of a galaxy in Section~\ref{sec: discussion}. The processes we consider rely on galactic-scale asymmetries to transport angular momentum inwards, such as bar structures and interaction. However, detailed kinematical analysis would be required to identify which, if any, of these mechanisms operate within our objects. We also consider the effects of the physical scale our CO(1\textrightarrow0) emission maps resolves and whether we can accurately analyse gas in GMCs or if our maps more effectively trace gas in a diffuse phase that does not contribute to star-formation. 

\end{itemize}

\par Our primary conclusion from this work is the apparent role of centralised molecular gas distributions in powering galactic-scale outflows of ionised gas. These well-defined nuclei of gas and star-formation (with high central gas fraction) may provide hints as to the driving mechanism behind large-scale outflows. However, we note that we do not detect elevation in gas or SFR density in our outflow-type objects (and discuss possible explanations for this in Section~\ref{sec: discussion}). The mechanisms driving molecular gas into the central regions of these galaxies can not be determined without a detailed kinematic analysis of our CO(1\textrightarrow0) data. This will form the basis of future work on this dataset. 

\section*{Acknowledgements}
\label{sec: acknowledgements}

We thank the anonymous referee for their comments that helped to improve this manuscript.

\par This paper makes use of the following ALMA data: \textcolor{red}{S/JAO.ALMA\#2011.0.01234.S.} ALMA is a partnership of ESO (representing its member states), NSF (USA) and NINS (Japan), together with NRC (Canada), MOST and ASIAA (Taiwan), and KASI (Republic of Korea), in cooperation with the Republic of Chile. The Joint ALMA Observatory is operated by ESO, AUI/NRAO and NAOJ.

\par The SAMI Galaxy Survey is based on observations made at the Anglo-Australian Telescope. The Sydney-AAO Multi-object Integral field spectrograph (SAMI) was developed jointly by the University of Sydney and the Australian Astronomical Observatory. The SAMI input catalogue is based on data taken from the Sloan Digital Sky Survey, the GAMA Survey and the VST ATLAS Survey. The SAMI Galaxy Survey is supported by the Australian Research Council Centre of Excellence for All Sky Astrophysics in 3 Dimension (ASTRO 3D), through project number CE170100013, the Australian Research Council Centre of Excellence for All-sky Astrophysics (CAASTRO), through project number CE110001020, and other participating institutions. The SAMI Galaxy Survey website is \url{http://sami-survey.org/}.

\par Part of this research was supported by the Australian Research Council Centre of Excellence for All Sky Astrophysics in 3 Dimensions (ASTRO 3D), through project number CE170100013. Luca Cortese is the recipient of an Australian Research Council Future Fellowship (FT180100066) funded by the Australian Government.

\par GAMA is a joint European-Australasian project based around a spectroscopic campaign using the Anglo-Australian Telescope. The GAMA input catalogue is based on data taken from the Sloan Digital Sky Survey and the UKIRT Infrared Deep Sky Survey. Complementary imaging of the GAMA regions is being obtained by a number of independent survey programmes including GALEX MIS, VST KiDS, VISTA VIKING, WISE, Herschel-ATLAS, GMRT and ASKAP providing UV to radio coverage. GAMA is funded by the STFC (UK), the ARC (Australia), the AAO, and the participating institutions. The GAMA website is \url{http://www.gama-survey.org/}.

\section{Data Availability}

The data underlying this article are available in the SAMI public database (\url{https://sami-survey.org/data}) and the ALMA science archive (\url{https://almascience.eso.org/asax/}).

\begin{dedication}

Dedicated to the memory of Cliff Rickett (1932--2020)

\end{dedication}

%%%%%%%%%%%%%%%%%%%%%%%%%%%%%%%%%%%%%%%%%%%%%%%%%%

%%%%%%%%%%%%%%%%%%%% REFERENCES %%%%%%%%%%%%%%%%%%

\bibliography{refs.bib}
\bibliographystyle{mnras}

%%%%%%%%%%%%%%%%% APPENDICES %%%%%%%%%%%%%%%%%%%%%

\appendix

\section{IRAM-30m CO spectra (online-only)}
\label{sec: IRAMspectra}

Presenting all the IRAM CO(1\textrightarrow0) spectra. 

\begin{figure*}
	\includegraphics[scale=0.95]{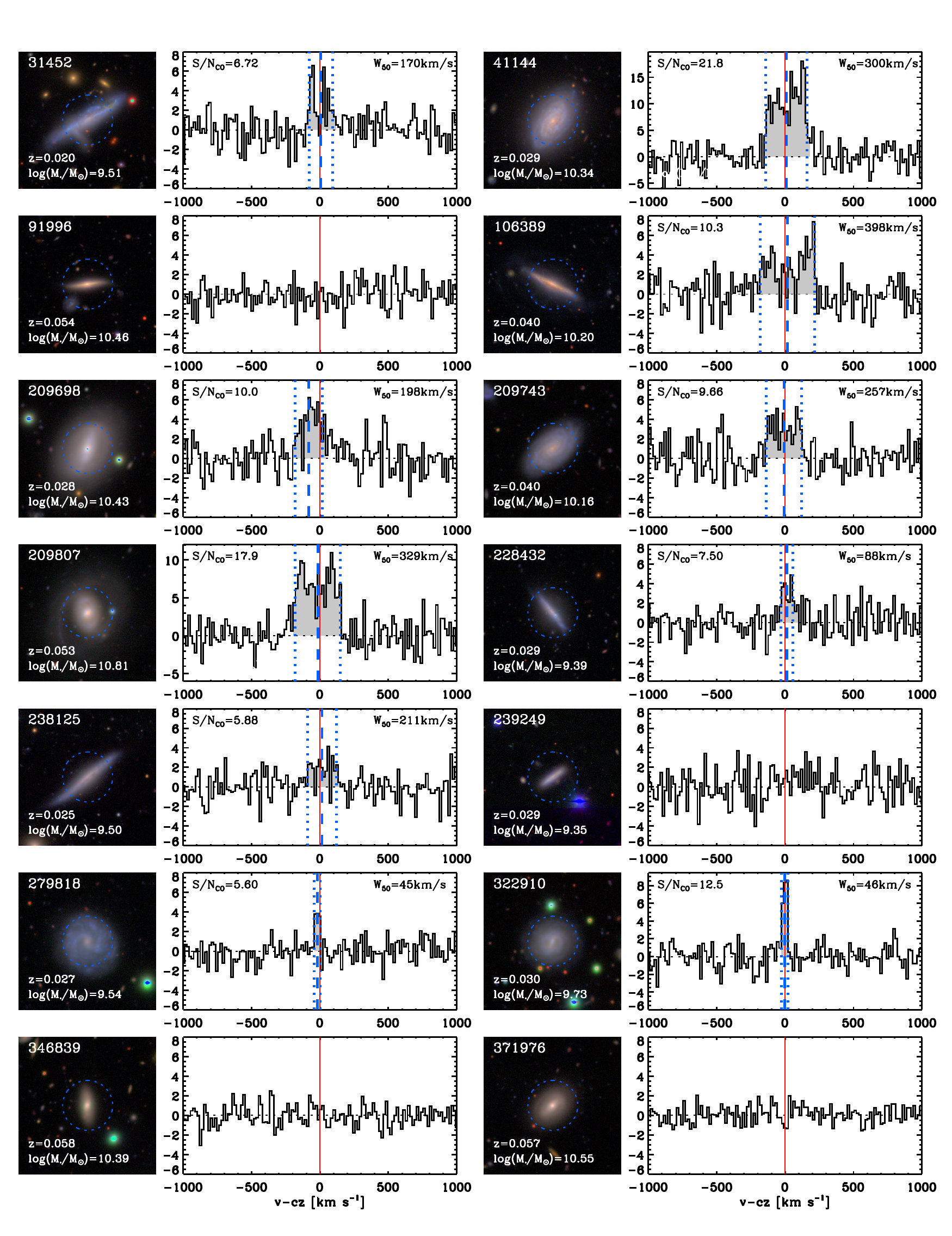}
    \caption{For each of the galaxies in the IRAM sample, {\it Left:} SDSS image (1\arcmin$\times$1\arcmin), with the position of the 22\arcsec\ IRAM beam shown as the dashed line (the GAMA ID, z and M* are noted in each image), and {\it Right:} IRAM spectrum, centered on the optical redshift of each galaxy (red line). The dashed and dotted blue lines represent the measured centre and width of the CO line, respectively. The gray shading represents the area which was integrated to calculate the total line flux of each galaxy.}
    \label{IRAMspectra1}
\end{figure*}

\begin{figure*}
	\includegraphics[scale=0.95]{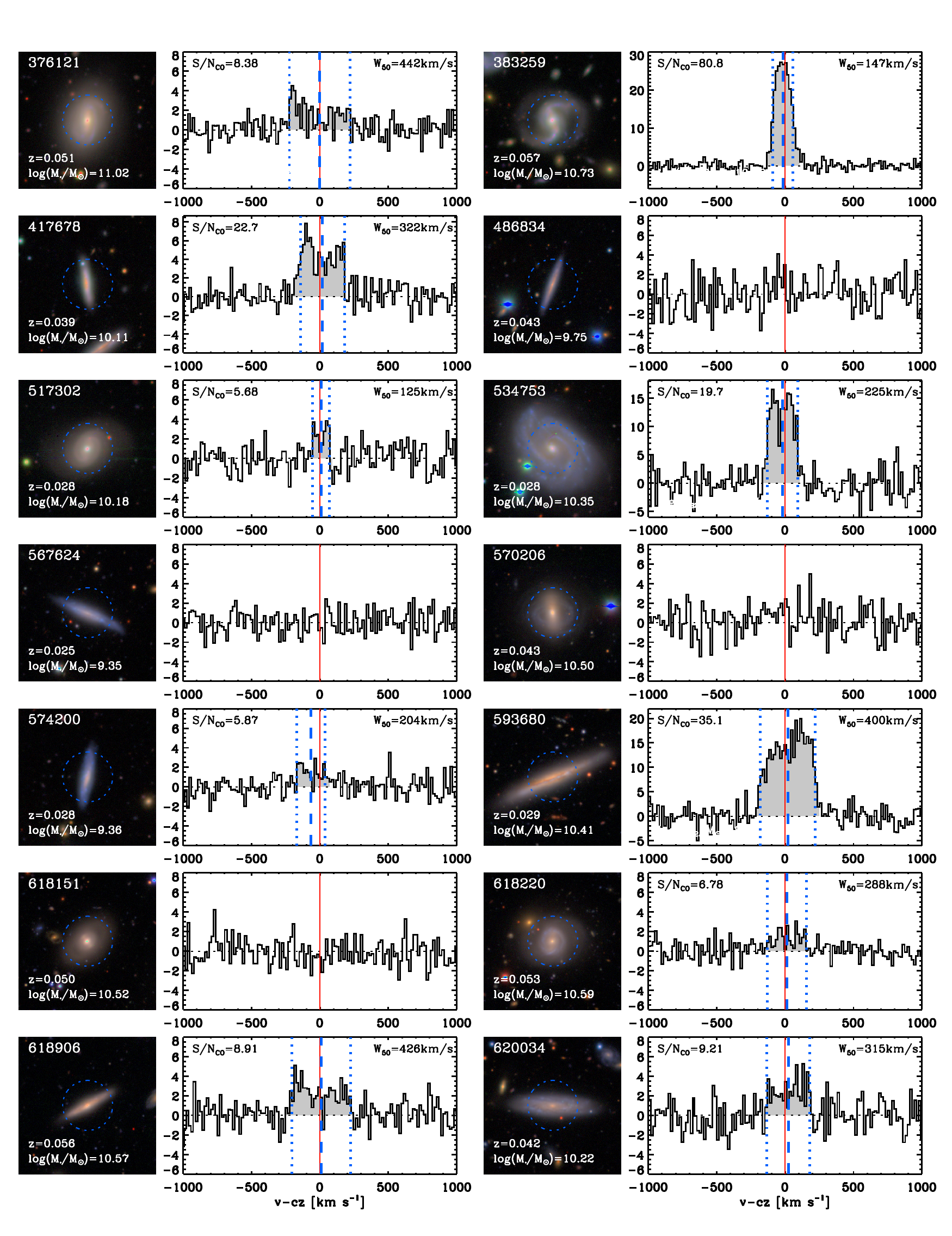}
    \caption{ continued from Fig. \ref{IRAMspectra1} }
    \label{IRAMspectra2}
\end{figure*}

\section{ALMA Moment Maps (online-only)}
\label{sec: ALMAmaps}

Presenting all the ALMA CO(1\textrightarrow0) moment maps (i.e. zeroth, first and second moments). 

\begin{figure*}
\centering
\includegraphics[scale=0.4]{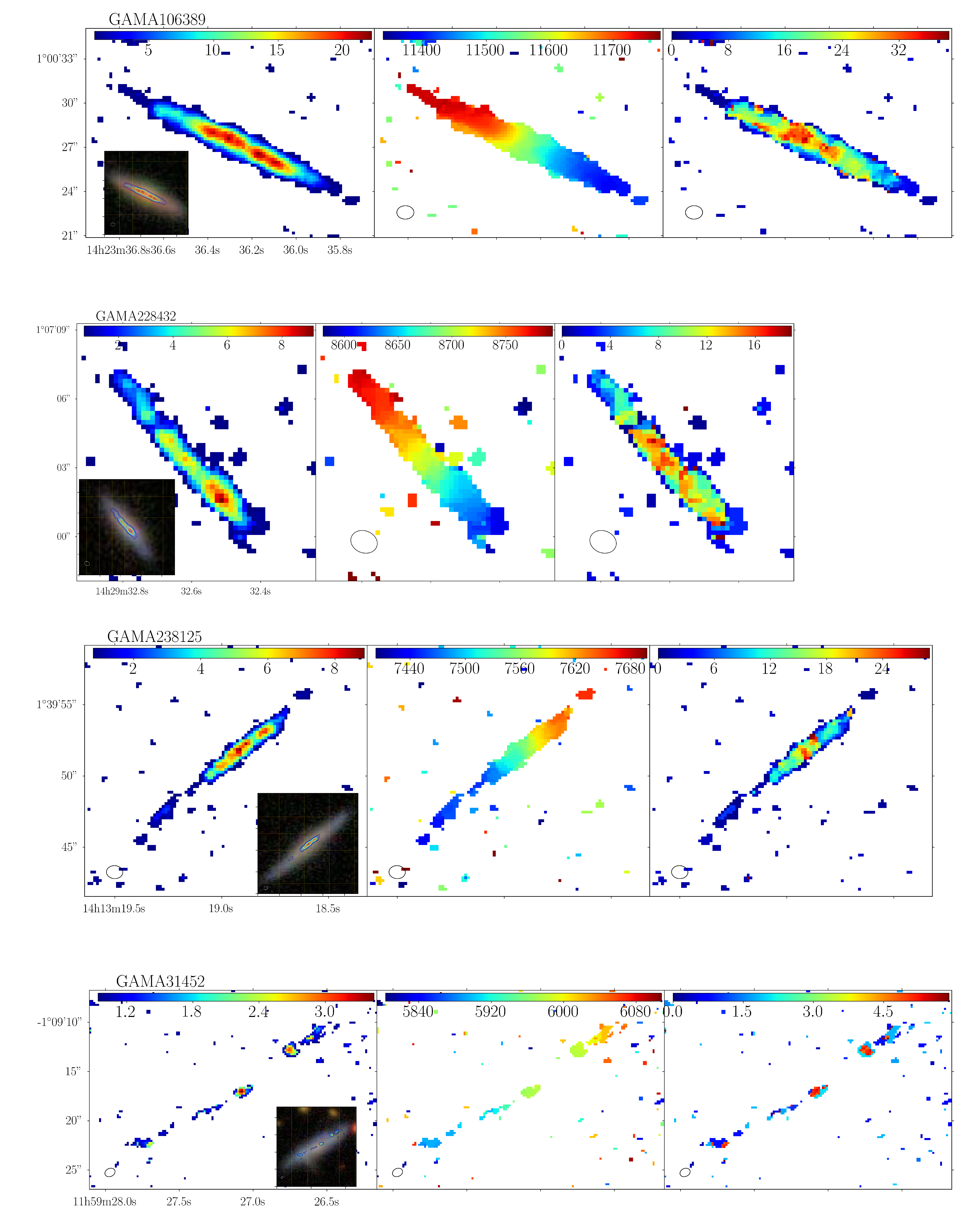}
\caption{CO(1\textrightarrow0) moment maps of ALMA galaxies with positive detection. Each row represents the maps derived from a different galaxy in the sample. From left to right, the maps represent the zeroth, first and second moments of each respective object. The colour bar in each panel indicates the magnitude of intensity, velocity and velocity dispersion in the zeroth, first and second moments respectively.}
\label{fig: moments}
\end{figure*}

\begin{figure*}
\centering
\includegraphics[scale=0.4]{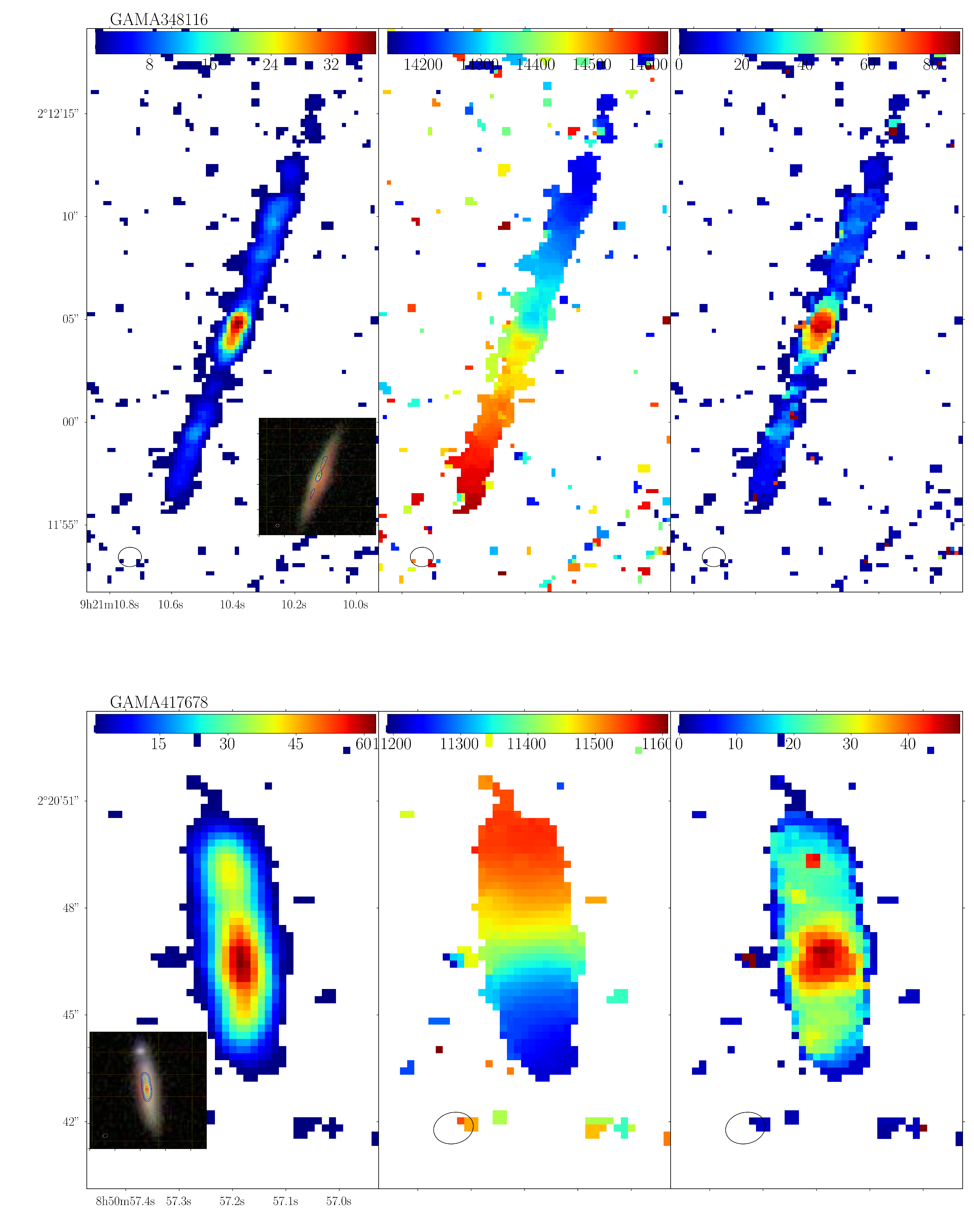}
\caption{continued from Figure~\ref{fig: moments}}
\end{figure*}

\begin{figure*}
\centering
\includegraphics[scale=0.4]{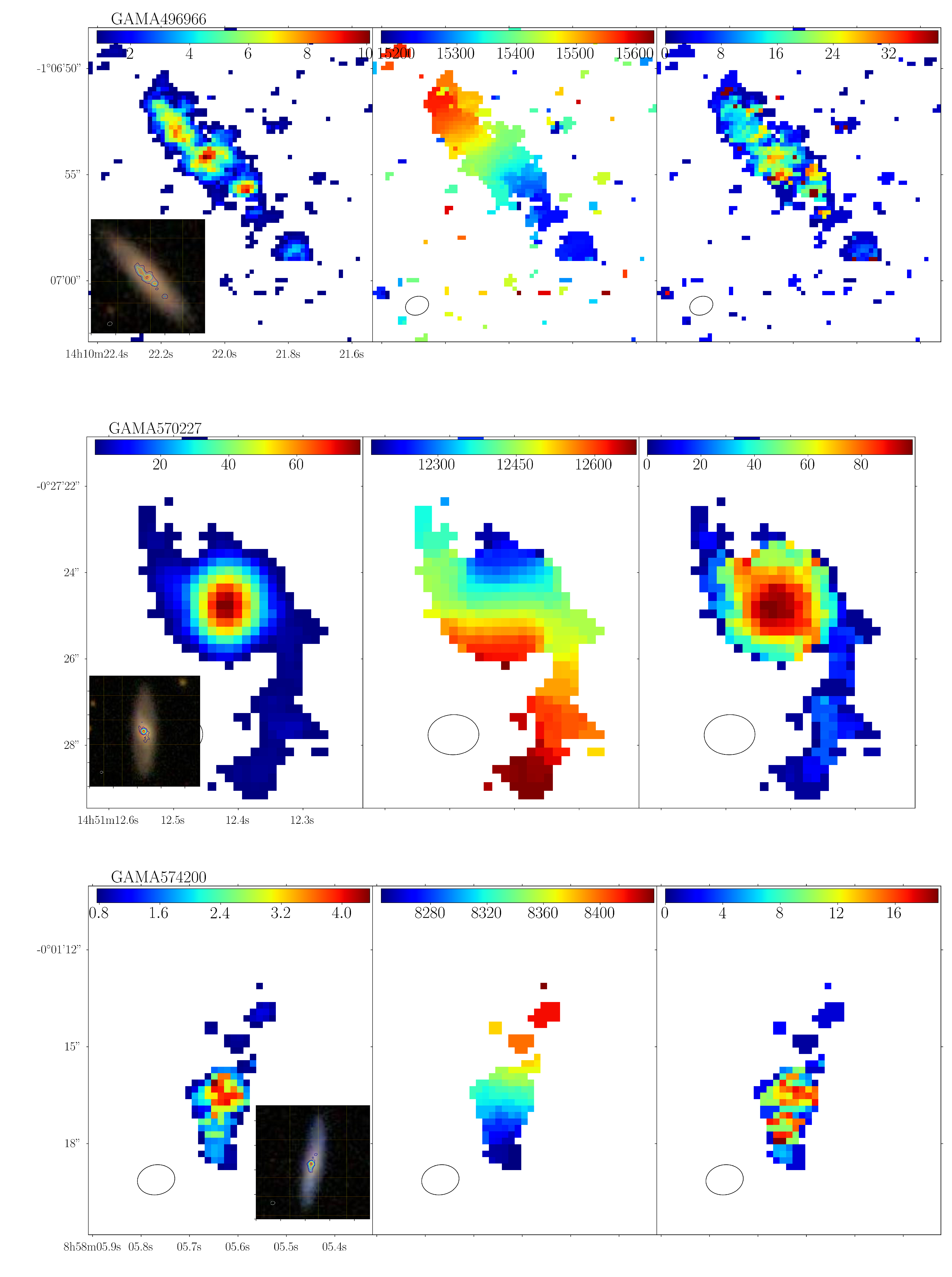}
\caption{continued from Figure~\ref{fig: moments}}
\end{figure*}

\begin{figure*}
\centering
\includegraphics[scale=0.4]{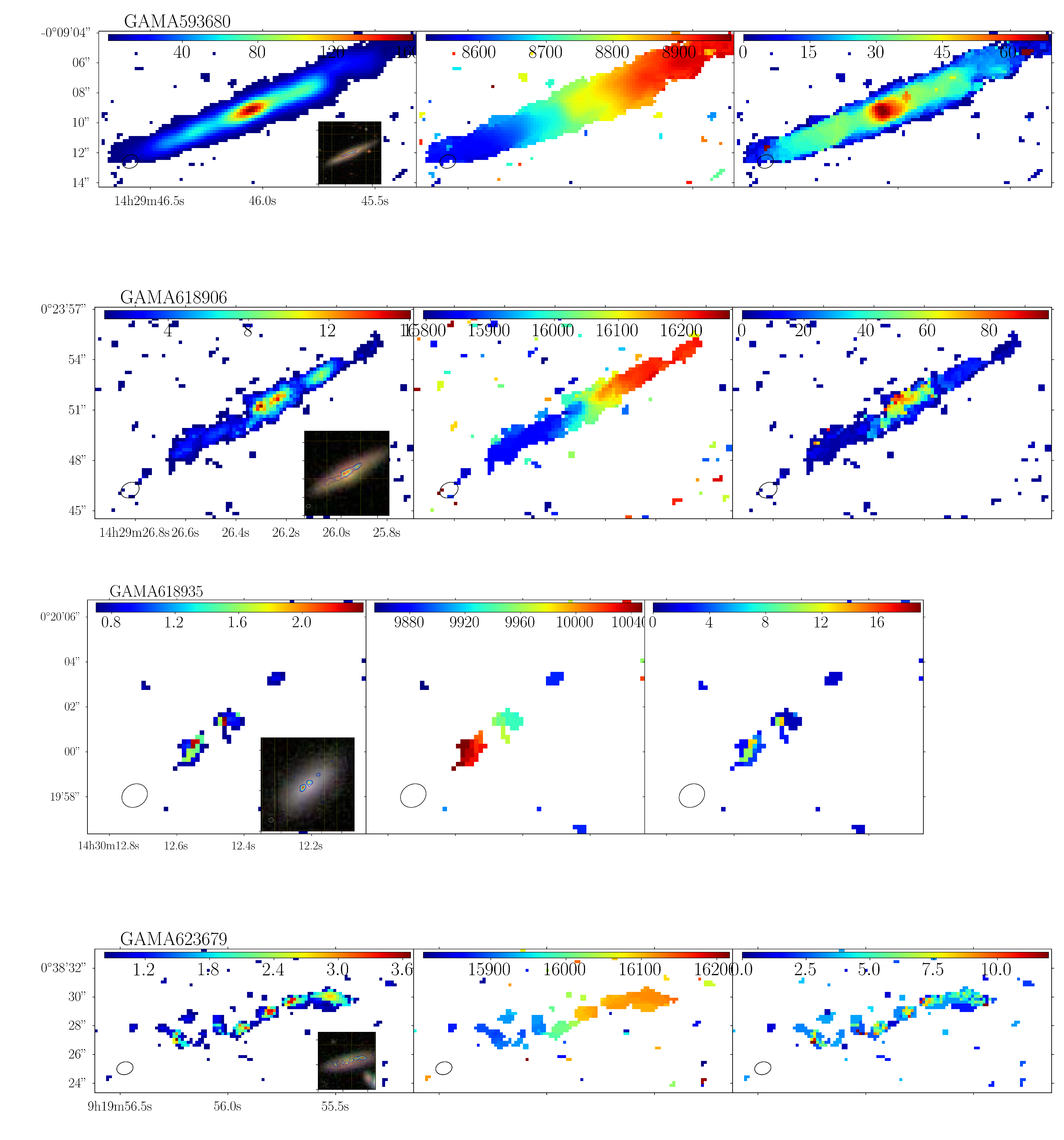}
\caption{continued from Figure~\ref{fig: moments}}
\end{figure*}

\section{ALMA Position-Velocity Maps (online-only)}
\label{sec: PVDs}

\begin{figure*}
\centering
\begin{minipage}{.5\textwidth}
  \centering
  \includegraphics[width=.9\linewidth]{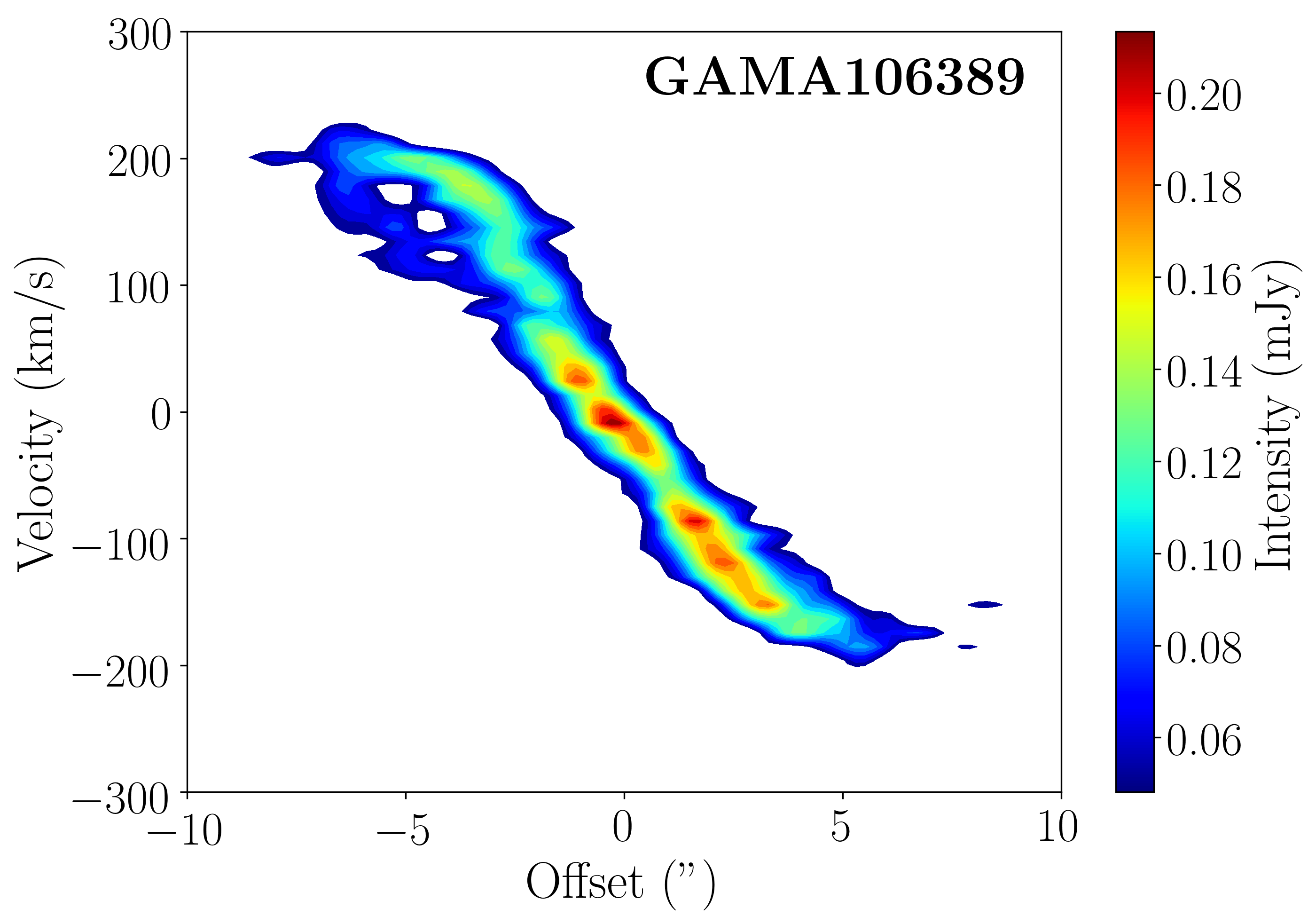}
\end{minipage}%
\begin{minipage}{.5\textwidth}
  \centering
  \includegraphics[width=.9\linewidth]{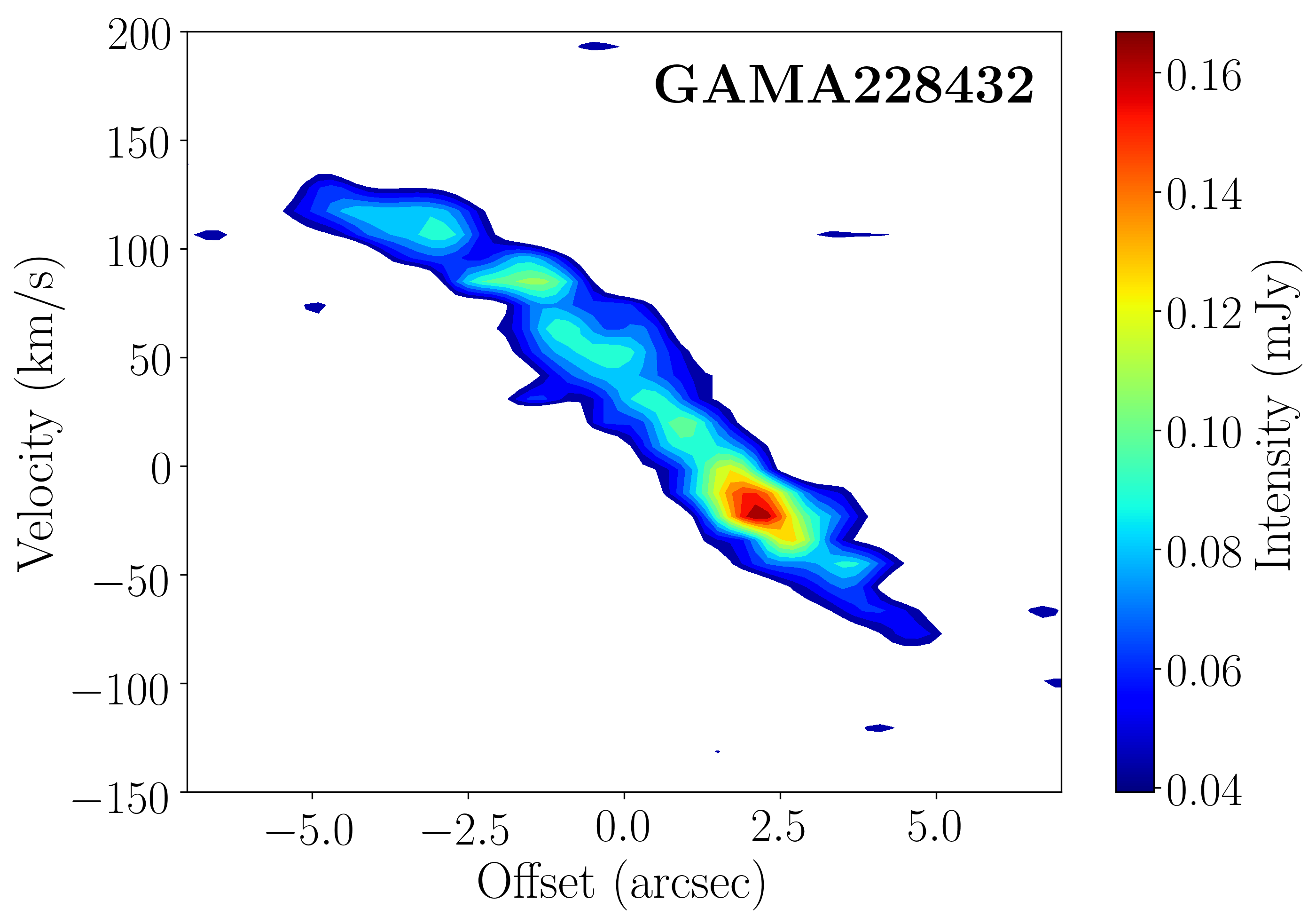}
\end{minipage}
\centering
\begin{minipage}{.5\textwidth}
  \centering
  \includegraphics[width=.9\linewidth]{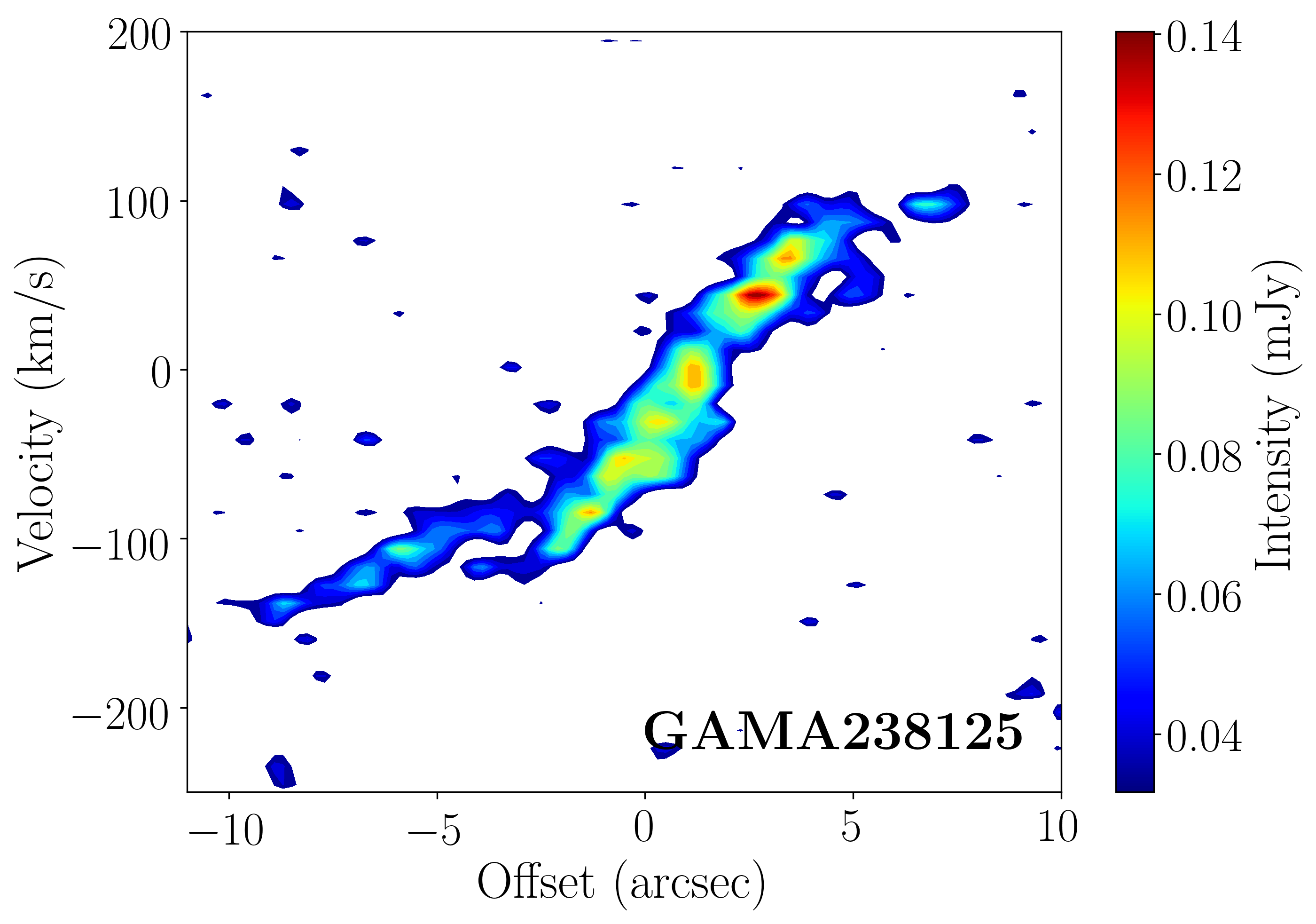}
\end{minipage}%
\begin{minipage}{.5\textwidth}
  \centering
  \includegraphics[width=.9\linewidth]{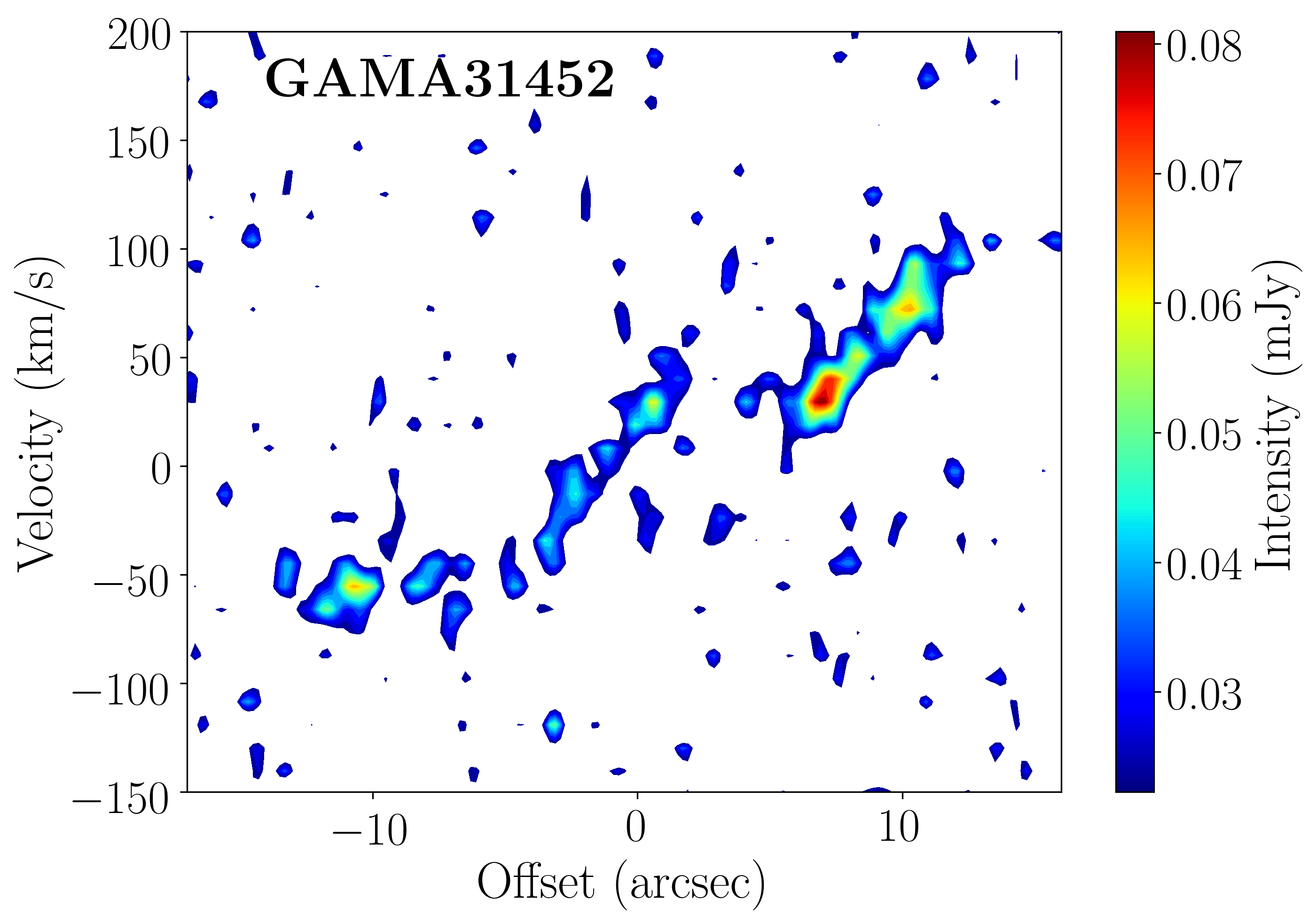}
\end{minipage}
\newline
\centering
\begin{minipage}{.5\textwidth}
  \centering
  \includegraphics[width=.9\linewidth]{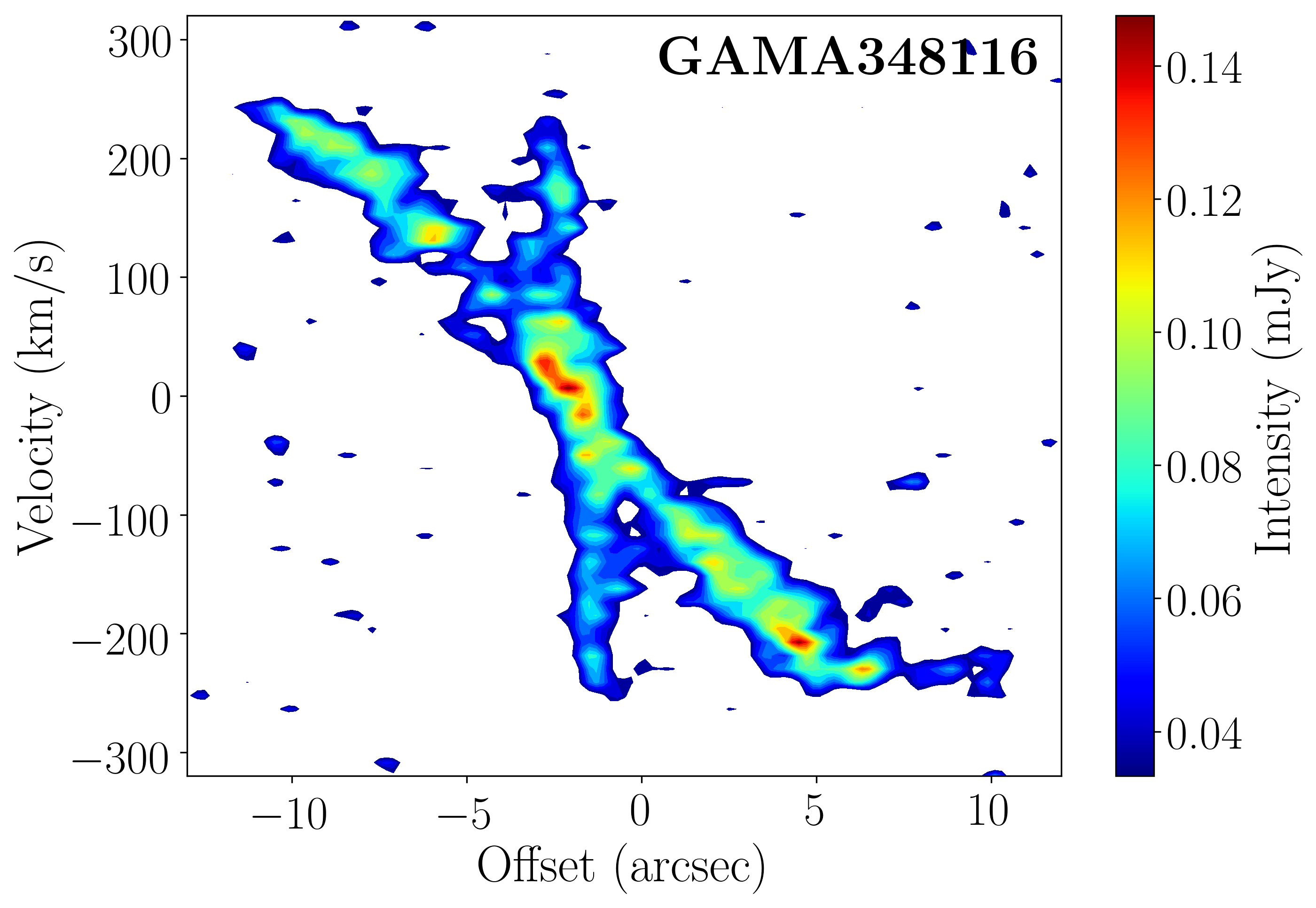}
\end{minipage}%
\begin{minipage}{.5\textwidth}
  \centering
  \includegraphics[width=.9\linewidth]{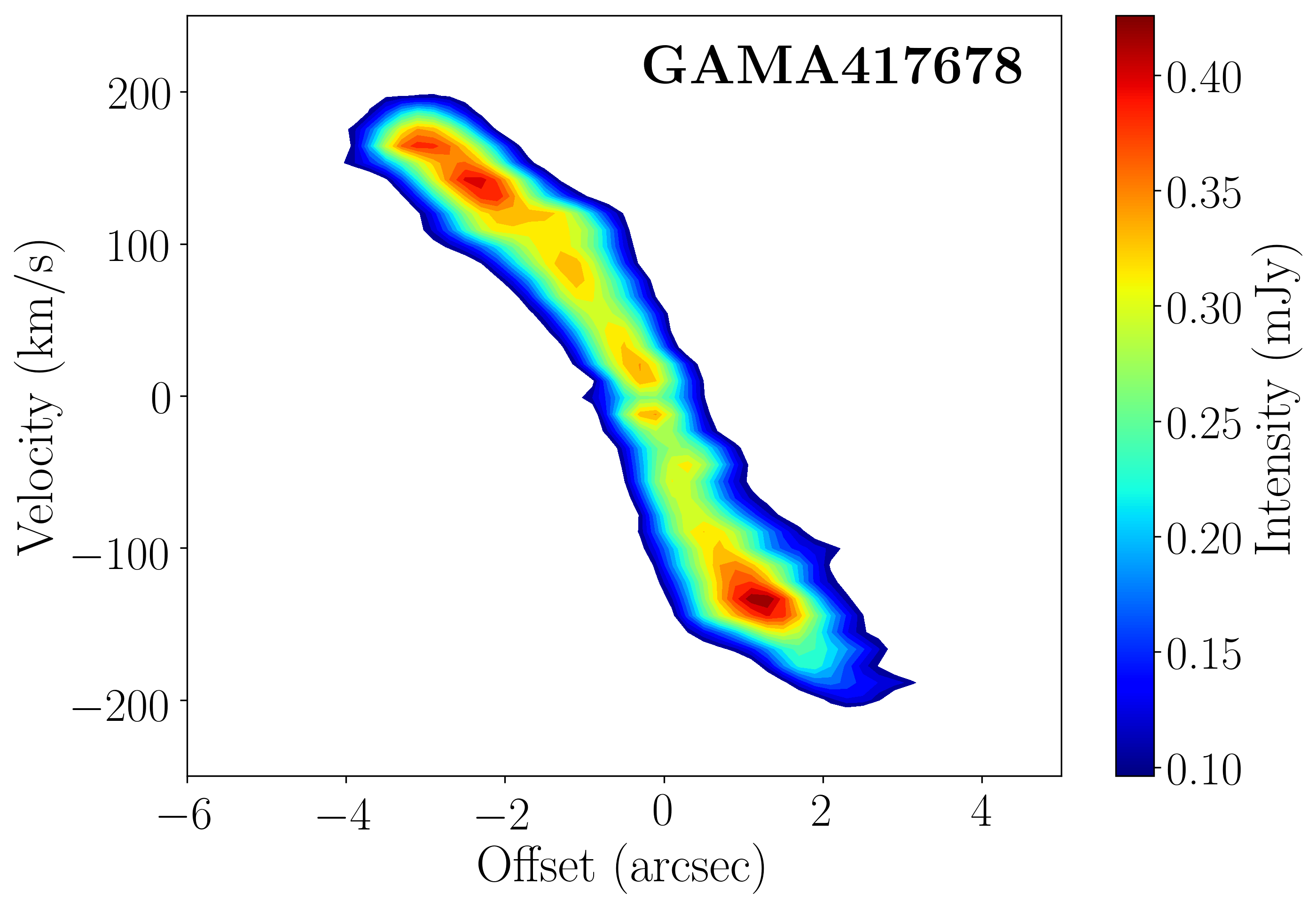}
\end{minipage}
\newline
\centering
\begin{minipage}{.5\textwidth}
  \centering
  \includegraphics[width=.9\linewidth]{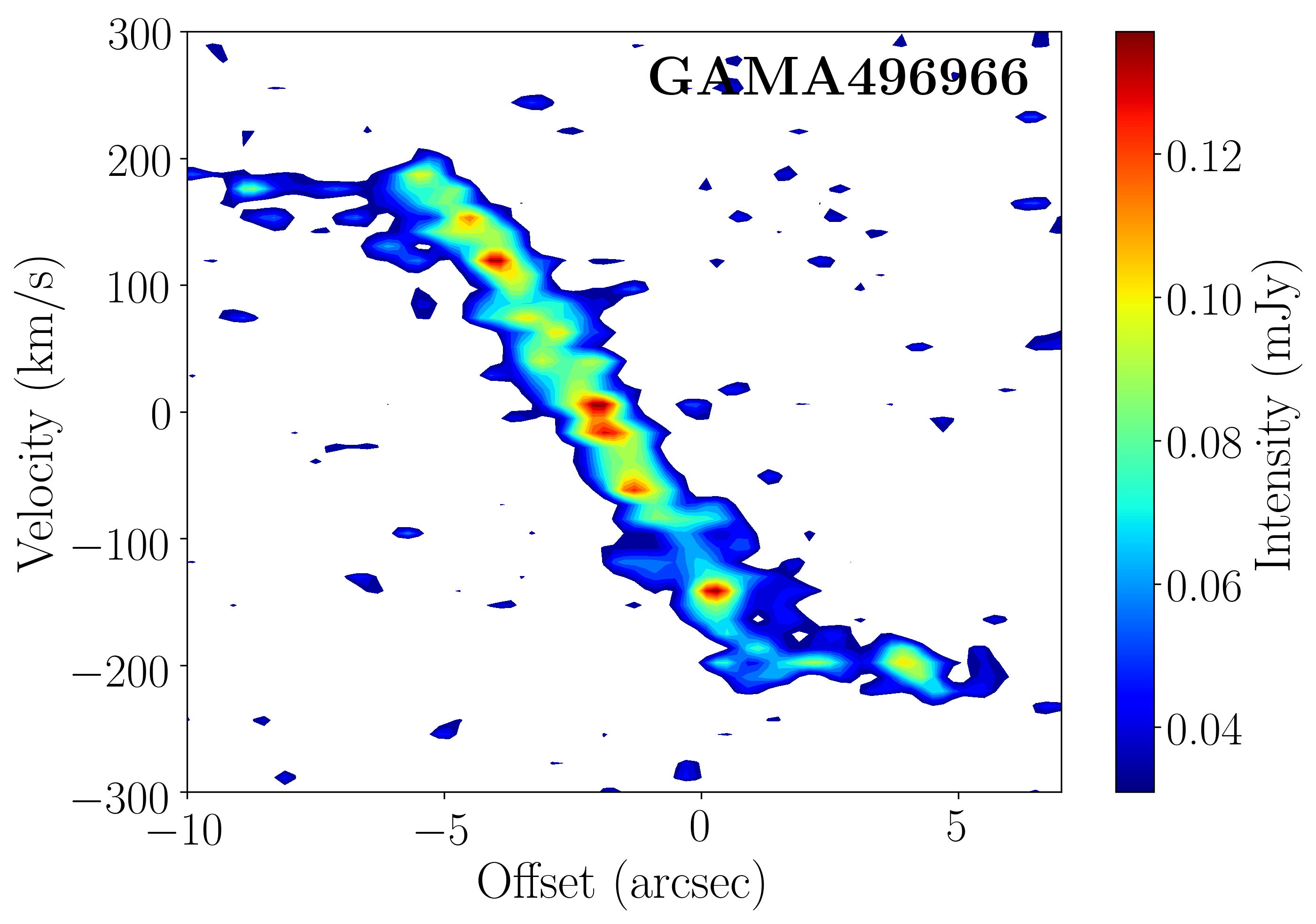}
\end{minipage}%
\begin{minipage}{.5\textwidth}
  \centering
  \includegraphics[width=.9\linewidth]{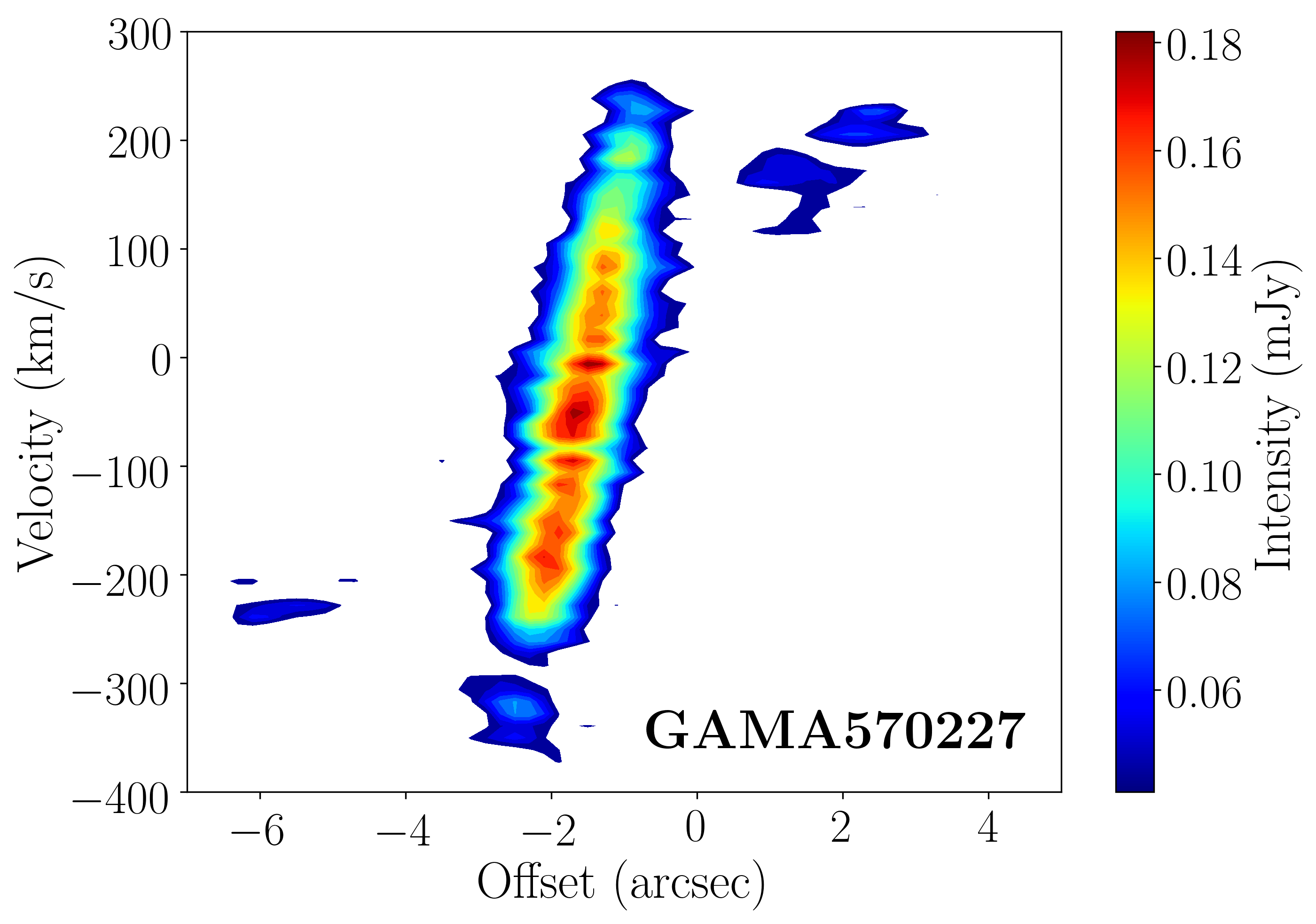}
  
\end{minipage}

\caption{CO(1\textrightarrow0) position-velocity diagrams (PVDs) of ALMA galaxies with positive detection.}
\label{fig: PVDs_all}
\end{figure*}

\begin{figure*}
\centering
\begin{minipage}{.5\textwidth}
  \centering
  \includegraphics[width=.9\linewidth]{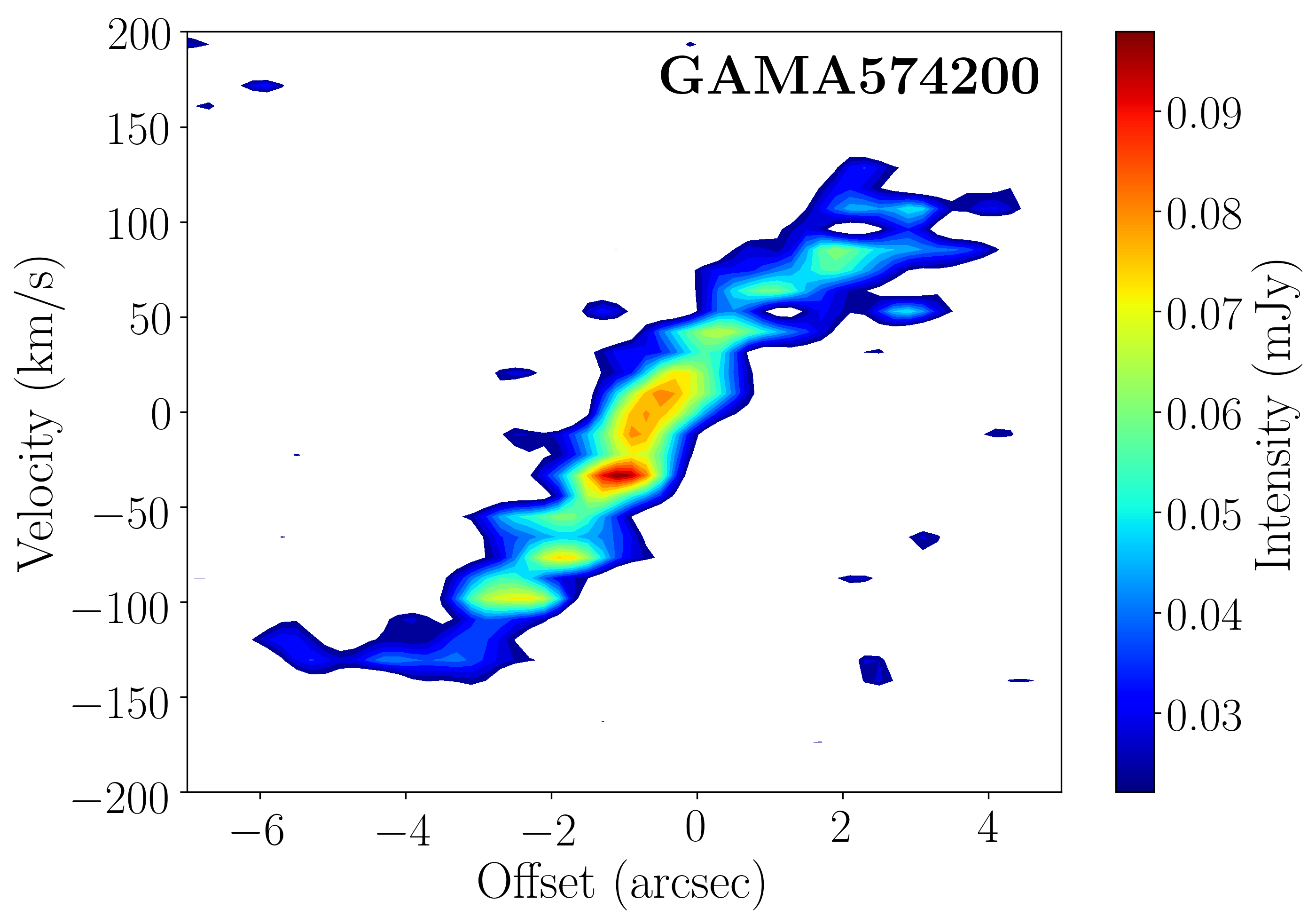}
\end{minipage}%
\begin{minipage}{.5\textwidth}
  \centering
  \includegraphics[width=.9\linewidth]{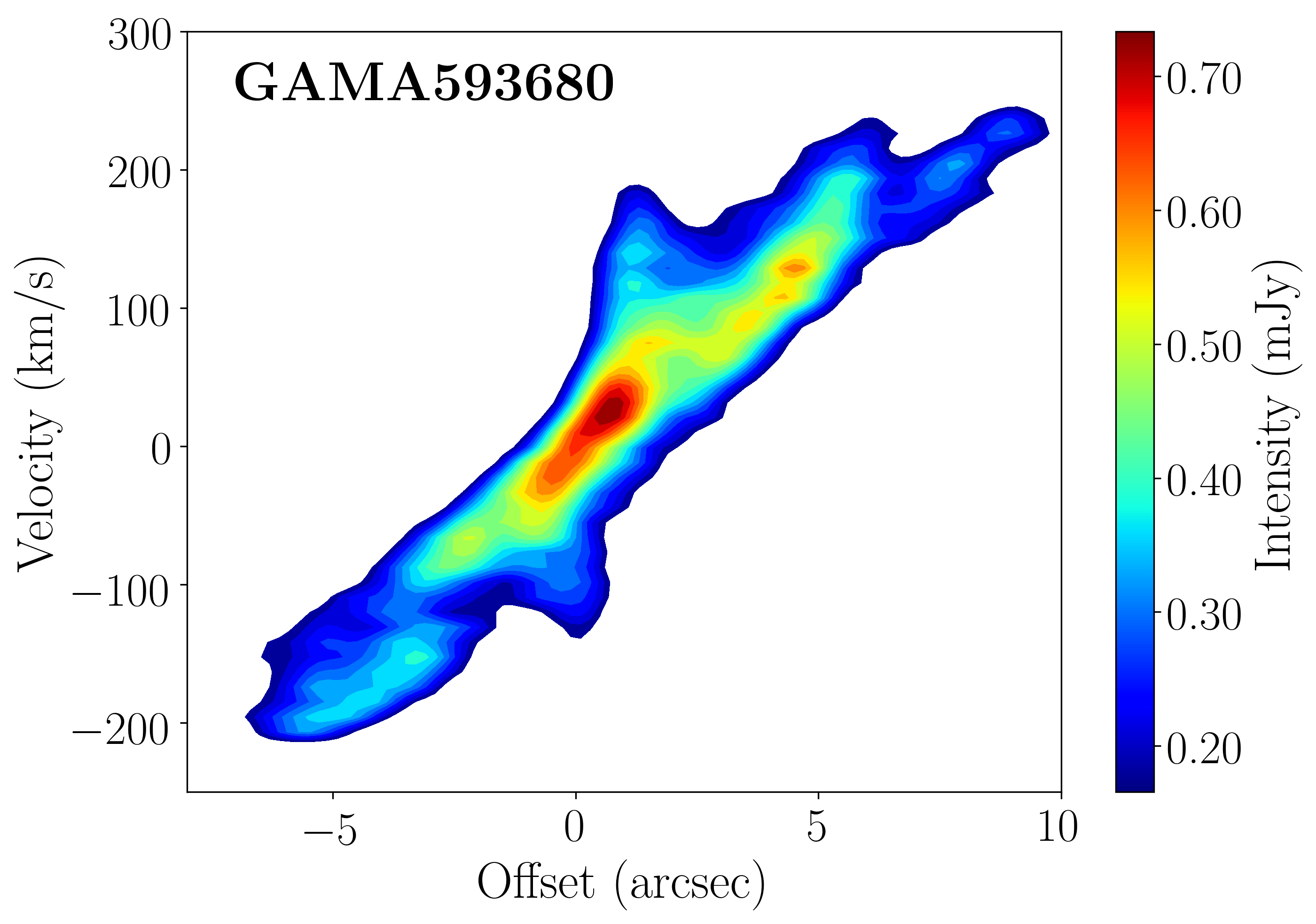}
\end{minipage}
\newline
\centering
\begin{minipage}{.5\textwidth}
  \centering
  \includegraphics[width=.9\linewidth]{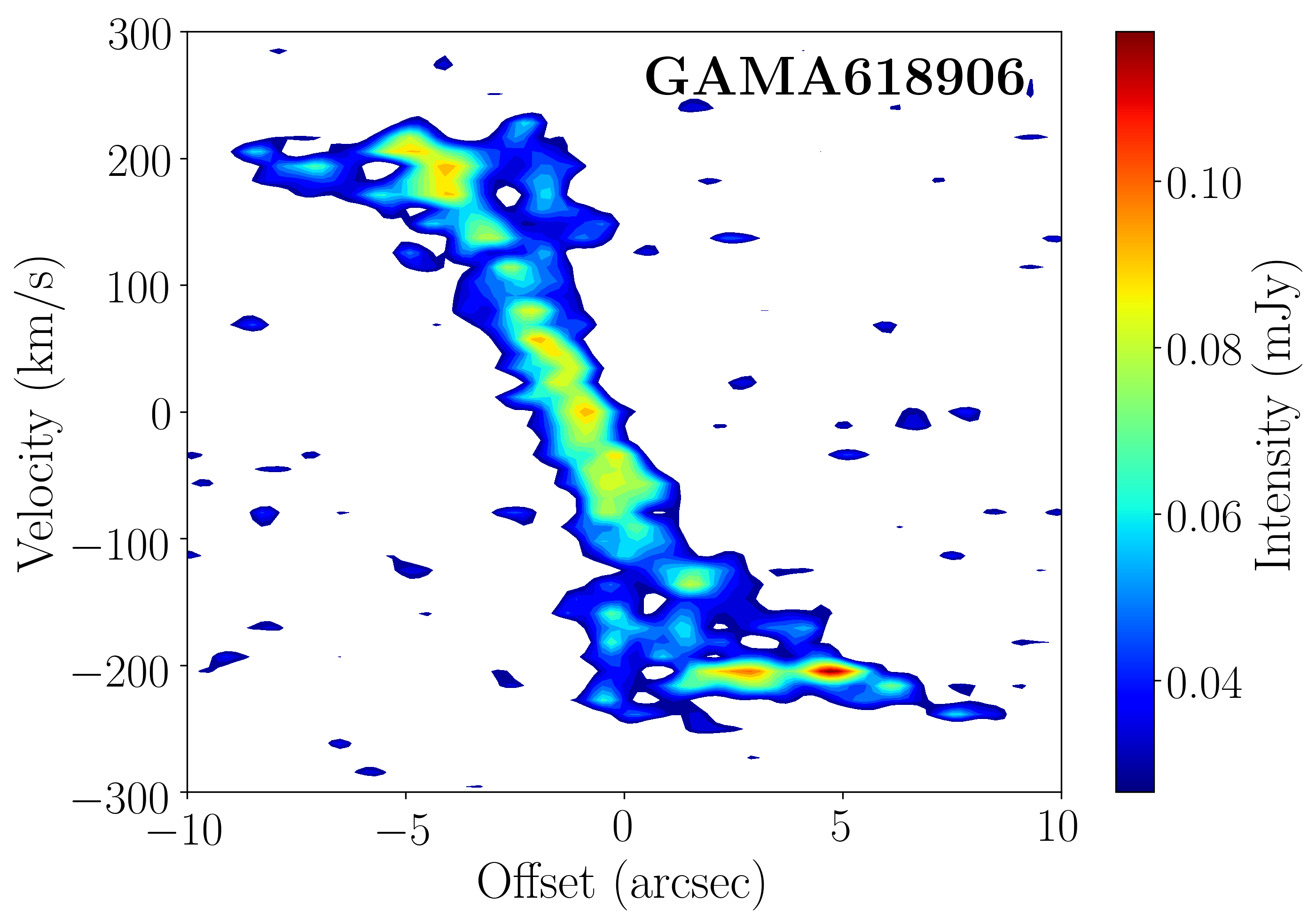}
\end{minipage}%
\begin{minipage}{.5\textwidth}
  \centering
  \includegraphics[width=.9\linewidth]{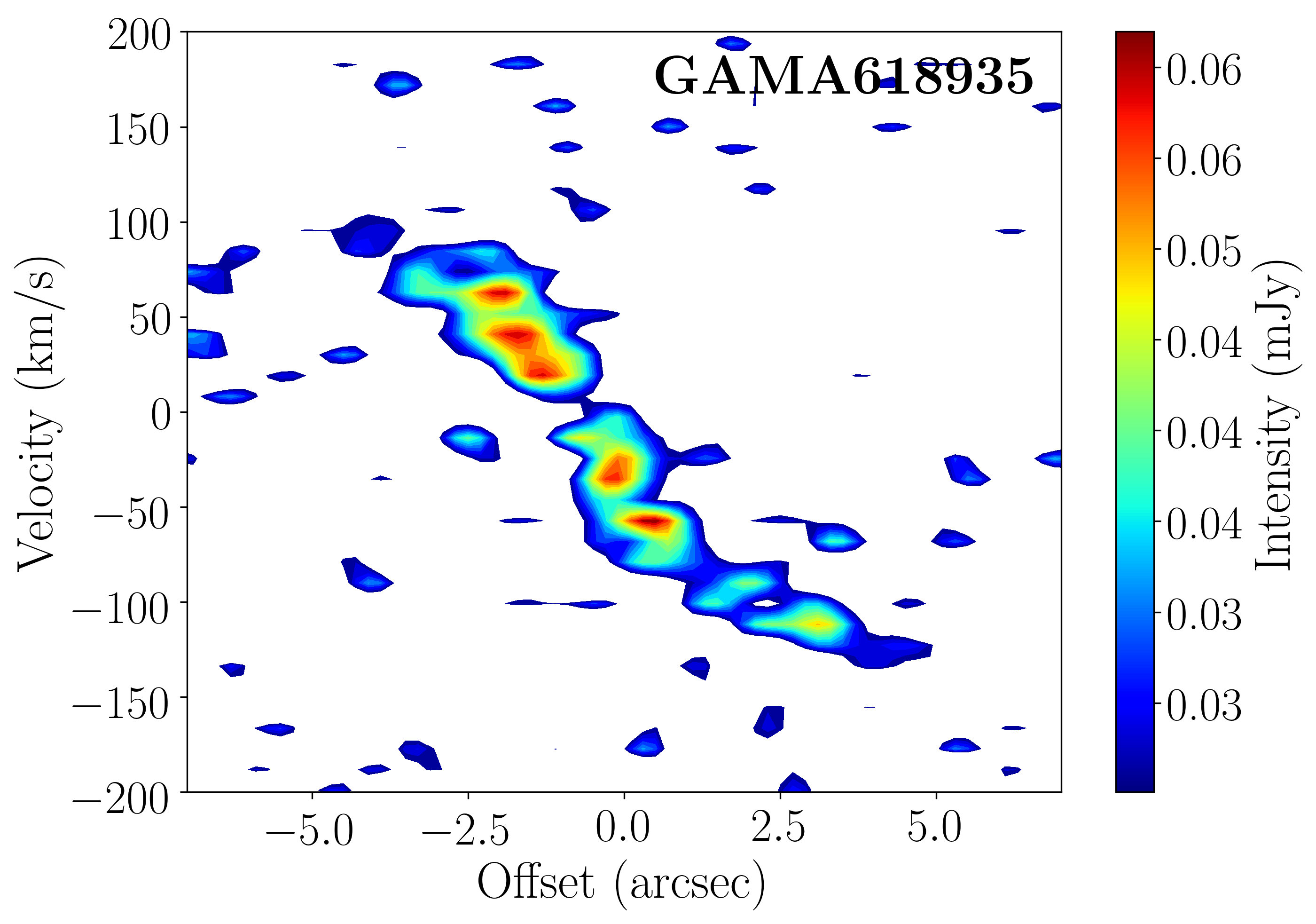}
\end{minipage}
\newline
\centering
\begin{minipage}{.5\textwidth}
  \centering
  \includegraphics[width=.9\linewidth]{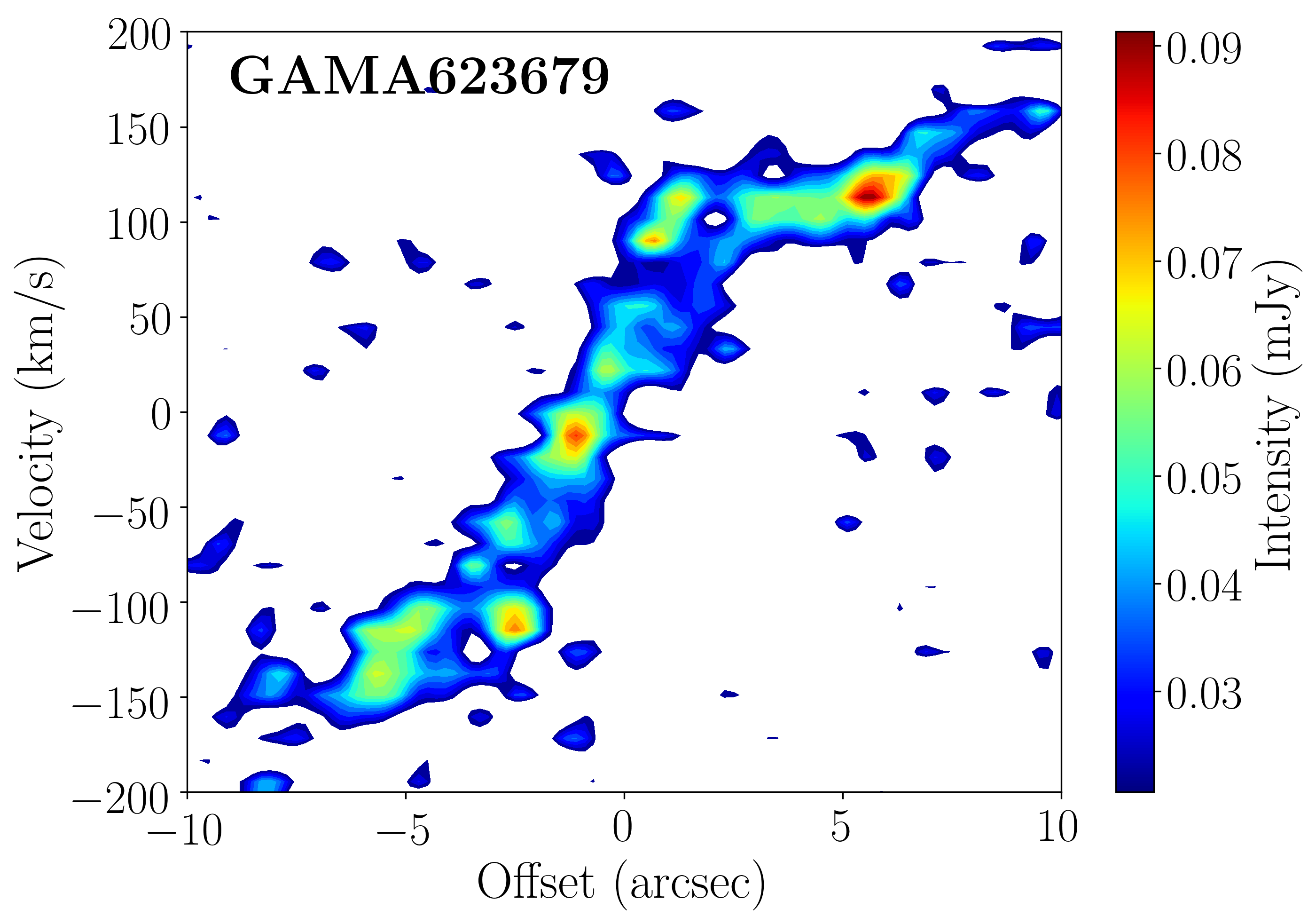}
\end{minipage}%
\caption{continued from Figure~\ref{fig: PVDs_all}}
\end{figure*}

%%%%%%%%%%%%%%%%%%%%%%%%%%%%%%%%%%%%%%%%%%%%%%%%%%

% Don't change these lines
\bsp	% typesetting comment
\label{lastpage}
\end{document}